\documentclass[pra,twocolumn, showpacs, showkeys, secnumarabic, aps, amsmath, amssymb, nofootinbib, superscriptaddress, longbibliography, floatfix, table-of-contents, dblfloatfix]{revtex4-2}

\usepackage[utf8]{inputenc}
\usepackage[pdftex]{graphicx}
\usepackage{mathrsfs}
\usepackage[colorlinks, breaklinks, urlcolor={blue}, linkcolor={blue}, citecolor={blue}]{hyperref}
\usepackage{array}
\usepackage{amsmath}
\usepackage{type1cm}
\usepackage[english]{babel}
\usepackage{lmodern}
\usepackage{microtype}
\usepackage{booktabs}
\usepackage{caption}
\usepackage{braket}
\usepackage{xcolor}
\usepackage{orcidlink}
\usepackage{tikz}
\usepackage{hyperref}
\usepackage{booktabs}
\usepackage{subcaption}
\begin{document}
\title{Phase-sensitive framed-ribbon representation of single-qubit
Pauli measurements in linear cluster states}
\author{Sougata Bhattacharyya\orcidlink{0009-0002-9803-4198}}
\email[]{msc2503121011@iiti.ac.in}
\affiliation{Dept. of Astronomy, Astrophysics and Space Engineering, Indian Institute of Technology, Indore 453552, India}
\author{Sovik Roy \orcidlink{0000-0003-4334-341X} }
\email[]{s.roy2.tmsl@ticollege.org}
\affiliation{Dept. of Mathematics, Techno Main Salt Lake (Engg. Colg.), \\Techno India Group, EM 4/1, Sector V, Salt Lake, Kolkata  700091, India}
\affiliation{Centre of Advanced Studies and Innovation Lab (CASILAB), 18/27 Kali Mohan Road,
		Tarapur, Silchar 788003, INDIA}
\begin{abstract}
\noindent We provide an explicit geometric classification of single-qubit projective measurements on one-dimensional linear cluster states within a topological framework. By establishing an explicit geometrical correspondence between local measurements and topological surgery operations on an associated link model i.e. a \textit{ measurement–surgery correspondence}, we represent the cluster state as a linear Hopf chain. Within this model, measurements in the computational ($Z$) basis act as \textit{topological severance} in case of bulk measurements while \textit{boundary pruning} happens for end measurements of qubits. In contrast, transverse ($X$) basis measurements remove the measured qubit and, rather than fusing its neighbours into a single splice, induce a \textit{geometric stratification} of the residual state into a superposition of two disjoint, correlated segments joined only through real-valued classical correlations. We show that lateral ($Y$) basis measurements instead preserve a single continuous spliced chain while generating intrinsically complex phase factors that are not captured by unframed link models. The unframed linking pattern already distinguishes the $X$- and $Y$-basis outcomes from one another by shape; what it cannot resolve is which of the two possible outcomes occurred within a fixed basis, since both members of either outcome pair share an identical unframed diagram. To resolve this residual ambiguity, we introduce a \textit{framed ribbon representation} in which quantum phases are encoded as geometric twists, with \textit{chiral $\pm 90^\circ$ twists} corresponding to the phases $\pm i$.  This framing yields a phase-sensitive, outcome-resolved \emph{geometric} description of single-shot Pauli measurements on linear cluster states. We stress that the twist angles introduced here are \emph{not} topological invariants: they are geometric labels fixed operationally by the measurement outcome and by the resulting by-product operator, whereas the linking pattern of the underlying unframed link is a genuine topological datum. The present work is deliberately restricted to single-shot single-qubit Pauli measurements on one-dimensional linear cluster states; composition rules for sequential measurements and for classical feedforward are not established here and are identified as the principal open problem.
\end{abstract}
\maketitle
\section{Introduction}\label{sec:intro}
\noindent Measurement-Based Quantum Computation (MBQC) is also known as one-way quantum computing (OQC) as the computation proceeds in a single irreversible direction driven  entirely by measurements. It constitutes an alternative paradigm for quantum information processing (QIP) in which computation is implemented solely through local measurements performed on a highly entangled many-body resource state \cite{r2001,r2003}. In this framework, entanglement is consumed rather than generated during computation, and the logical evolution is encoded in the adaptive choice of measurement bases together with classical feedforward of outcomes. Classical feedforward in QIP is a control strategy where measurement outcomes from one part of a quantum system are used to deterministically select and apply correction operations to another part of the system
\cite{hein2004,nielsen2006,danos2007,browne2005,knill2001}.\\
\noindent A canonical resource for MBQC is the cluster state, a highly entangled quantum state of multiple qubits arranged in a lattice or graph structure \cite{hein2004,hein2006,nest2004,r2001,walther2005}. These states are prepared by initializing qubits in the $|+\rangle$ state and applying Controlled-Z ($CZ$) or (a specific type of Controlled phase gate) gates between nearest neighbours, following the pattern of a graph where vertices represent qubits and edges represent entanglement \cite{nielsen2006,walther2005,prevedel2007}. While the stabilizer formalism and measurement calculus provide a complete algorithmic framework for describing measurements on graph states via graph transformations and Pauli by-products \cite{gottesman1998,schlingemann2002,hein2004,danos2007,
danos2006,nest2004,nielsen2006,browne2005}, they offer limited physical insights into the consequent restructuring of entanglement, especially when measurements introduce complex phases. \\
\noindent We posit that the underlying dynamics of measurements, particularly the emergence of complex phases, are inherently geometric phenomena. These may be characterized not merely through graph-theoretic updates, but through the evolution of the state's intrinsic topological structure. Topological methods have proven valuable in diverse areas of quantum theory, including topological quantum computation \cite{kitaev2003,nayak2008,preskill1999}, knot invariants in quantum field theory \cite{witten1989}, and the geometric classification of quantum phases \cite{wen1990,wen2002,hasan2010,berry1984,zanardi1999,budich2013,vijay2016}. Qualitative analogies between multipartite entanglement and linked/knotted structures have been explored \cite{aravind1997,kauffman2004}, establishing knot theory as a promising language for describing nonlocal correlations and their phase reorganization. However, a precise operational correspondence between quantum measurements and topological surgery has not been systematically established. Recently, we undertook a series of studies investigating the role of topology and measurement-induced phases in quantum systems, emphasizing structural features that go beyond stabilizer and graph-theoretic updates \cite{roy2025(1),roy2025(2)}.\\
\noindent While the measurement calculus and stabilizer graph-transformation rules already provide a complete algebraic description of measurement effects on cluster states, these frameworks do not offer a phase-sensitive geometric representation of measurement-induced correlations. In particular, conventional link based topological analogies capture connectivity changes but remain insensitive to complex phase structure. The principal contribution of the present work is to introduce a framed topological representation in which measurement-induced quantum phases are encoded geometrically as ribbon twists, yielding a phase resolved topological description of single-qubit measurements on linear cluster states (LCS) \citep{r2001,r2003,danos2007,coecke2011,coeckebook}.\\
\noindent In this work, we develop a phase-sensitive framed-topological representation that provides a geometric classification of single-qubit projective measurements on one-dimensional (1D) LCS. By establishing an explicit operational correspondence (explicitly \emph{not} an isomorphism; its domain, codomain and preserved structure are stated in Sec.~IV.1) between local measurement choices and topological surgery operations on an associated link model, we introduce the \textit{measurement-cutting correspondence}. Within this framework, the 1D linear cluster state is represented by a linear Hopf chain, in which each qubit corresponds to a closed loop and each nearest neighbour $CZ$ interaction is encoded as a Hopf link. We emphasize two types of measurements viz. bulk-qubit measurements and end-qubit measurements of LCS. With respect to the structure of $N$-qubit LCS that would be defined in later sections, (i) a bulk-qubit measurement refers to measuring a qubit located inside the chain ($1<k<N$) that has two neighbours and (ii) an end-qubit measurement refers to measuring a qubit located at the boundary of linear cluster (e.g. $Q_1$ or $Q_N$) which has only one neighbour.\\
\noindent We show that projective measurement in the computational ($Z$) basis corresponds to a topological cutting operation that removes the measured component and, for measurements in the bulk qubits, disconnects the chain into independent segments. We call this \textit{topological severance} or simply \textit{severance}. On the other hand, the end-qubit measurements yield \textit{unlinking} the terminal ring.\\ 
\noindent In contrast, measurement in the transverse ($X$) basis removes the measured component and, at the level of the residual quantum state, induces a \textit{geometric stratification}: rather than fusing its neighbouring components into a single splice, the residual state separates into a superposition of two disjoint, correlated segments joined only through real-valued classical correlations (Sec.~VII). This construction furnishes a direct geometric representation of the standard graph-state measurement update rules, and corrects a simpler ``single splice'' picture that does not survive explicit computation (Sec.~VII.1).\\
\noindent A central result of our present work concerns measurements performed in the lateral ($Y$) basis. Unlike $X$-basis measurements, which stratify the chain into two disjoint branches, $Y$-basis measurements preserve a single continuous spliced chain while generating intrinsically complex phase factors of $\pm i$ that are not captured within conventional unframed link models. The unframed linking pattern therefore already distinguishes the $X$- and $Y$-basis outcomes from one another by shape; what it cannot resolve is which of the two possible outcomes occurred \emph{within} a fixed basis -- $|+i\rangle_k$ versus $|-i\rangle_k$ for $Y$, or the two branch-labelling patterns for $X$ -- since both members of either pair share an identical unframed diagram. This is the residual ambiguity in unframed topological representations that motivates the present construction. To resolve it, we introduce a \textit{framed ribbon representation}, drawing on the mathematical theory of framed knots \cite{adams2004,kauffman2004}, in which quantum correlations are encoded not only through connectivity but also via geometric twist. Within this extended framework, real-valued correlations correspond to untwisted or orientation-flipped ribbons, while complex phases are represented by \textit{chiral $\pm 90^\circ$ twists}. This mapping from quantum phase to ribbon twist parallels the correspondence between quantum evolution and geometric phases \cite{berry1984,aharonov1987}, and yields a complete, phase-sensitive geometric description of all single-qubit projective measurements on the LCS. The present results should therefore be viewed as a geometric reformulation of known measurement update rules, augmented by a new phase-sensitive framed-ribbon structure that encodes measurement-induced Pauli frame corrections.\\
\noindent \textit{Scope and contributions.} Because the framework proposed here is geometric rather than algorithmic, we state at the outset precisely what is and what is not claimed.\\
\noindent \textit{What this work contributes.} (i) An outcome-resolved dictionary that assigns to each single-shot Pauli measurement on a $1$D LCS a surgery operation on a framed link, together with a geometric label (a ribbon twist) for the phase of the induced by-product operator. (ii) The observation, made precise in Sec.~IX, that an \emph{unframed} link model retains only the pairwise linking pattern of the residual state and is therefore blind to the \emph{sign} of the induced by-product operator: the two possible outcomes of a fixed-basis bulk measurement share an identical unframed diagram, even though $X$- and $Y$-basis measurements are themselves already distinguishable from one another by shape (geometric stratification into two branches versus a single spliced chain). (iii) A fully explicit set of derivations for all six single-shot Pauli outcomes (Appendices A--C), including the exact residual state produced by an $X$-basis bulk measurement, which is \emph{not} the linear cluster state obtained by naively joining the two neighbours (Sec.~VII.1).\\
\noindent \textit{What this work does not claim.} The results are a geometric re-encoding of measurement-update rules that are already complete within the stabilizer formalism and the measurement calculus \cite{gottesman1998,danos2007,hein2004}. We do not derive new verifiable predictions, new invariants, or new protocol constraints, and we make no claim of computational advantage. In particular, relative to the ZX-calculus \cite{coecke2011,coeckebook} -- which is a complete algebraic rewriting system in which phases appear as scalar labels attached to spiders -- the framed-ribbon picture offers no additional rewriting power. Its only distinguishing feature is representational: the same phase datum is realised as an \emph{embedded geometric attribute} (the handedness of a ribbon in $\mathbb{R}^3$) rather than as a symbolic label, so that Pauli-frame corrections become visible as spatial handedness. Whether this representation carries any computational benefit is left open.\\
\noindent \textit{What would be required to generalise.} A framed-ribbon description beyond the present setting would need, at minimum: (a) a composition rule specifying how the framings induced by successive measurements combine, together with a geometric analogue of classical feedforward -- neither of which is established here (see the \emph{Limitations} paragraph of Sec.~XII); (b) a replacement for the ribbon at lattice sites of degree greater than two, since a $2$D or $3$D cluster state has vertices with three or more incident edges, so that the connecting object is a branched surface rather than a band; and (c) a continuous, rather than quantised, framing parameter in order to accommodate non-Pauli measurement angles. The parts of the present construction that are purely local to the measured vertex and its neighbourhood -- the twist dictionary of Sec.~X.1 and the $Z$/$X$/$Y$ trichotomy -- are expected to survive verbatim at degree-two vertices of higher-dimensional lattices; the parts that rely on the chain being one-dimensional, in particular the identification of the residual state as a single spliced chain, are not.\\
\noindent The remainder of this paper is organized as follows. Sec.~$II$ introduces the foundational concepts required for the analysis, including projective quantum measurements, Schmidt rank based entanglement characterization, and essential notions from knot and link topology. In Sec.~$III$ we define the one-dimensional linear cluster state and establish the notation used throughout the work. Sec.~$IV$ develops the operational topological framework by introducing the measurement--surgery correspondence and the linear Hopf chain representation of cluster states. Sec.~$V$ analyses the topological effects of single-qubit measurements in the $Z$ basis, Sec.~$VI$ those in the $Y$ basis, and Sec.~$VII$ those in the $X$ basis, highlighting the severing, splicing, and stratifying character of each. Sec.~$VIII$ summarizes the mathematical outcomes of all six measurement scenarios in tabular form, while Sec.~$IX$ discusses the residual ambiguity that persists between the two outcomes of a fixed measurement basis in conventional, unframed link representations. Sec.~$X$ introduces the framed ribbon model that resolves this ambiguity through a phase-sensitive geometric description, Sec.~$XI$ applies this framework to classify all measurement-induced transformations in detail, and Sec.~$XII$ concludes with implications of the proposed geometric framework and directions for future research.
\section{Foundational Concepts}
\subsection{Quantum Projective Operations}
\noindent A projective quantum operation is characterized by an \textit{observable}, represented by a Hermitian operator $\mathcal{M}$ acting on the Hilbert space of the system. This operator admits a spectral resolution $\mathcal{M} = \sum_m m P_m$, where the eigenvalues $m$ correspond to distinct measurement outcomes, and each $P_m$ is the orthogonal projector onto the eigenspace associated with $m$. These projectors satisfy Hermiticity $P_m = P_m^\dagger$, idempotence $P_m^2 = P_m$, and the completeness relation $\sum_m P_m = I$ \cite{nielsen2010}.\\
\noindent For a tripartite quantum state $\mid \psi_{ABC} \rangle$, a projective measurement performed exclusively on subsystem $A$ in its standard basis $\lbrace { \mid 0 \rangle, \mid 1 \rangle }\rbrace $ is defined by the set of operators ${P^A_k}$ with $k \in \lbrace 0, 1\rbrace$, where $P^A_k = \mid k \rangle_A\langle k \mid \otimes I_B \otimes I_C$. According to the Born rule, the probability of recording outcome $k$ is
\begin{eqnarray}
\label{pkprojector}
p(k) = \langle \psi_{ABC} \mid P^A_k \mid \psi_{ABC} \rangle.
\end{eqnarray}
Conditioned on this outcome, the physical state of the remaining subsystems $B$ and $C$ collapses to the post-measurement state
\begin{eqnarray}
\label{subsyststate}
\mid \psi^{(k)}_{BC} \rangle = \frac{P^A_k \mid \psi_{ABC} \rangle}{\sqrt{p(k)}}.
\end{eqnarray}
The entanglement characteristics of this resultant bipartite state $\mid \psi^{(k)}_{BC} \rangle$ form the central object of our investigation.
\subsection{The Schmidt Rank and Entanglement Quantification}
\noindent For any pure bipartite state $\mid \psi_{BC} \rangle$ shared between parties $B$ and $C$, there exists a canonical representation known as the Schmidt decomposition \cite{peres1995} given as
\begin{eqnarray}
\label{schmidt}
\mid \psi_{BC} \rangle = \sum_{i=1}^R \sqrt{\lambda_i} \mid \phi^i_B \rangle \otimes \mid \chi^i_C \rangle,
\end{eqnarray}
where the sets ${ \mid \phi^i_B \rangle }$ and ${ \mid \chi^i_C \rangle }$ constitute orthonormal bases for the respective subsystems. The coefficients $\lambda_i$ are non-negative real numbers satisfying $\sum_i \lambda_i = 1$, and the integer $R$ denotes the \textit{Schmidt rank}. The Schmidt rank serves as a fundamental discrete measure of bipartite entanglement \cite{nielsen2010,peres1995}. Thus we have, 
\begin{itemize}
\item If $R=1$, the state is separable, representing a simple product state with no entanglement.
\item If $R>1$, the state is entangled. The specific distribution of the $\lambda_i$ further quantifies the degree of entanglement.
\item A state with $R=2$ and $\lambda_1 = \lambda_2 = \frac{1}{2}$ corresponds to a maximally entangled bipartite state, such as a Bell pair.
\end{itemize}
\noindent In our protocol, we compute the Schmidt rank of the conditional state $\mid \psi^{(k)}_{BC} \rangle$ to classify the entanglement status between qubits $B$ and $C$ following the local measurement on the qubit $A$.
\subsection{Topology of Links and Knots}
\noindent Knot theory provides a mathematical framework for studying the embedding of closed curves within three-dimensional space. A \textit{knot} is defined as a single, non-self-intersecting closed loop, considered non-trivial if it cannot be smoothly deformed into a simple circle (the \textit{unknot}). A \textit{link} generalises this concept to a collection of one or more such knots, which may be interwoven \cite{adams2004}. Two links are deemed topologically equivalent if one can be continuously transformed into the other without any cutting or piercing of strands. The topological analogue to a local quantum measurement is the \textit{deletion} of a single link component. 
\subsection{The Hopf Link}
\noindent The \textit{Hopf link} (See Fig.~$1$) is the minimal non-trivial link comprising of two circles. Its defining feature is that each circle is linked with the other. If one component is cut and removed, the remaining component becomes a trivial \textit{unknot}, representing a state of complete \textit{unlinking} \cite{adams2004}.
\begin{figure}[t]
\centering
\includegraphics[width=0.3\textwidth]{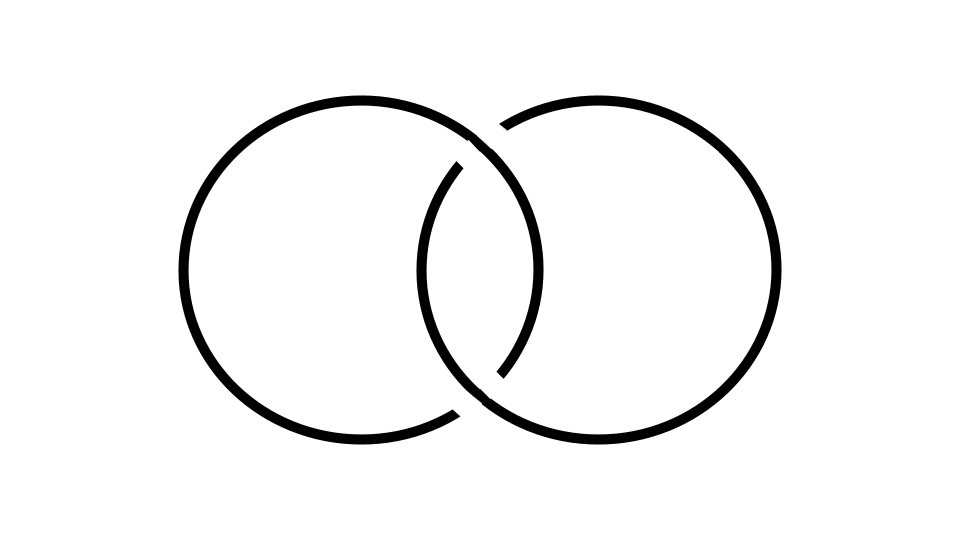}
\caption{The Hopf link}
\label{fig:hopf_link}
\end{figure}
\section{The Linear Cluster State}
\subsection{Definition}
\noindent The $N$-qubit linear cluster state $|C_N\rangle$ is a specific instance of a graph state corresponding to a 1D line graph. It is canonically defined by initializing a register of $N$ qubits in the product state $|+\rangle^{\otimes N}$ and applying a Controlled-Phase ($CZ$) gate between all adjacent pairs $(i, i+1)$, where the suffix $i$ denotes the $i^{th}$ qubit of the state. Explicitly, the state can be written in the computational basis as
\begin{eqnarray}
\label{cluster}
|C_{N}\rangle &= \frac{1}{2^{N/2}} \sum_{x_{1}, \dots, x_{N} \in \{0,1\}} (-1)^{\sum_{i=1}^{N-1} x_{i}x_{i+1}} |x_{1} \dots x_{N}\rangle, \nonumber\\
\end{eqnarray}
where the exponent $\sum_{i=1}^{N-1} x_{i}x_{i+1}$ encodes the phase correlations characteristic of the 1D Ising interaction. 1D Ising interaction refers to a nearest neighbour two qubit interaction whose phase structure is mathematically identical to the interaction term in the 1D Ising model. In the classical (where spins take classical values $s_i = \pm 1$) or in the quantum (where spins are represented by quantum operators such as Pauli matrices) Ising model, the interaction energy between neighbouring spins $i$ and $i+1$ is proportional to $\sigma_z^{(i)}\sigma_z^{(i+1)}$ for nearest neighbours along a 1D chain. In LCS, qubits produce phase factors $(-1)^{x_ix_{i+1}}$ which has the same mathematical structure as the Ising nearest neighbour coupling term. Thus, the cluster state phase pattern can be viewed as arising from an effective 1D Ising type interaction between adjacent qubits.\\
\noindent The LCS state defined in Eq.(\ref{cluster}) possesses maximal connectedness and persistent entanglement, making it the universal resource for OQC \cite{hein2006,schlingemann2002}. For our further topological analysis of the state, we shall use notation such as $\vert C_L\rangle$ and $\vert C_R\rangle$ which are the left and right segments of cluster state $\vert C_N\rangle$ in the following sections. 
\section{Operational Topology: The Measurement-Cutting Correspondence}
\noindent To rigorously classify the measurement dynamics of the cluster state, we adopt an operational-topological framework that bridges quantum information theory with knot theory. In this paradigm, the abstract algebraic process of quantum measurement is mapped directly to a concrete geometric operation called \textit{Topological Surgery}.
\subsection{The measurement--surgery correspondence}
\noindent The fundamental correspondence relies on the \textit{destructive nature} of projective measurement. When a qubit within an entangled system is measured in a standard basis, it is projected onto a product state, effectively disentangling it from the rest of the wavefunction.
\begin{itemize}
    \item \textit{Quantum Operation:} A local projective measurement $\mathcal{M}_{proj}$ on qubit $Q_k$ collapses the superposition, factoring the global state $|\Psi\rangle$ into a product of the measured qubit and the residual system i.e. $|\phi\rangle_k \otimes |\Psi'\rangle_{\text{rest}}$. Consequently, the qubit $Q_k$ ceases to be a resource for entanglement.
    \item \textit{Topological Analogue:} This corresponds to the physical \textsc{Cutting (or Deletion)} of the ring $k$ from the topological chain. The ring is removed from the manifold, and the stability of the remaining structure depends entirely on how the surviving end-points are connected.
\end{itemize}
\noindent Thus we define a map $\mathcal{T}$,
\begin{equation}
\label{eq:corr}
    \mathcal{T}:\;\; \mathcal{M}^{(k)}_{\mathcal{B},s} \;\longmapsto\; \mathcal{S}\big(\mathcal{B},k,s\big),
\end{equation}
\noindent and we state explicitly what is, and what is not, being asserted for $\mathcal{T}$.
\begin{itemize}
\item \textit{Domain.} The finite set of single-shot single-qubit measurement events $\mathcal{M}^{(k)}_{\mathcal{B},s}$ on $|C_N\rangle$, labelled by a Pauli basis $\mathcal{B}\in\lbrace X,Y,Z\rbrace$, a site $k\in\lbrace 1,\dots,N\rbrace$ and an outcome $s=\pm 1$. It contains $6N$ elements and no composite (sequential) events.
\item \textit{Codomain.} The set $\mathcal{S}$ of surgery operations on the framed link $\mathcal{L}(C_N)$ associated with $|C_N\rangle$: deletion of a component, splicing of two components, assignment of a framing $\theta\in\lbrace 0^\circ,\pm 90^\circ,180^\circ\rbrace$ to a ribbon, and formation of a branch-superposed pair of link diagrams (\emph{stratification}), used when a single-shot outcome does not uniquely fix the residual connectivity (Sec.~VII).
\item \textit{What is preserved.} $\mathcal{T}$ reproduces (i) the pairwise linking pattern of the residual state, in the coarse-grained sense that two components are declared linked whenever the corresponding qubits remain quantum-mechanically correlated across the relevant cut, (ii) once the framing is included, the \emph{phase} of the outcome-dependent by-product operator, i.e.\ its diagonal part $\{I,Z,S,S^\dagger\}$, and (iii) for a stratifying outcome, the branch-conditioned linking pattern of each sector $b\in\{0,1\}$ taken separately, together with the relative weight and Pauli dressing distinguishing the two branches.
\item \textit{What is not preserved, and not claimed.} $\mathcal{T}$ is not asserted to be injective, surjective, or invertible; it is not functorial, since composition of measurement events is not defined on the domain; and it does not determine the local-Clifford class of the residual graph state, which is strictly finer information than the linking pattern (the \emph{Limitations} paragraph of Sec.~XII makes this explicit). A by-product operator with a non-diagonal Pauli component is not representable by framing alone.
\end{itemize}
\noindent Accordingly, $\mathcal{T}$ is an \emph{operational dictionary} between measurement events and surgery operations, and we use the words ``correspondence'' and ``dictionary'' for it throughout, rather than ``isomorphism''.
\subsection{Measurement as a Stress Test}
\noindent This correspondence allows us to treat quantum measurement as a \textit{topological stress test}. By mathematically performing the measurement and analyzing the residual state, we can determine the connectivity of the underlying topology.
However, the LCS exhibits a richer phenomenology than simple binary connectivity. While the \textit{Cut} is the fundamental operation for all measurements, the \textit{basis} of the measurement ($Z, X,$ or $Y$) \textit{imposes different boundary conditions} on the surgery. These conditions determine whether the cut-ends (a) remain severed, (b) are fused back together, or (c) acquire additional geometric structure. Accordingly, distinguishing the topological effects of different measurement bases requires examining not only whether the chain remains connected but also \textit{how} the geometry of the link is modified by the operation.
\subsection{The Topological Analogy: Linear Hopf Chain}
\begin{figure}[t]
\centering
\begin{subfigure}{0.3\textwidth}
    \centering
    \includegraphics[width=\linewidth]{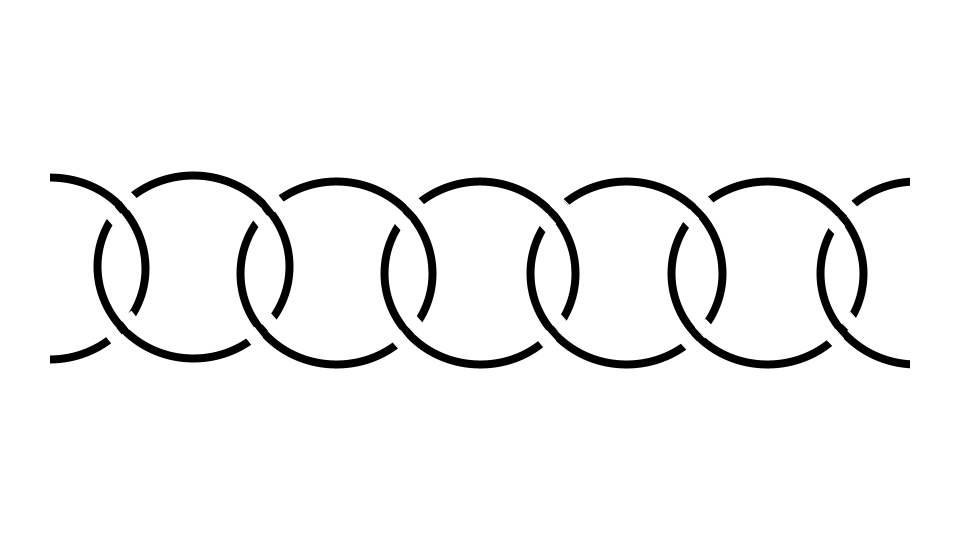}
    \caption{2D Schematic}
    \label{fig:hopf_chain_2d}
\end{subfigure}
\hfill
\begin{subfigure}{0.3\textwidth}
    \centering
    \includegraphics[width=\linewidth]{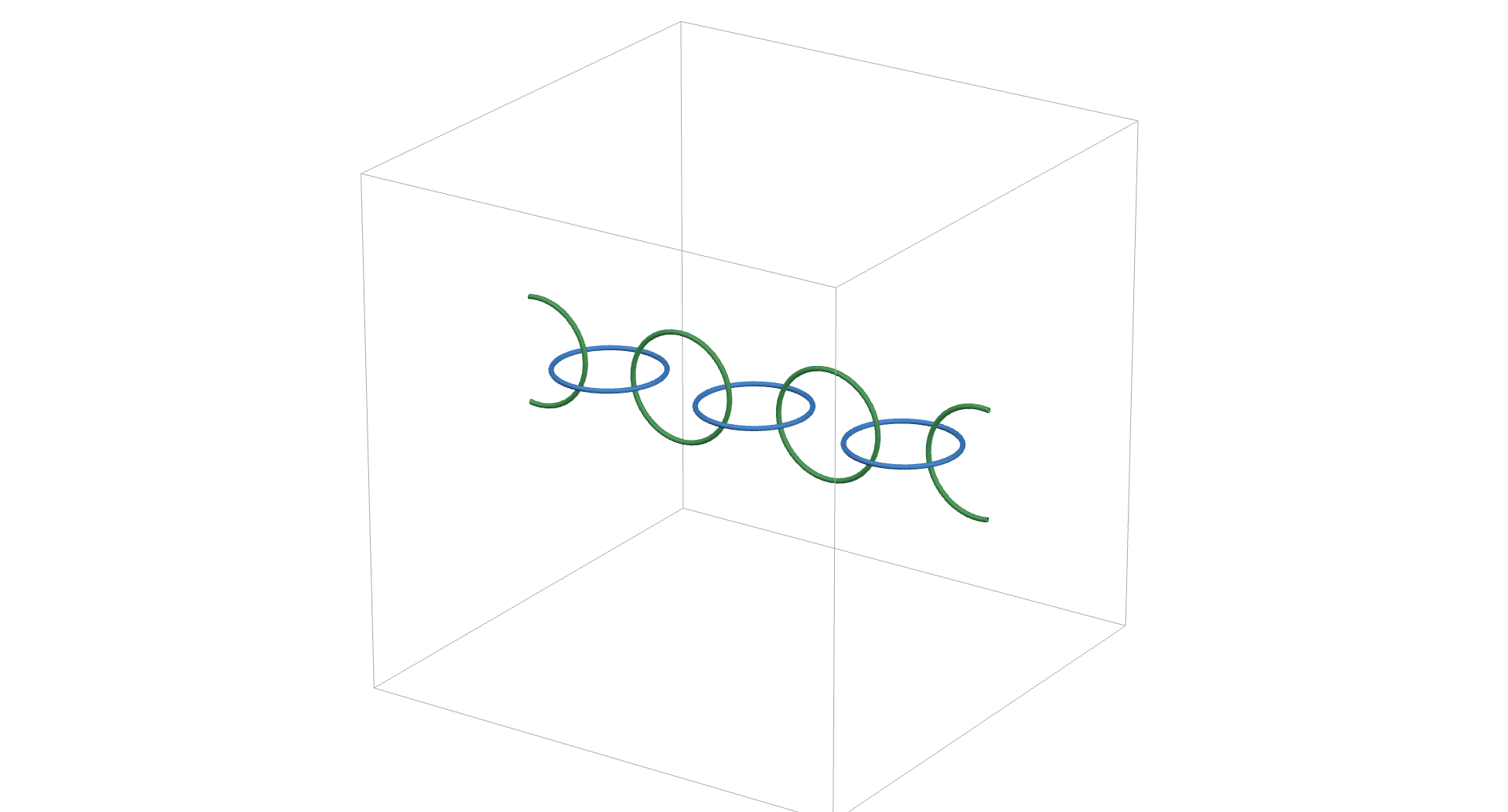}
    \caption{3D Visualization}
    \label{fig:hopf_chain_3d}
\end{subfigure}
\caption{\textbf{The Linear Hopf Chain Model.}
\textbf{(a)} 2D diagrammatic representation of the 1D cluster state connectivity.
\textbf{(b)} 3D visualization of the same state as a series of physically interlinked rings. In this dual picture, quantum measurements correspond to topological operations (cutting or splicing) performed on these links.}
\label{fig:hopf_chain}
\end{figure}

\noindent We propose a direct topological visualization for the 1D LCS $|C_N\rangle$ of Eq.~(\ref{cluster}), which we refer to as the \textit{Linear Hopf Chain}, drawing inspiration from earlier topological approaches to quantum entanglement \cite{aravind1997,kauffman2004}. Rather than viewing the state purely as an abstract graph, we map the quantum resource onto a \textit{physical assembly of interlocked components}.\\
\noindent In this framework, the quantum correlations are translated into topological constraints as follows:
\begin{itemize}
    \item \textit{Qubits as Topological Loops:} We view each physical qubit $Q_i$ as a distinct and closed topological ring.
    \item \textit{Interactions as Hopf Links:} The entanglement bond between two neighbouring qubits is represented physically as a \textsc{Hopf Link}. The two rings pass through each other exactly once, establishing a robust connection (as shown in Fig.~\ref{fig:hopf_link}).
    \item \textit{Linear Connectivity:} The strictly local nature of the cluster state means that only adjacent rings are linked. Distant rings do not interact, resulting in a linear chain-like topology.
\end{itemize}
This model provides a concrete interpretation of the quantum system, as visualized in Fig. \ref{fig:hopf_chain}.\\
\noindent  The \textit{flow of information} along the cluster state is in direct correspondence with the \textit{mechanical connectivity} of the chain. Consequently, a local measurement on a qubit is topologically equivalent to \textit{cutting} and \textit{removing} that specific ring from the assembly.
% ==========================================
% FIGURE: LINEAR HOPF CHAIN - SIDE BY SIDE
% ==========================================

\section{The $Z$-Basis Measurement: Topological Severance} 
\noindent We begin by analyzing projective measurements on the LCS in the computational ($Z$) basis. The simplest operation in the measurement calculus is the projection $\mathcal{M}_Z = \{|0\rangle, |1\rangle\}$. While other measurements reshape entanglement connectivity, the $Z$-basis measurement is fundamentally destructive. We classify this operation as a \textit{Topological Cut}, serving as the baseline for topological surgery. By performing this removal, one can observe how the remaining structure changes, thereby revealing the role that the specific node played in maintaining the connectivity. 

\subsection{Internal Severance (Bulk Measurement)}
\begin{figure}[t]
\centering
\begin{subfigure}{0.3\textwidth}
    \centering
    \includegraphics[width=\linewidth]{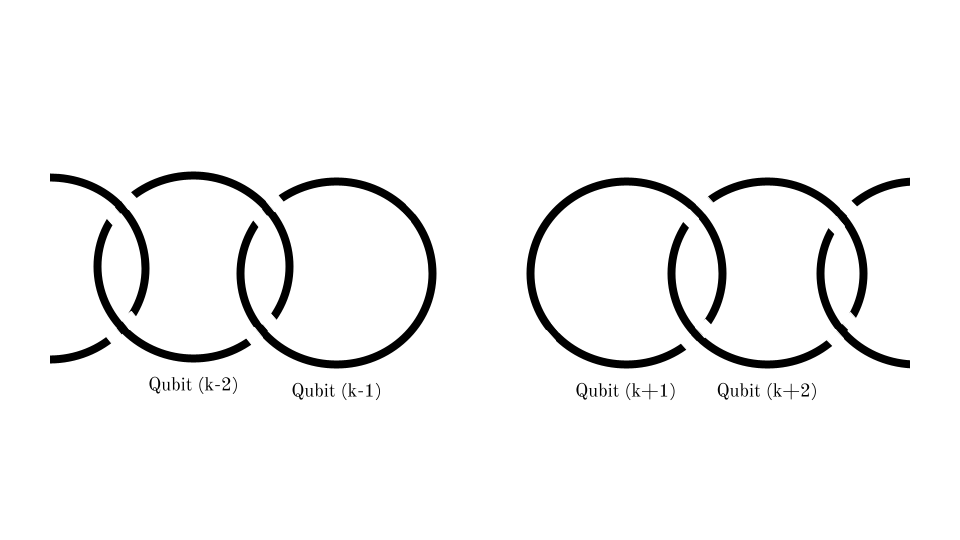}
    \caption{Bulk Measurement (2D Schematic)}
    \label{fig:z_bulk_2d}
\end{subfigure}
\hfill
\begin{subfigure}{0.3\textwidth}
    \centering
    \includegraphics[width=\linewidth]{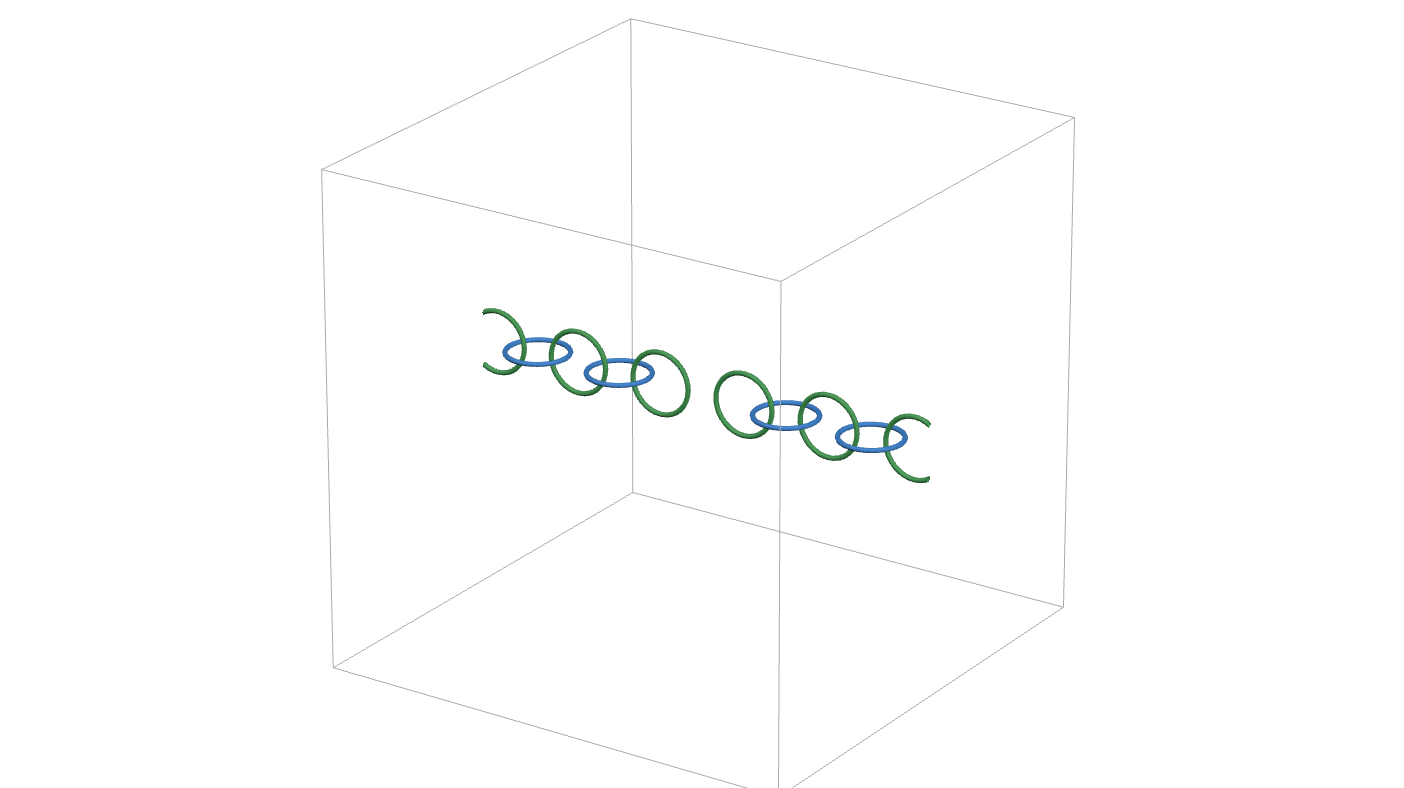}
    \caption{Bulk Measurement (3D Visualization)}
    \label{fig:z_bulk_3d}
\end{subfigure}
\caption{\textbf{Topological Cutting in the Z-Basis - Internal Severance.}
Measuring a bulk qubit ($Q_k$) physically removes the central ring. Unlike Splicing (which fuses neighbours), Severance destroys the path, leaving two unlinked, independent segments ($R=1$).}
\label{fig:z_basis_bulk}
\end{figure}

\noindent Mathematically, measuring a bulk qubit $k$ ($1 < k < N$) in the $Z$-basis disentangles it entirely from the graph state. As derived explicitly in \textsc{Appendix A.3}, the post-measurement state factorizes into a tensor product of two independent cluster states:
\begin{eqnarray}
\label{zbulk}
|\Psi_{\text{final}}\rangle = \sigma_{\text{bound}} \left( |C_{L}\rangle_{1\dots k-1} \otimes |C_{R}\rangle_{k+1\dots N} \right),
\end{eqnarray}
where $\sigma_{\text{bound}}$ represents the Pauli $Z$ byproduct operators $Z_{k-1}Z_{k+1}$ that arise only for the outcome $|1\rangle_k$. Crucially, the Schmidt Rank $R$ across the partition collapses to unity (i.e., $R = 1$), indicating a total loss of quantum correlations between the left ($C_L$) and right ($C_R$) segments.\\
\noindent Topologically, this maps to the physical removal of the central ring $k$ from the Hopf chain, as illustrated in Fig. \ref{fig:z_basis_bulk}. The $Z$ measurement completely deletes the entanglement bridge. 

% ==========================================
% FIGURE: BULK Z-BASIS CUTTING - SIDE BY SIDE
% ========================================== 

\subsection{Boundary Pruning (End Measurement)}
\begin{figure}[t]
\centering
\begin{subfigure}{0.3\textwidth}
    \centering
    \includegraphics[width=\linewidth]{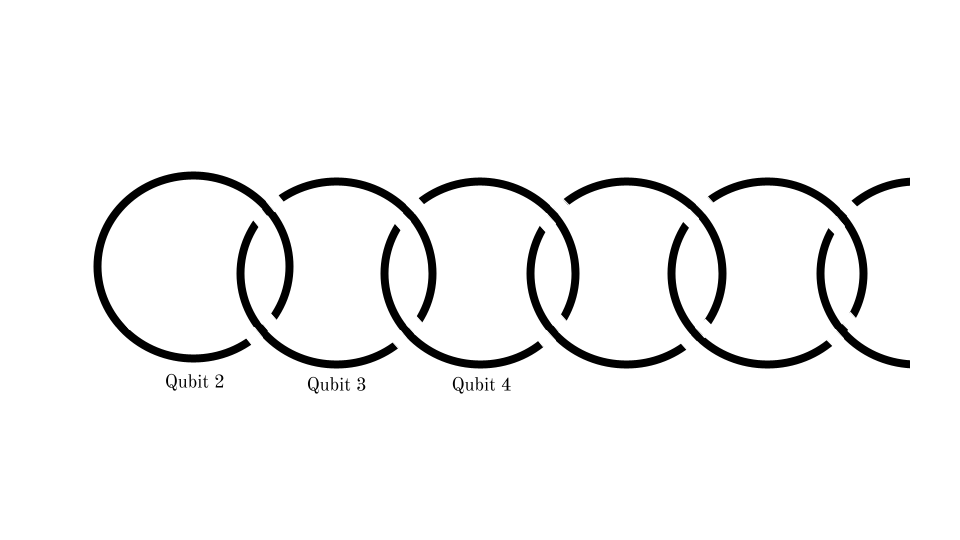}
    \caption{End Measurement (2D Schematic)}
    \label{fig:z_end_2d}
\end{subfigure}
\hfill
\begin{subfigure}{0.3\textwidth}
    \centering
    \includegraphics[width=\linewidth]{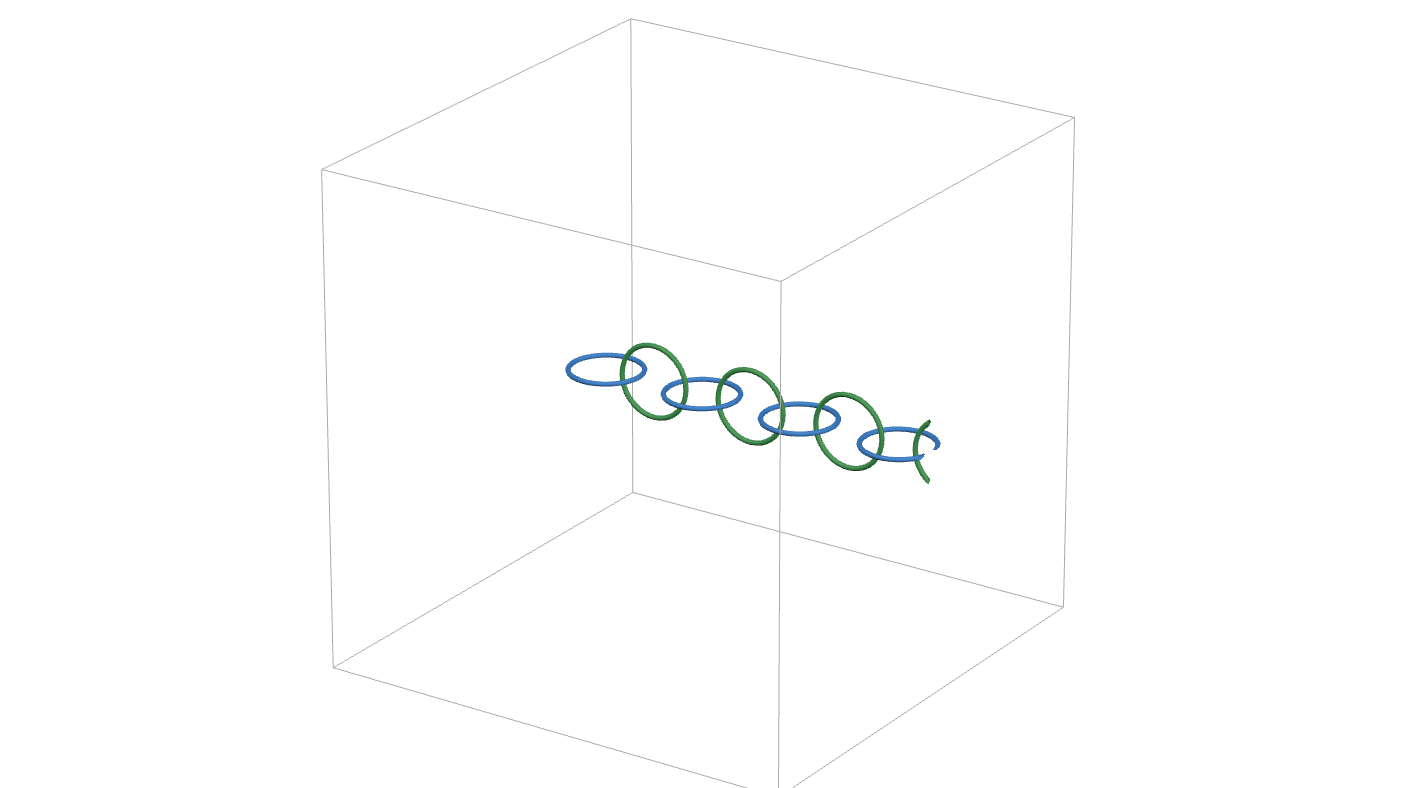}
    \caption{End Measurement (3D Visualization)}
    \label{fig:z_end_3d}
\end{subfigure}
\caption{\textbf{Topological Cutting in the $Z$-Basis - Boundary Pruning.} 
Measuring an end qubit ($Q_1$) removes the terminal ring. The chain remains connected but shortens ($N \to N-1$).}
\label{fig:z_basis_end}
\end{figure}

\noindent When the measurement targets an end-point qubit ($k=1$), the topology is not severed into two, but rather \textit{shortened}. As shown in \textsc{Appendix A.2}, measuring qubit $1$ yields a residual state on the remaining $N-1$ qubits:
\begin{equation}
|\Psi_{\text{final}}\rangle = \sigma_{2}^{\mu} |C_{N-1}\rangle_{2\dots N},
\end{equation}
where $\sigma_{2}^{\mu}$ denotes an outcome dependent Pauli operator acting on qubit $2$. For a $Z$-basis measurement, only two of the four Pauli labels occur: $\mu = 0$ ($\sigma^{0}=I$) for $|0\rangle_1$, and $\mu = z$ ($\sigma^{z}=Z$) for $|1\rangle_1$. \\
\noindent Topologically, this corresponds to unlinking and discarding the terminal ring. The chain remains continuous but is reduced in length by one unit ($N \to N-1$). We refer to this operation as \textit{topological pruning}, depicted in Fig. \ref{fig:z_basis_end}.

% ==========================================
% FIGURE: END Z-BASIS CUTTING - SIDE BY SIDE
% ==========================================

\section{The $Y$-Basis Measurement: Topological Splicing and Complex Phases}
\noindent Having established the destructive nature of the $Z$-basis, we now turn to the lateral basis, $\mathcal{M}_Y = \{|+i\rangle, |-i\rangle\}$. This measurement is fundamentally constructive: it preserves the one-dimensional topological continuity of the chain while embedding complex phases via \textit{twisted splicing}.

\subsection{Internal Connectivity (Bulk Measurement)}
\noindent Mathematically, measuring a bulk-qubit $k$ in the $Y$-basis preserves the entanglement chain perfectly. As derived in \textsc{Appendix C.3}, this operation removes site $k$, but fuses the neighbours $k-1$ and $k+1$ into a connected state. The residual state \emph{is} exactly the linear cluster state on $N-1$ qubits with the new bond $k-1\leftrightarrow k+1$, dressed by local phase gates:
\begin{eqnarray}
\label{ybulk}
|\Psi_{s}\rangle = \big(S_{k-1}^{\,s} \otimes S_{k+1}^{\,s}\big) \, |C_{N-1}^{(k-1 \leftrightarrow k+1)}\rangle,\qquad s=\pm 1,
\end{eqnarray}
where $S = \text{diag}(1, i)$. The exponent $s=+1$ corresponds to the outcome $|+i\rangle_k$ and $s=-1$ to $|-i\rangle_k$. Crucially, the Schmidt Rank ($R$) across the partition is $2$, confirming that the 1D chain remains physically intact.

% ==========================================
% FIGURE: Y-BASIS AMBIGUITY - SIDE BY SIDE
% ==========================================
\begin{figure}[t]
\centering
\begin{subfigure}{0.3\textwidth}
    \centering
    \includegraphics[width=\linewidth]{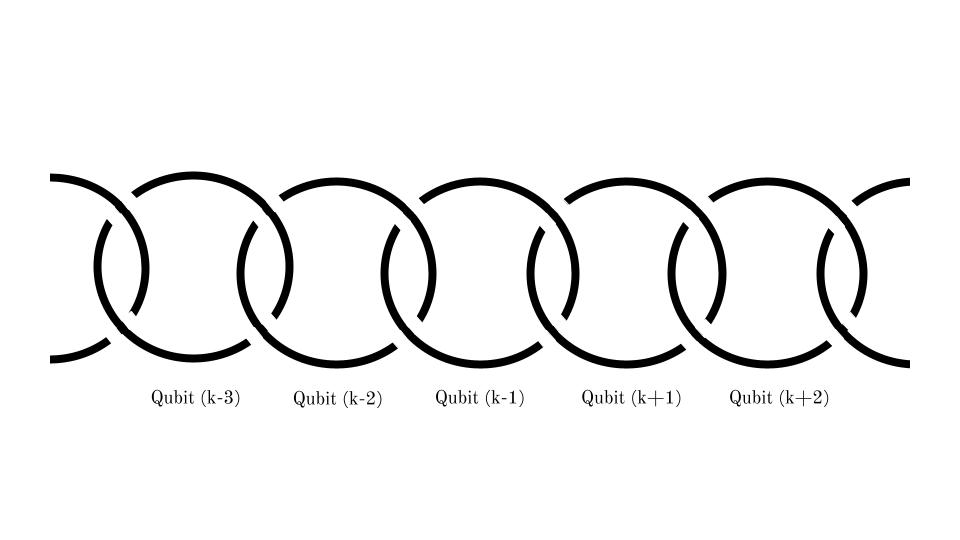}
    \caption{Bulk Measurement (2D Schematic)}
    \label{fig:y_amb_bulk_2d}
\end{subfigure}
\hfill
\begin{subfigure}{0.3\textwidth}
    \centering
    \includegraphics[width=\linewidth]{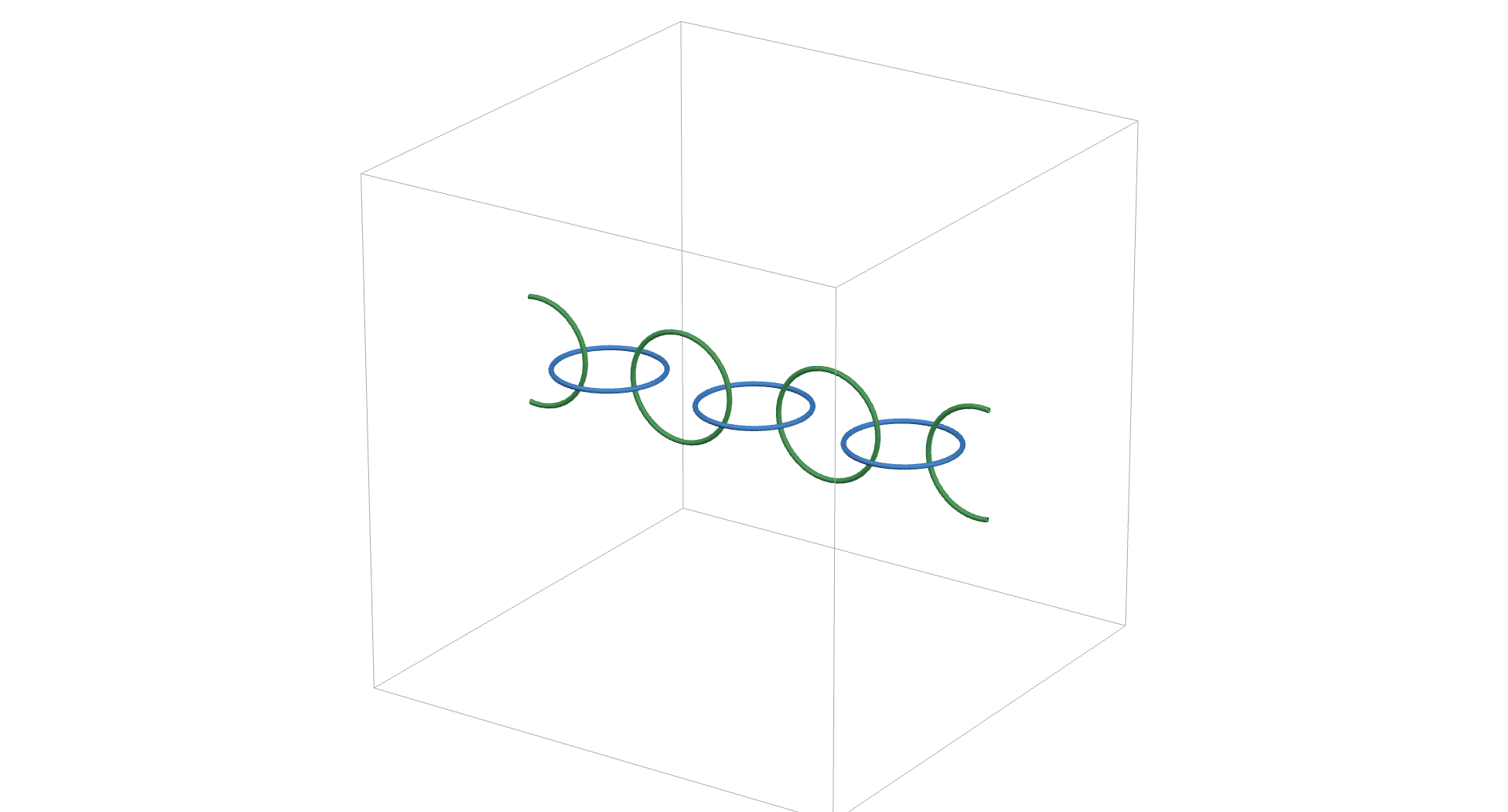}
    \caption{Bulk Measurement (3D Visualization)}
    \label{fig:y_amb_bulk_3d}
\end{subfigure}
\caption{\textbf{The $Y$-Basis Bulk Splice.} Measuring a bulk qubit in the $Y$ basis fuses the neighbours ($k-1,k+1$) into a single continuous spliced chain (Eq.~(\ref{ybulk})). At the level of unframed connectivity, this diagram is identical for both outcomes $|+i\rangle_k$ and $|-i\rangle_k$: the model is \textit{phase-blind} to the sign of the induced twist, a residual ambiguity resolved only by the framed-ribbon representation of Sec.~X.}
\label{fig:y_basis_ambiguity_bulk}
\end{figure}

\subsection{Boundary Phase (End Measurement)}
\noindent When measuring an endpoint qubit ($Q_1$), the operation behaves like a \textit{boundary twist}. As shown in \textsc{Appendix C.2}, the measurement shortens the chain but imprints a local phase gate onto the new boundary qubit $Q_2$:
\begin{eqnarray}
\label{s2boundaryphase}
|\Psi_{s}\rangle = S_{2}^{\,s} |C_{N-1}\rangle_{2\dots N},\qquad s=\pm 1 .
\end{eqnarray}
This contrasts with the $Z$-basis (which simply removes $Q_1$ without phase) and the $X$-basis (which decouples $Q_2$ entirely, as we will see next).

% ==========================================
% FIGURE: Y-BASIS END AMBIGUITY - SIDE BY SIDE
% ==========================================
\begin{figure}[t]
\centering
\begin{subfigure}{0.3\textwidth}
    \centering
    \includegraphics[width=\linewidth]{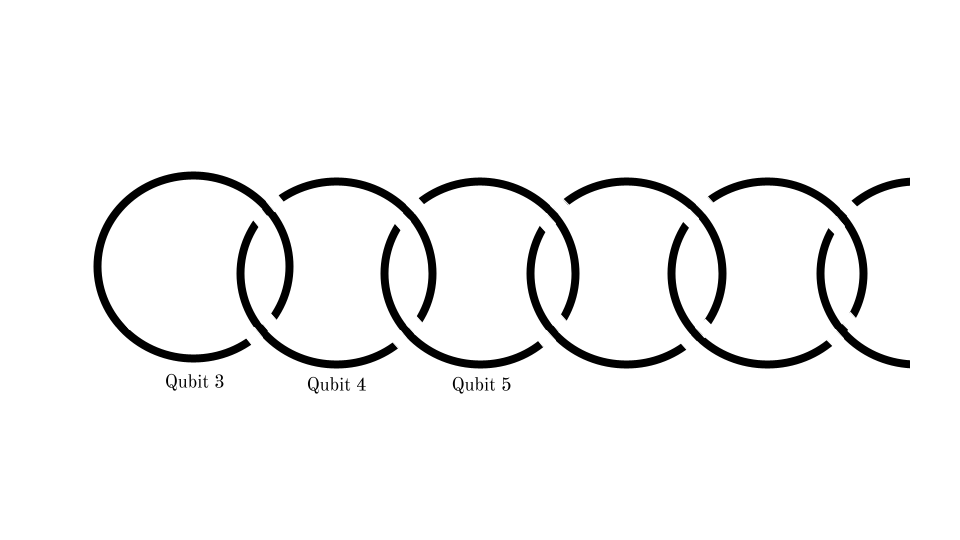}
    \caption{End Measurement (2D Schematic)}
    \label{fig:y_amb_end_2d}
\end{subfigure}
\hfill
\begin{subfigure}{0.3\textwidth}
    \centering
    \includegraphics[width=\linewidth]{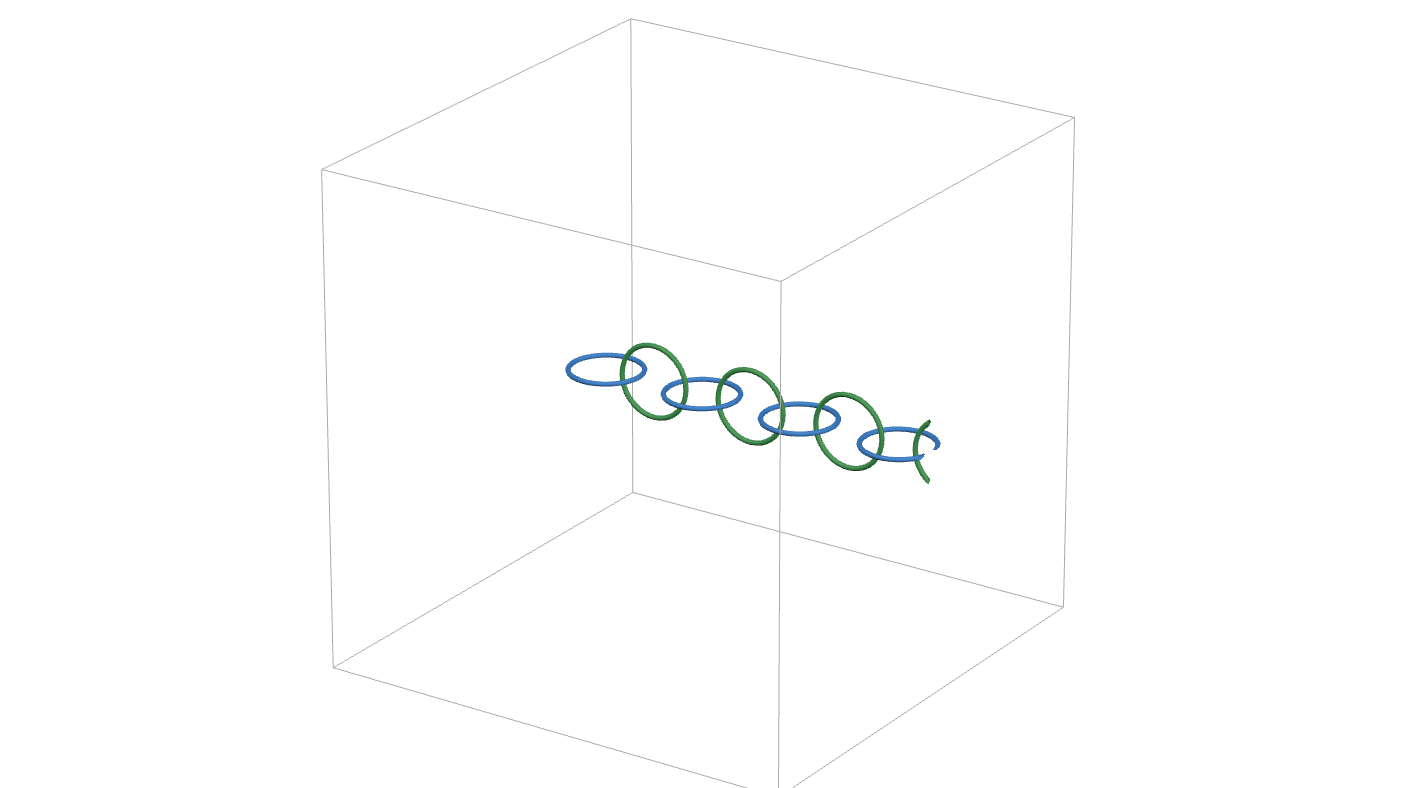}
    \caption{End Measurement (3D Visualization)}
    \label{fig:y_amb_end_3d}
\end{subfigure}
\caption{\textbf{The Boundary Ambiguity.} In the standard link model, measuring the end qubit ($Q_1$) is visualized simply as pruning the terminal ring. The model fails to capture the physical distinction: $Y$-measurements maintain connectivity but induce a chiral phase twist ($S$ gate) on the boundary.}
\label{fig:y_basis_ambiguity_end}
\end{figure}

\section{The $X$-Basis Measurement: Topological Stratification}
\noindent We now turn to the transverse ($X$) basis, $\mathcal{M}_X = \{|+\rangle, |-\rangle\}$. While it shares the property of preserving connectivity ($R=2$) with the $Y$-basis, its topological mechanism is vastly different. It does not perfectly splice the 1D chain into a single continuous path; instead, it fundamentally restructures the connectivity into parallel strata.

\subsection{Internal Stratification (Bulk Measurement)}
\noindent Unlike the $Y$-basis measurement which yields a simple linear cluster state, an $X$-basis measurement removes the physical site while inducing a complex fusion of its nearest neighbours. Carrying out the projection explicitly (Appendix~B.3) gives, for the outcome $|+\rangle_k$:
\begin{eqnarray}
\label{splice1}
|\Psi_{+}\rangle = \frac{1}{2^{(N-2)/2}}\!\sum_{\mathbf{x}_{\neq k}}\!
\delta_{x_{k-1},x_{k+1}}(-1)^{\Phi_{\mathrm{rest}}}|\mathbf{x}_{\neq k}\rangle ,
\end{eqnarray}
and for the outcome $|-\rangle_k$:
\begin{eqnarray}
\label{splice1b}
|\Psi_{-}\rangle \;=\; X_{k-1}Z_{k-2}\,|\Psi_{+}\rangle \;=\; X_{k+1}Z_{k+2}\,|\Psi_{+}\rangle .
\end{eqnarray}
\noindent The projection imposes a \emph{parity constraint} ($x_{k-1}\oplus x_{k+1}=0$ or $1$) on the surviving configurations. It does \emph{not} reinstate a quadratic phase $(-1)^{x_{k-1}x_{k+1}}$. Consequently, the residual state is \emph{not} the linear cluster state $|C_{N-1}^{(k-1\leftrightarrow k+1)}\rangle$. 

\noindent Instead of a simple topological splice, the $X$-measurement acts as a quantum switch that stratifies the state into parallel tracks. The residual state separates into a macroscopic superposition of two non-intersecting topological strata composed of the remaining cluster segments:
\begin{eqnarray}
\label{splice1c}
|\Psi_{+}\rangle = \frac{1}{\sqrt{2}} \Big[ |00\rangle_{k-1,k+1}|\chi_0\rangle + |11\rangle_{k-1,k+1}|\chi_1\rangle \Big],
\end{eqnarray}
where $|\chi_0\rangle = |C_{k-2}\rangle \otimes |C_{N-k-1}\rangle$ represents the unperturbed cluster sectors, and $|\chi_1\rangle = Z_{k-2}Z_{k+2}|\chi_0\rangle$ represents the sectors modified by local Pauli corrections.

\begin{figure}[t]
\centering
\begin{subfigure}{0.3\textwidth}
    \centering
    \includegraphics[width=\linewidth]{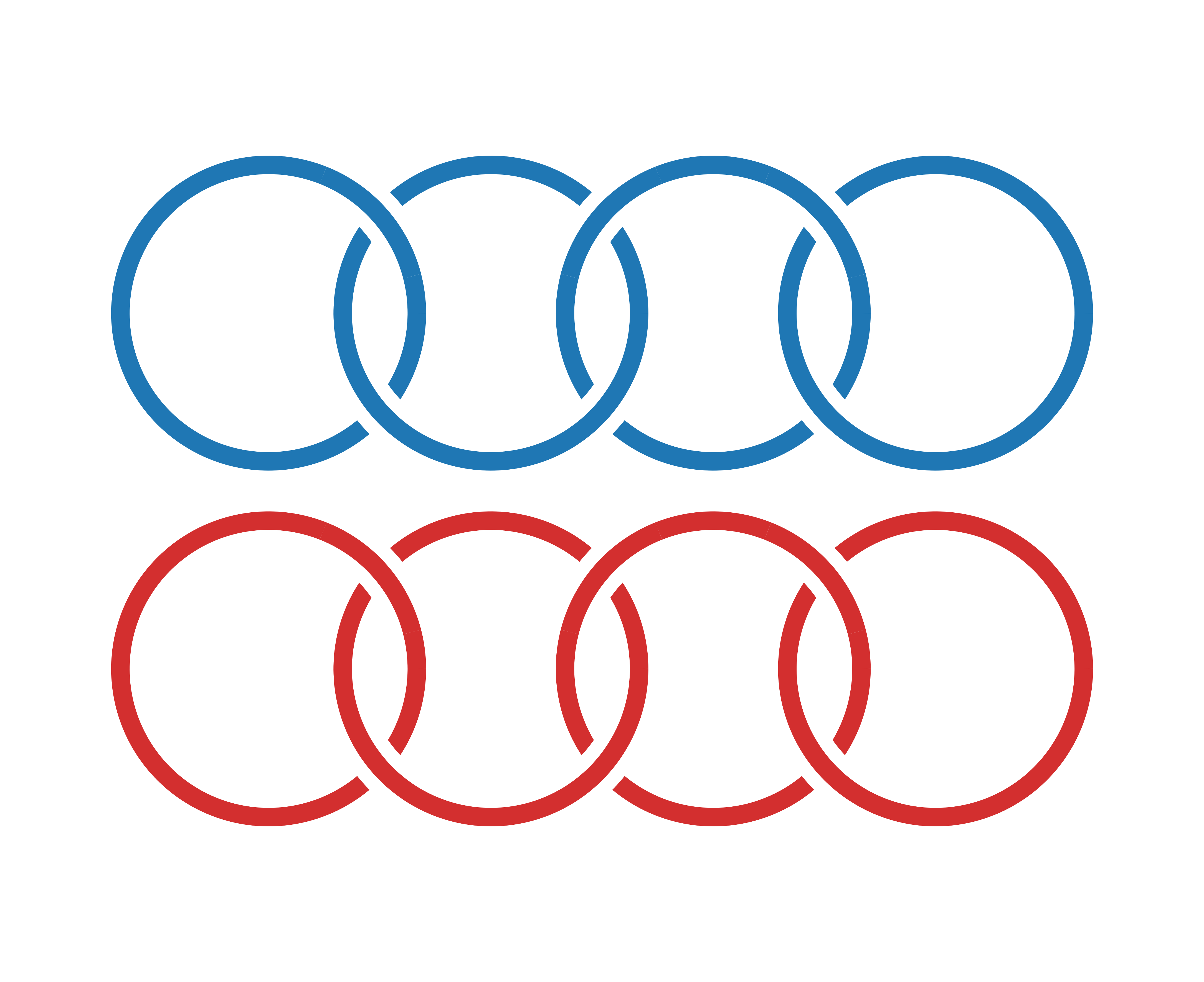}
    \caption{Bulk Measurement (2D Schematic)}
    \label{fig:x_bulk_2d}
\end{subfigure}
\hfill
\begin{subfigure}{0.3\textwidth}
    \centering
    \includegraphics[width=\linewidth]{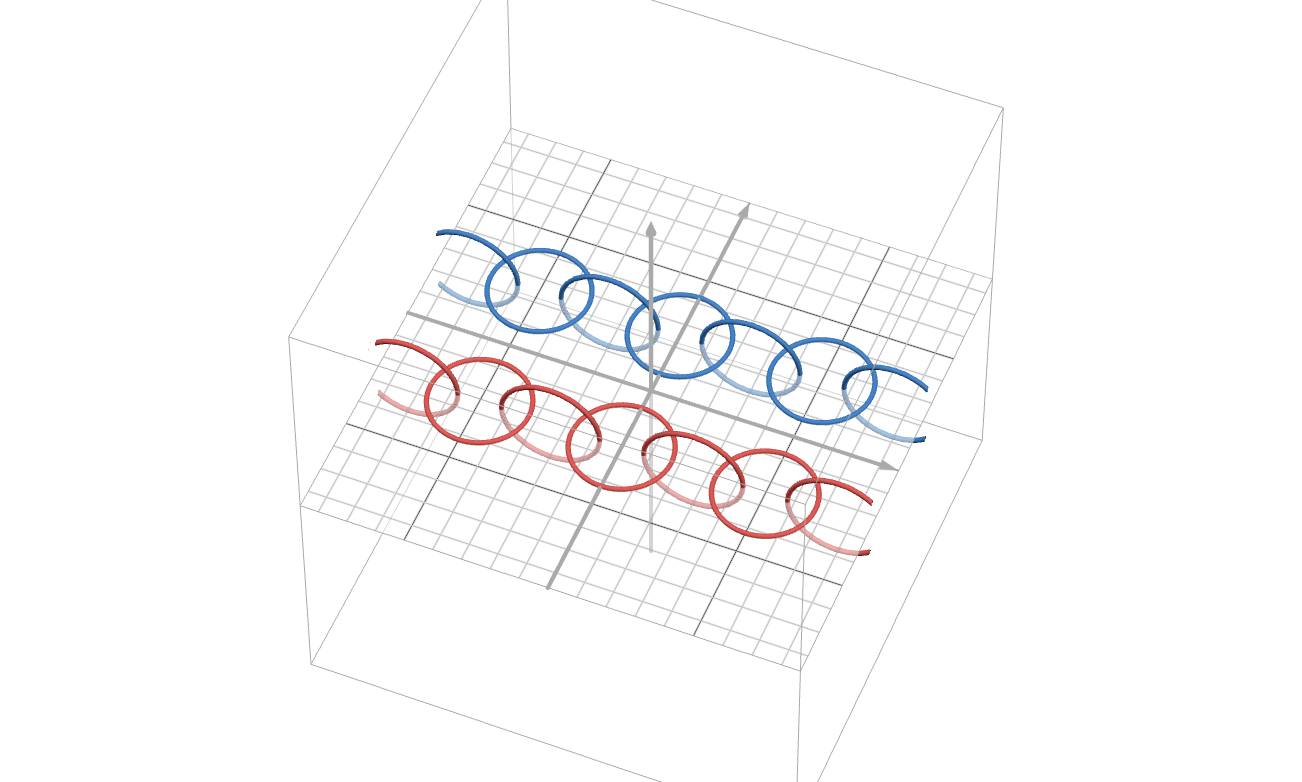}
    \caption{Bulk Measurement (3D Visualization)}
    \label{fig:x_bulk_3d}
\end{subfigure}
\caption{\textbf{Topological Stratification in the X-Basis.}
Measuring $Q_k$ preserves connectivity ($R=2$) but acts as a quantum switch. As shown in Eqs.~(\ref{splice1})--(\ref{splice1c}), the measurement stratifies the state into parallel, non-intersecting tracks representing related cluster sectors, rather than forming a single spliced linear chain.}
\label{fig:x_basis_bulk}
\end{figure}

The two-branch structure of Eq.~(\ref{splice1c}) is a direct consequence of the parity-sector decomposition derived in Appendix~B.3; the ``parallel strata'' language used in Fig.~\ref{fig:x_basis_bulk} and throughout this section is a \emph{geometric rendering} of this algebraic superposition, not an independent topological derivation.

% ==========================================
% FIGURE: X-BASIS STRATIFICATION - SIDE BY SIDE
% ==========================================

\subsection{Boundary Propagation (End Measurement)}
\noindent An $X$-measurement on end-qubit $1$ triggers a \textit{skipping} effect. As shown in \textsc{Appendix B.2}, the measurement effectively disentangles the \textit{next} qubit ($Q_2$), forcing it into a fixed computational state. The active boundary of the cluster effectively jumps to $Q_3$:
\begin{eqnarray}
\label{splice2}
|\Psi_{+}\rangle = |0\rangle_2 \otimes |C_{N-2}\rangle_{3\dots N},\nonumber\\
|\Psi_{-}\rangle = |1\rangle_2 \otimes Z_3|C_{N-2}\rangle_{3\dots N}.
\end{eqnarray}
This demonstrates the information propagation capability of the $X$-basis, actively pushing the logical information forward along the chain.

% ==========================================
% FIGURE: X-BASIS END - SIDE BY SIDE
% ==========================================
\begin{figure}[t]
\centering
\begin{subfigure}{0.3\textwidth}
    \centering
    \includegraphics[width=\linewidth]{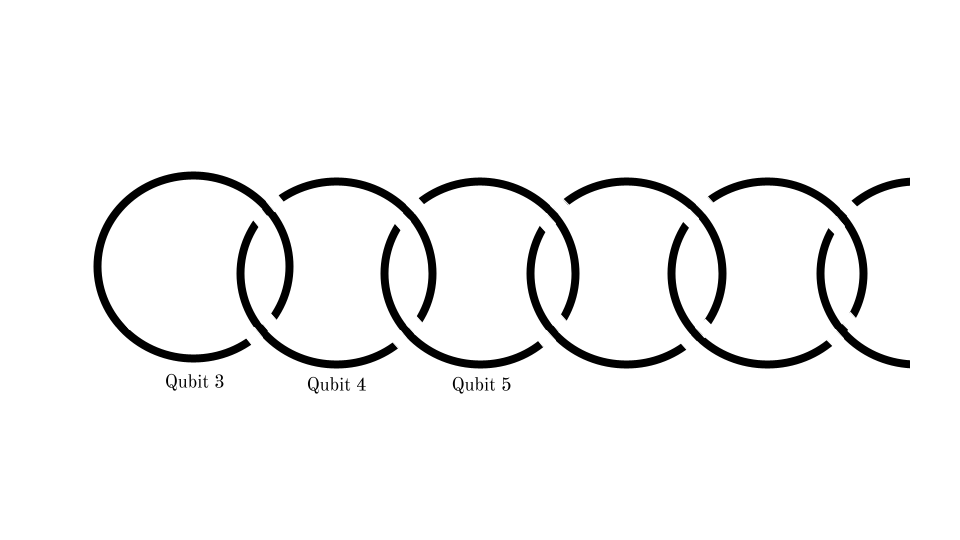}
    \caption{End Measurement (2D Schematic)}
    \label{fig:x_end_2d}
\end{subfigure}
\hfill
\begin{subfigure}{0.3\textwidth}
    \centering
    \includegraphics[width=\linewidth]{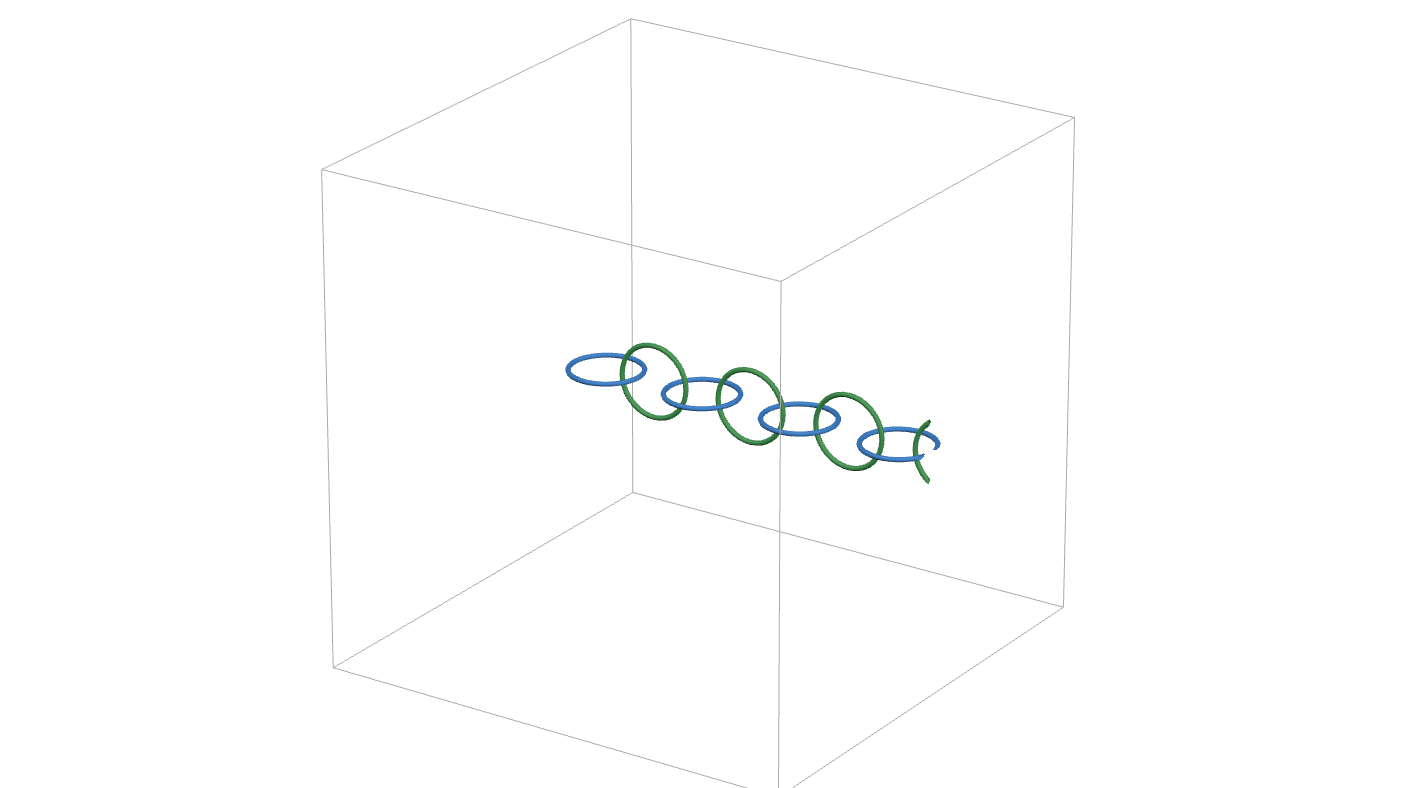}
    \caption{End Measurement (3D Visualization)}
    \label{fig:x_end_3d}
\end{subfigure}
\caption{\textbf{Topological Splicing in X-Basis - Boundary Propagation.} 
Unlike Z-pruning, measuring $Q_1$ in X decouples $Q_2$, effectively transporting the boundary to $Q_3$.}
\label{fig:x_basis_end}
\end{figure}

\section{Summary of Single-Qubit Measurements}
\noindent We summarize below (in Table~\ref{tab:measurement_summary}) the exact mathematical results of projective measurements on $|C_N\rangle$. The table condenses the derivations provided in Appendices A, B, and C, detailing the residual state and entanglement properties. 
\begin{table*}[t] % Note the asterisk

\centering

\renewcommand{\arraystretch}{1.5}

\begin{tabular}{@{}l l l l l@{}}

\toprule

\textbf{Basis} & \textbf{Target} & \textbf{Outcome} & \textbf{Post-Measurement State} & \textbf{Entanglement Properties} \\ \midrule

% --- Z-BASIS ---

\textbf{Z} & End ($Q_1$) & $|0\rangle_1$ & $|C_{N-1}\rangle_{2\dots N}$ & Chain Shortened ($N \to N-1$). \\

(Comp.) & & $|1\rangle_1$ & $Z_2 |C_{N-1}\rangle_{2\dots N}$ & Chain Shortened + Pauli $Z$ on $Q_2$. \\ \cmidrule{2-5}

& Bulk ($Q_k$) & $|0\rangle_k$ & $|C_L\rangle \otimes |C_R\rangle$ & \textbf{Severed} ($R=1$). Disconnected. \\

& & $|1\rangle_k$ & $(Z_{k-1}Z_{k+1})(|C_L\rangle \otimes |C_R\rangle)$ & Severed + Boundary $Z$ corrections. \\ \midrule

% --- X-BASIS ---

\textbf{X} & End ($Q_1$) & $|+\rangle_1$ & $|0\rangle_2 \otimes |C_{N-2}\rangle_{3\dots N}$ & $Q_2$ Decouples. Boundary skips to $Q_3$. \\

(Trans.) & & $|-\rangle_1$ & $|1\rangle_2 \otimes Z_3 |C_{N-2}\rangle_{3\dots N}$ & $Q_2$ Decouples. Boundary skips + $Z_3$. \\ \cmidrule{2-5}

& Bulk ($Q_k$) & $|+\rangle_k$ & $\frac{1}{\sqrt{2}} \big[ |00\rangle_{k-1,k+1}|\chi_0\rangle + |11\rangle_{k-1,k+1}|\chi_1\rangle \big]$ & \textbf{Stratified} ($R=2$). Correlated branch. \\

& & $|-\rangle_k$ & $\frac{1}{\sqrt{2}} \big[ |01\rangle_{k-1,k+1} Z_{k+2}|\chi_0\rangle + |10\rangle_{k-1,k+1} Z_{k-2}|\chi_0\rangle \big]$ & Stratified ($R=2$). Anti-correlated branch. \\ \midrule

% --- Y-BASIS ---

\textbf{Y} & End ($Q_1$) & $|+i\rangle_1$ & $S_2 |C_{N-1}\rangle_{2\dots N}$ & Shortened + Phase Gate $S$ on $Q_2$. \\

(Lat.) & & $|-i\rangle_1$ & $S^\dagger_2 |C_{N-1}\rangle_{2\dots N}$ & Shortened + Phase Gate $S^\dagger$ on $Q_2$. \\ \cmidrule{2-5}

& Bulk ($Q_k$) & $|+i\rangle_k$ & $(S_{k-1} S_{k+1}) |C_{N-1}^{(k-1 \leftrightarrow k+1)}\rangle$ & \textbf{Twisted Splice} ($R=2$). Chiral ($+i$). \\

& & $|-i\rangle_k$ & $(S^\dagger_{k-1} S^\dagger_{k+1}) |C_{N-1}^{(k-1 \leftrightarrow k+1)}\rangle$ & Twisted Splice. Anti-Chiral ($-i$). \\ \bottomrule

\end{tabular}

\caption{Mathematical summary of single-qubit measurements on the 1D Linear Cluster State. The table explicitly details the post-measurement state for each outcome, highlighting the specific Pauli or Phase corrections induced on the remaining system. For the bulk $X$-measurement, $|\chi_0\rangle = |C_{k-2}\rangle \otimes |C_{N-k-1}\rangle$ and $|\chi_1\rangle = Z_{k-2}Z_{k+2}|\chi_0\rangle$ represent the unperturbed and modified sectors of the stratified state. All states are given up to an irrelevant global phase, and $s=\pm1$ labels the $Y$-basis outcome as in Eq.~(\ref{ybulk}).}

\label{tab:measurement_summary}

\end{table*}
\section{Outcome Ambiguity Within a Fixed Measurement Basis}
\noindent The corrected bulk-$X$ derivation of Sec.~VII already resolves the coarser ambiguity considered in earlier work on this topic: the unframed linking pattern alone is sufficient to distinguish the three measurement bases by qualitative shape. A $Z$-basis measurement severs the chain into two disconnected components (Sec.~V); a $Y$-basis measurement preserves a single continuous spliced chain (Sec.~VI); an $X$-basis measurement instead produces two parallel, non-intersecting branches sharing a common break at the measured site (Sec.~VII). These are three structurally distinct residual topologies, and no framing is required to tell them apart.

\noindent A finer ambiguity remains, however, \emph{within} each basis. The two possible outcomes of a $Y$-basis measurement, $|+i\rangle_k$ and $|-i\rangle_k$, produce residual states with identical connectivity -- a single spliced chain in both cases -- differing only in the sign of the induced phase gate, $S_{k-1}\otimes S_{k+1}$ versus $S_{k-1}^\dagger\otimes S_{k+1}^\dagger$. Likewise, the two outcomes of a bulk $X$-basis measurement, $|+\rangle_k$ and $|-\rangle_k$, both produce the same two-branch stratified topology; they differ only in which qubit pairing carries the induced Pauli-$Z$ correction (Sec.~VII.1). In the conventional unframed picture, both members of each outcome pair are represented by an identical diagram, so the diagram alone cannot report which outcome was actually recorded. This is the precise sense in which the standard model remains phase-blind: not across measurement bases, which are already distinguishable by shape, but across the two possible outcomes of any single basis choice.
\subsection{Necessity of a Framed Description}
\noindent The origin of this residual ambiguity lies in the limited geometric vocabulary of an unframed link diagram: it can record which components are linked and how many disjoint pieces the residual state contains, but it has no further degree of freedom with which to record a sign. Two diagrams with the same connectivity are, by definition, the same diagram. Consequently, an unframed representation cannot distinguish $|+i\rangle_k$ from $|-i\rangle_k$, nor the two branch-labelling patterns of the $X$-basis outcomes, even though the underlying by-product operators are physically distinct.

\noindent To resolve this, the link model must be extended to a \emph{framed} representation in three-dimensional space, in which each ribbon segment carries an additional geometric degree of freedom -- its twist. Within this extended model, $Y$-basis measurements correspond to \emph{twisted splicing operations}, where the fused ribbon acquires a chiral twist to represent the complex phase $\pm i$; the two $Y$-basis outcomes are then distinguished by the handedness of that twist. Conversely, $X$-basis measurements manifest as a \emph{geometric stratification} into two parallel, non-intersecting tracks (Sec.~X.3), and the two $X$-basis outcomes are distinguished by \emph{where} along those tracks a $180^\circ$ flip is placed. This motivates the framed-ribbon formulation developed in the following section, which provides a unified, phase-sensitive geometric representation of all single-qubit measurements on linear cluster states.
\section{The Framed Ribbon Model: A Unified Topological Framework}
\noindent \textit{Status of the twist angle.} Before introducing the construction we state its status unambiguously, since the language of topological invariance is easily over-extended here. The linking pattern of the unframed link -- whether two components are linked, i.e.\ the pairwise linking number -- is a genuine topological invariant, unchanged under ambient isotopy, and it is the object that encodes connectivity in Secs.~V--VII. The twist angle $\theta$ introduced below is \emph{not} of this kind. It takes the discrete values $\lbrace 0^\circ,\pm 90^\circ,180^\circ\rbrace$, and quarter twists are not integer framings, so $\theta$ is not an invariant of framed links; it is not preserved under arbitrary continuous deformation of the ribbon, and it may be altered by a local rotation of the framing at a free end. As shown in Appendix~D, $\theta$ is instead an \emph{operationally fixed geometric label}: its value is pinned by the measurement outcome and by the by-product operator that outcome induces, not by the isotopy class of the ribbon. Throughout we therefore speak of a phase-sensitive \emph{geometric} encoding, and reserve the word ``topological'' for the connectivity/linking statements.\\
\noindent To resolve this ambiguity, we introduce a framed ribbon representation, drawing on the mathematical theory of framed knots \cite{adams2004,kauffman2004}. In this extended framework, quantum correlations are encoded not only through connectivity but also via geometric twist, with chiral $\pm 90^\circ$ twists corresponding to the phases $\pm i$. (A geometric object is said to be \textit{chiral} if it cannot be superposed onto its mirror image by any combination of rotations and translations. Equivalently, the object possesses a \textit{handedness}, right-handed or left-handed, that distinguishes it from its mirror reflected configuration. In the present framework, the $\pm 90^\circ$ ribbon twists are chiral because right-handed and left-handed twists correspond to physically distinct configurations representing the phases $\pm i$.)  This mapping from quantum phase to geometric torsion parallels the correspondence between quantum evolution and geometric phases \cite{berry1984,aharonov1987}. We define the ribbon $\mathcal{R}$ connecting neighbours $k-1$ and $k+1$ by its \textit{Twist Angle $\theta$}, which encodes the relative quantum phase $\phi$ of the spliced state according to the map $\theta = \phi$. In the \textsc{Appendix D} we define the framed ribbon and chirality.
\subsection{The Twist Dictionary}
\begin{figure}[t]
\centering
\begin{subfigure}{0.3\textwidth}
    \centering
    \includegraphics[width=\linewidth]{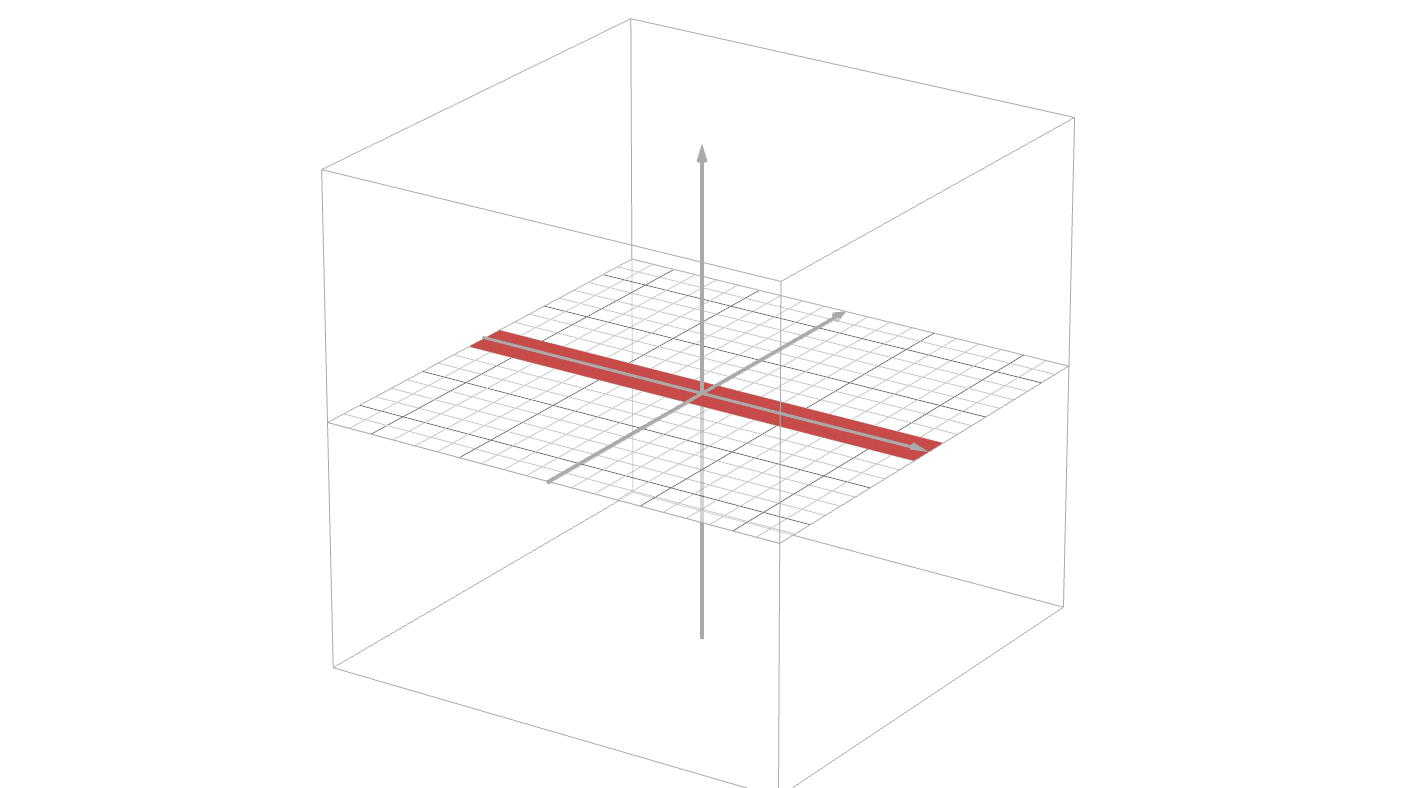}
    \caption{Flat ($0^\circ$)}
    \label{fig:flat_ribbon}
\end{subfigure}
\hfill
\begin{subfigure}{0.3\textwidth}
    \centering
    \includegraphics[width=\linewidth]{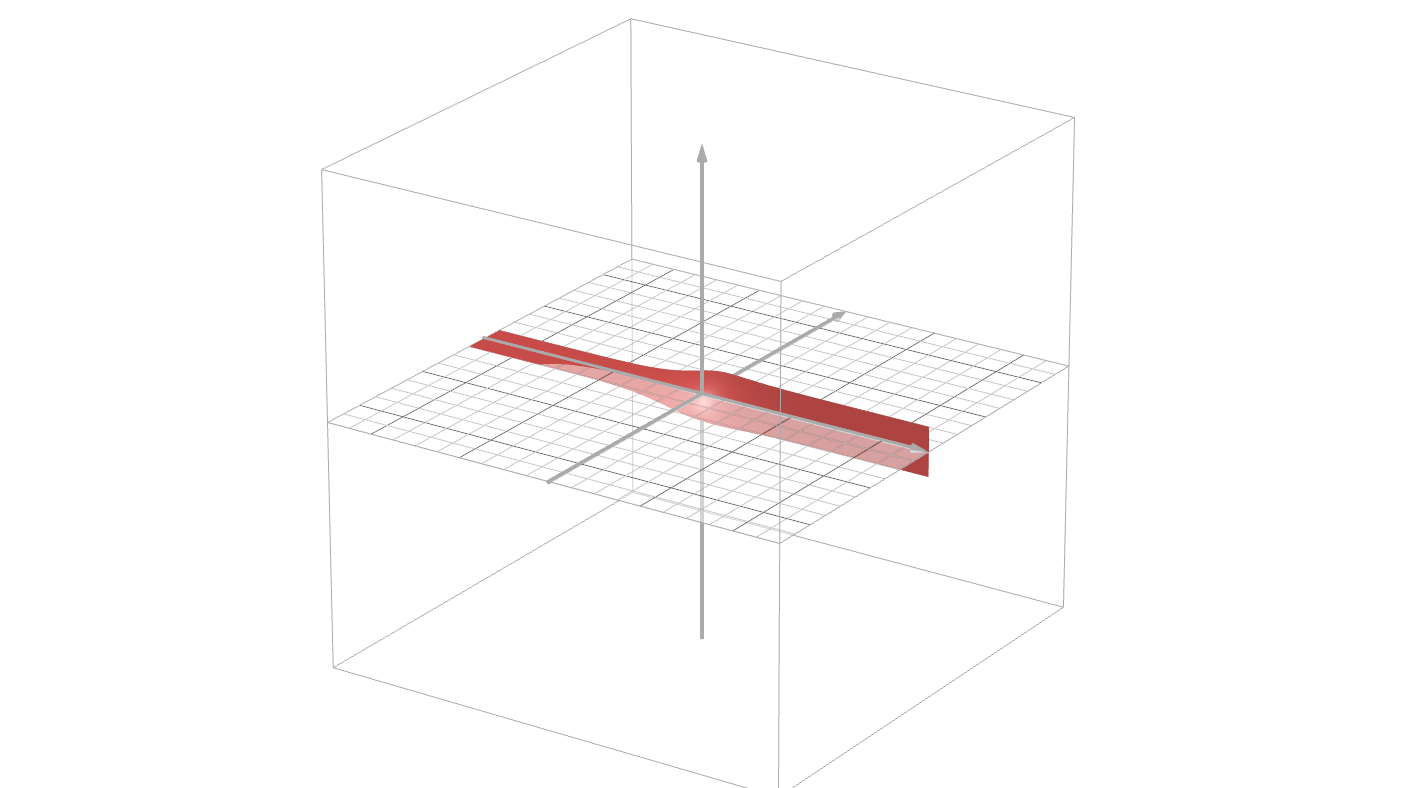}
    \caption{Chiral ($90^\circ$)}
    \label{fig:chiral_twist}
\end{subfigure}
\hfill
\begin{subfigure}{0.3\textwidth}
    \centering
    \includegraphics[width=\linewidth]{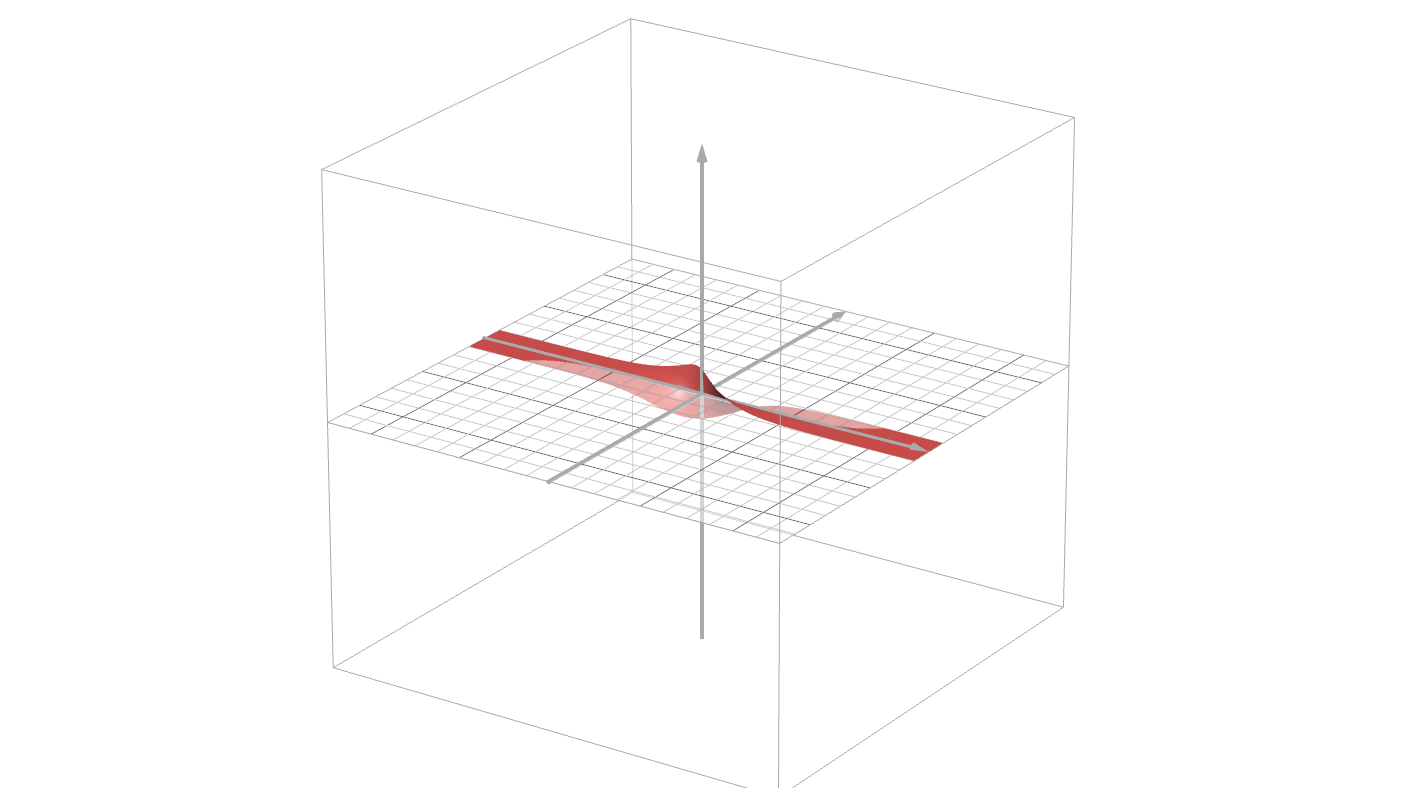}
    \caption{Full Flip ($180^\circ$)}
    \label{fig:full_flip}
\end{subfigure}
\caption{\textbf{The Twist Dictionary.} (a) Flat ribbon representing quantum phase $+1$.
(b) Chiral twist representing quantum phase $\pm i$. (c) Full flip representing quantum phase $-1$.}
\label{fig:twist_dictionary}
\end{figure}

\noindent Before analyzing the specific basis measurements, we establish the geometric dictionary of the ribbon.\\
\noindent  We consider the ribbon in isolation (qubits removed for clarity), as shown in Fig. \ref{fig:twist_dictionary}.
To establish the geometric correspondence between ribbon framing and quantum phase, we introduce a twist angle parameter $\theta$ that characterizes the relative rotation of a ribbon connecting neighbouring qubits. In the framed ribbon representation, this geometric twist encodes the phase factor associated with the corresponding quantum correlation. The mapping between twist angle and phase is defined as $\theta = \phi$, where $\phi$ denotes the relative quantum phase.\\
\noindent We distinguish the following canonical configurations.
\begin{enumerate}
   \item \textit{Flat ribbon ($\theta = 0^\circ$):}  
    The ribbon connects the neighbouring qubits without rotation.  
    This configuration corresponds to the trivial phase factor $+1$ and represents real, positive correlations.
    \item \textit{Chiral quarter-twist ($\theta = \pm 90^\circ$):}  
    The ribbon undergoes a quarter rotation, producing a chiral configuration that distinguishes the two possible orientations viz.
    \begin{itemize}
        \item Right-handed twist ($+90^\circ$): corresponds to the phase factor $+i$.
        \item Left-handed twist ($-90^\circ$): corresponds to the phase factor $-i$.
    \end{itemize}
    \item \textit{Half-twist ($\theta = 180^\circ$):}  
    The ribbon undergoes a half rotation, reversing its orientation.  
    This configuration corresponds to the phase factor $-1$, equivalent to the action of a Pauli-$Z$ operator on the associated correlation.
\end{enumerate}
This twist dictionary provides the geometric basis for representing measurement-induced phase transformations within the framed-ribbon model, enabling a unified topological description of both real and complex correlations generated by single-qubit measurements \cite{rolfsen2003}.
% ==========================================
% FIGURE: TWIST DICTIONARY - SIDE BY SIDE
% ==========================================
\subsection{Positive and Negative Rotation: Defining Chirality}
\noindent A critical feature of the Framed Ribbon Model is its ability to distinguish between conjugate complex phases, $+i$ and $-i$, through geometric chirality. Unlike the flat ($\theta=0^\circ$) or flipped ($\theta=180^\circ$) ribbons, which are \textit{achiral} (i.e. superposable on their mirror images), the quarter-twist ($\theta=90^\circ$) \textit{breaks spatial symmetry}. We define the sign of the rotation using the standard Right-Hand Rule along the propagation axis of the cluster.
\begin{enumerate}
    \item \textit{Positive Rotation (Right-Handed, $+90^\circ$):}
    As the ribbon propagates from the qubit $Q_{k-1}$ to the qubit $Q_{k+1}$, a clockwise rotation of the frame corresponds to a positive phase shift $+i$. Visually, this manifests as the right edge of the ribbon crossing \textsc{OVER} the left edge. This specific topology encodes the measurement outcome $|+i\rangle$ (Fig. \ref{fig:positive_rotation}).
    % ==========================================
    % FIGURE: POSITIVE ROTATION - SIDE BY SIDE
    % ==========================================
    \begin{figure}[t]
    \centering
    \begin{subfigure}{0.3\textwidth}
        \centering
        \includegraphics[width=\linewidth]{fulltwist_ribbon.png}
        \label{fig:pos_rotation_1}
    \end{subfigure}
    \hfill
    \begin{subfigure}{0.3\textwidth}
        \centering
        \includegraphics[width=\linewidth]{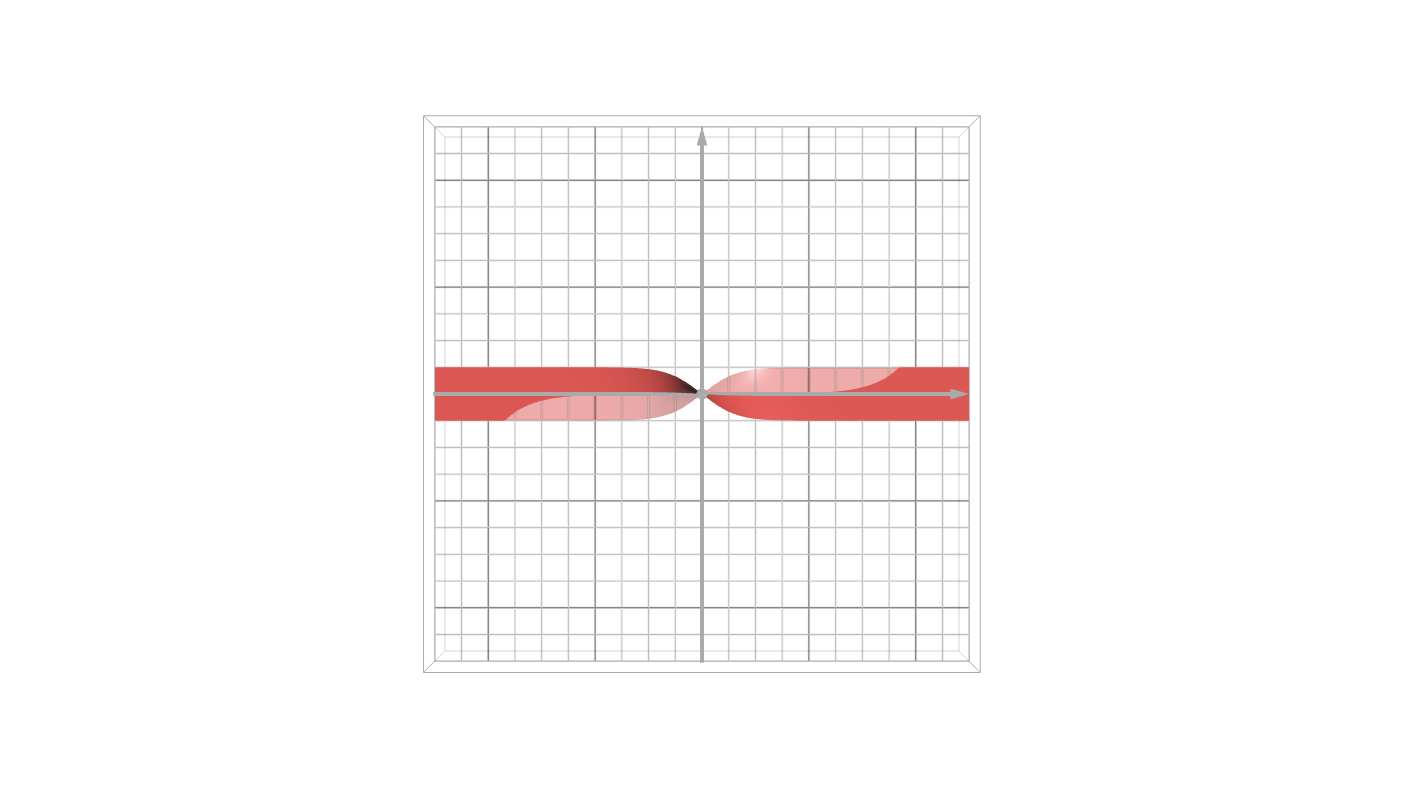}
        \label{fig:pos_rotation_2}
    \end{subfigure}
    \caption{\textbf{Positive (Right-Handed) Rotation.} The ribbon twists clockwise ($+90^\circ$), representing the phase factor $+i$. Note the specific over-crossing structure in both views.}
    \label{fig:positive_rotation}
    \end{figure}
    \item \textit{Negative Rotation (Left-Handed, $-90^\circ$):}
    A counter-clockwise rotation corresponds to a negative phase shift $-i$. Visually, this manifests as the right edge crossing \textsc{UNDER} the left edge. This specific topology encodes the measurement outcome $|-i\rangle$ (Fig. \ref{fig:negative_rotation}).
    % ==========================================
    % FIGURE: NEGATIVE ROTATION - SIDE BY SIDE
    % ==========================================
    \begin{figure}[t]
    \centering
    \begin{subfigure}{0.3\textwidth}
        \centering
        \includegraphics[width=\linewidth]{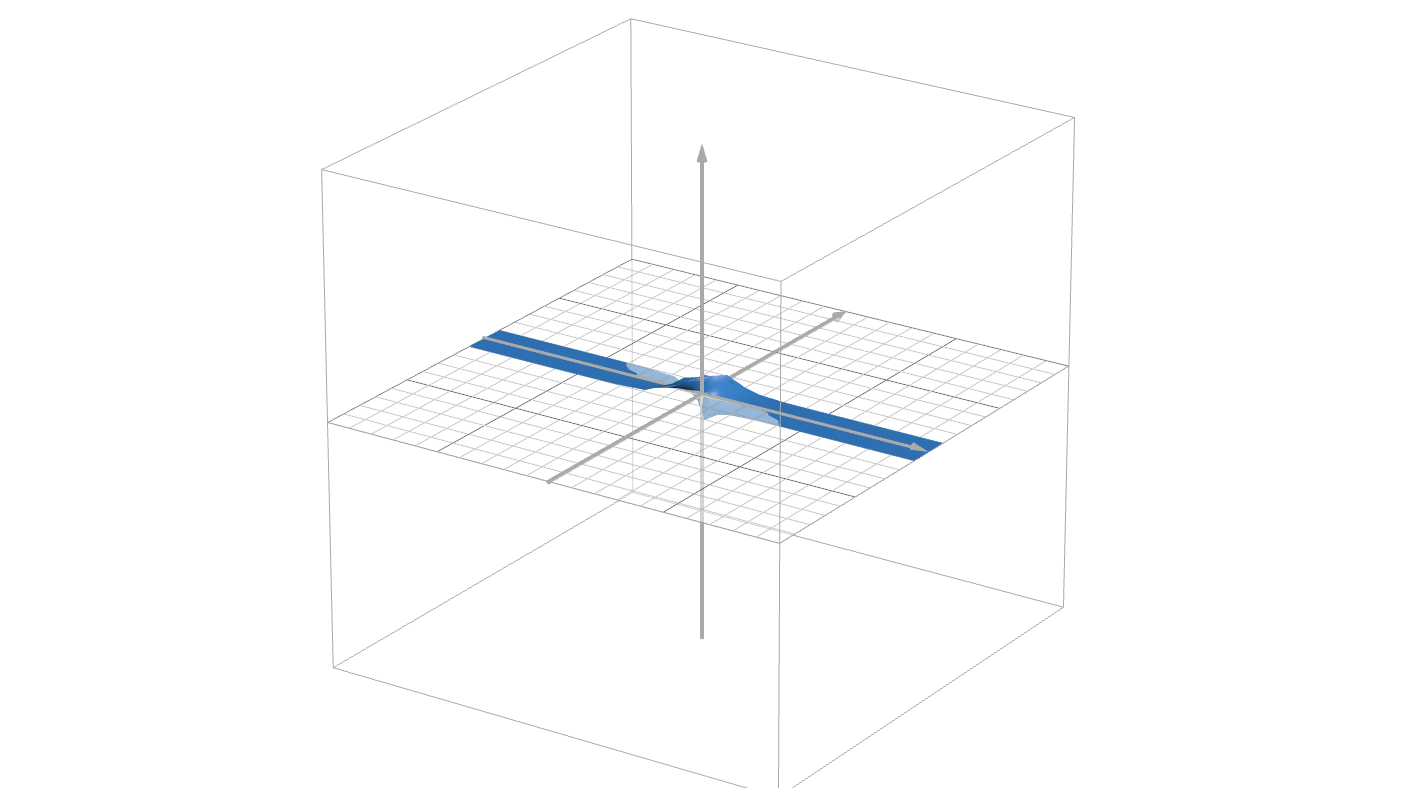}
        \label{fig:neg_rotation_1}
    \end{subfigure}
    \hfill
    \begin{subfigure}{0.3\textwidth}
        \centering
        \includegraphics[width=\linewidth]{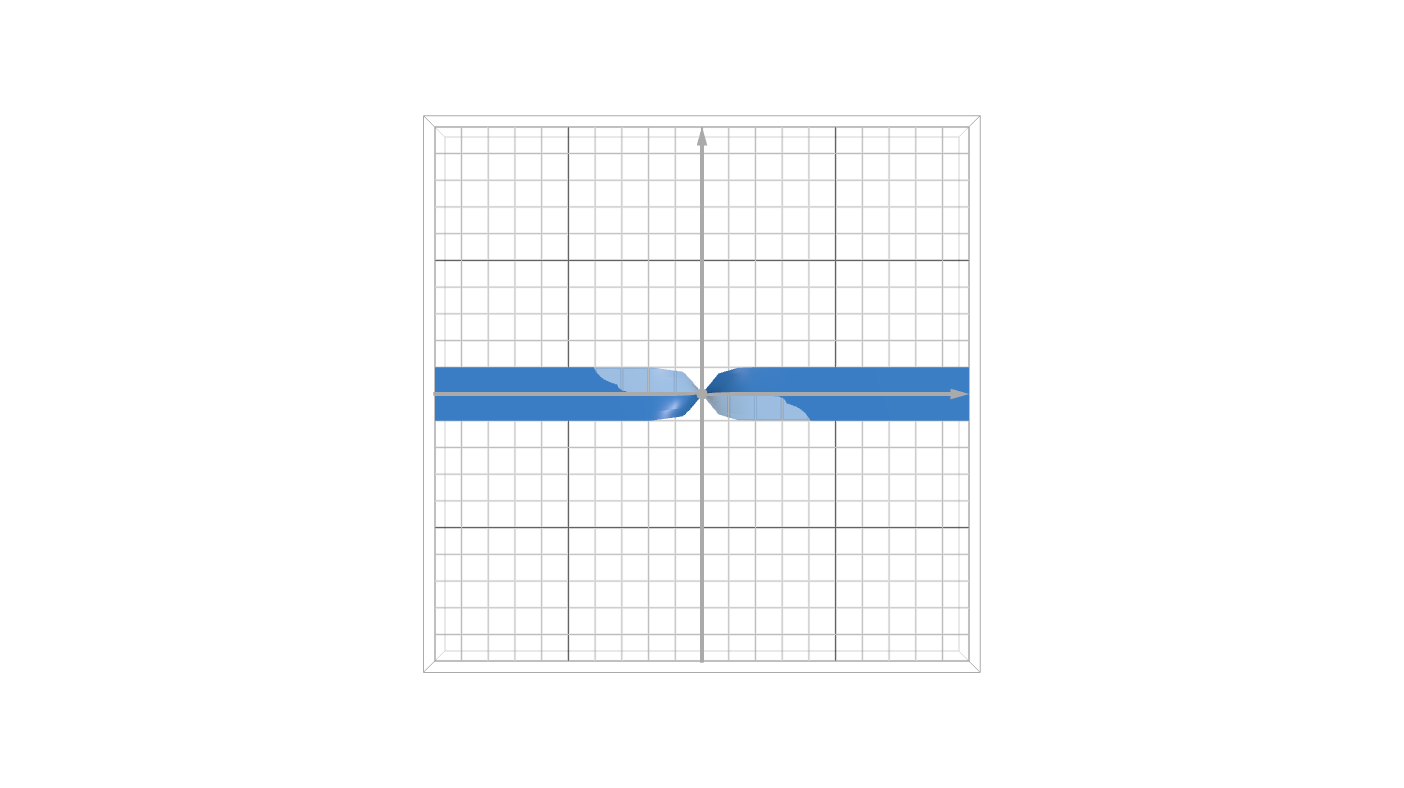}
        \label{fig:neg_rotation_2}
    \end{subfigure}
    \caption{\textbf{Negative (Left-Handed) Rotation.} The ribbon twists counter-clockwise ($-90^\circ$), representing the phase factor $-i$. Note the specific under-crossing structure in both views.}
    \label{fig:negative_rotation}
    \end{figure}
\end{enumerate}
This geometric encoding of quantum phases through ribbon framing parallels the well-established correspondence between quantum evolution and geometric phases in adiabatic and holonomic quantum computation \cite{berry1984,aharonov1987,zanardi1999}.

\subsection{Topological Stratification: The Parallel Strata}

The flat, chiral, and flipped ribbons defined above are sufficient to describe quantum operations that preserve a single, continuous one-dimensional topological trajectory. However, certain measurement operations act as a quantum switch, driving the remaining entanglement structure into a macroscopic superposition of distinct topological sectors. 

To capture this geometrically without implying that mutually exclusive quantum realities physically intersect, we extend the framed ribbon model to include \emph{Topological Stratification}. Rather than forming an intersecting Y-junction, this is represented by parallel, non-intersecting ribbon strata that span across the measurement site. Because the post-measurement state is a highly correlated superposition rather than a simple 1D chain, the geometry requires separate parallel tracks to represent all co-existing trajectories simultaneously.

Crucially, because each parallel stratum represents an independent sector of the quantum superposition, the separate tracks can undergo entirely separate geometric twists. The specific torsion of each stratum visually encodes the distinct local phase corrections applied to that particular sector of the wavefunction. 

\begin{figure}[t]
\centering
\begin{subfigure}{0.45\textwidth}
    \centering
    \includegraphics[width=\linewidth]{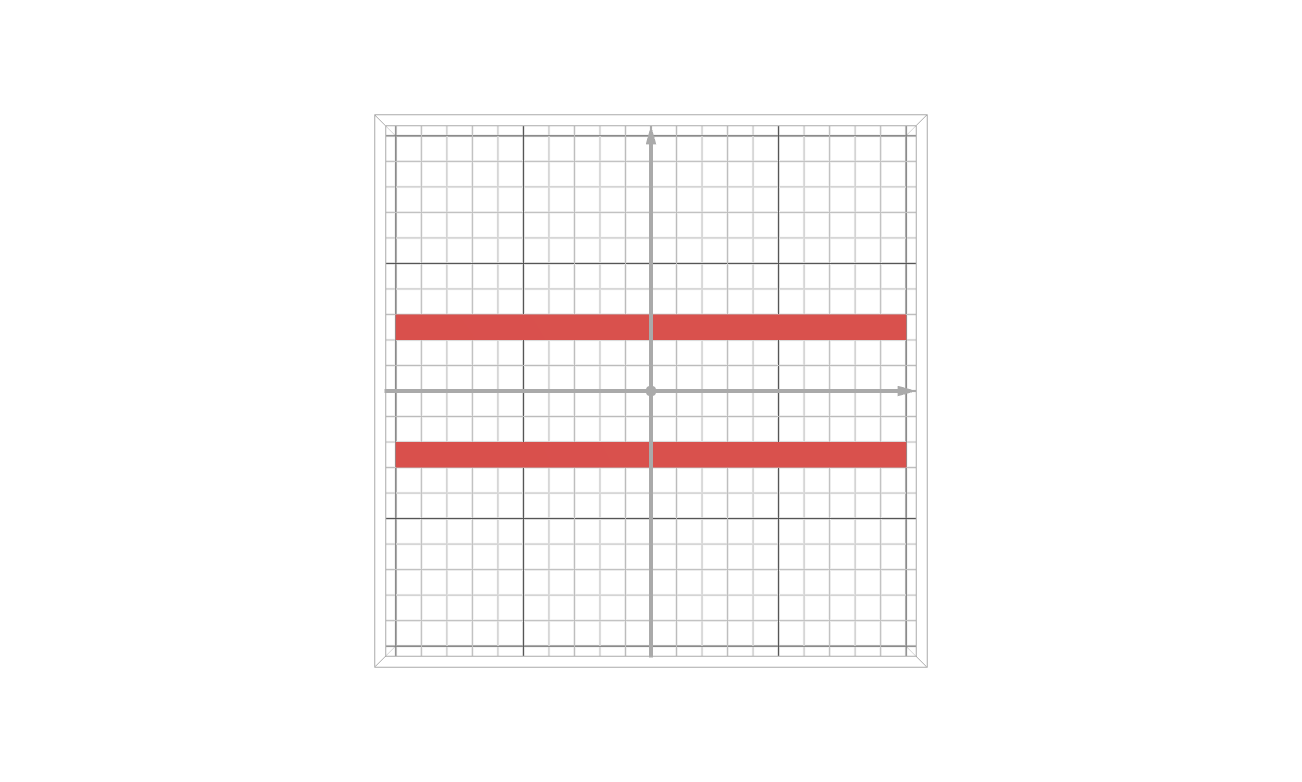}
    \label{fig:bifurcation_3d}
\end{subfigure}
\hfill
\begin{subfigure}{0.45\textwidth}
    \centering
    \includegraphics[width=\linewidth]{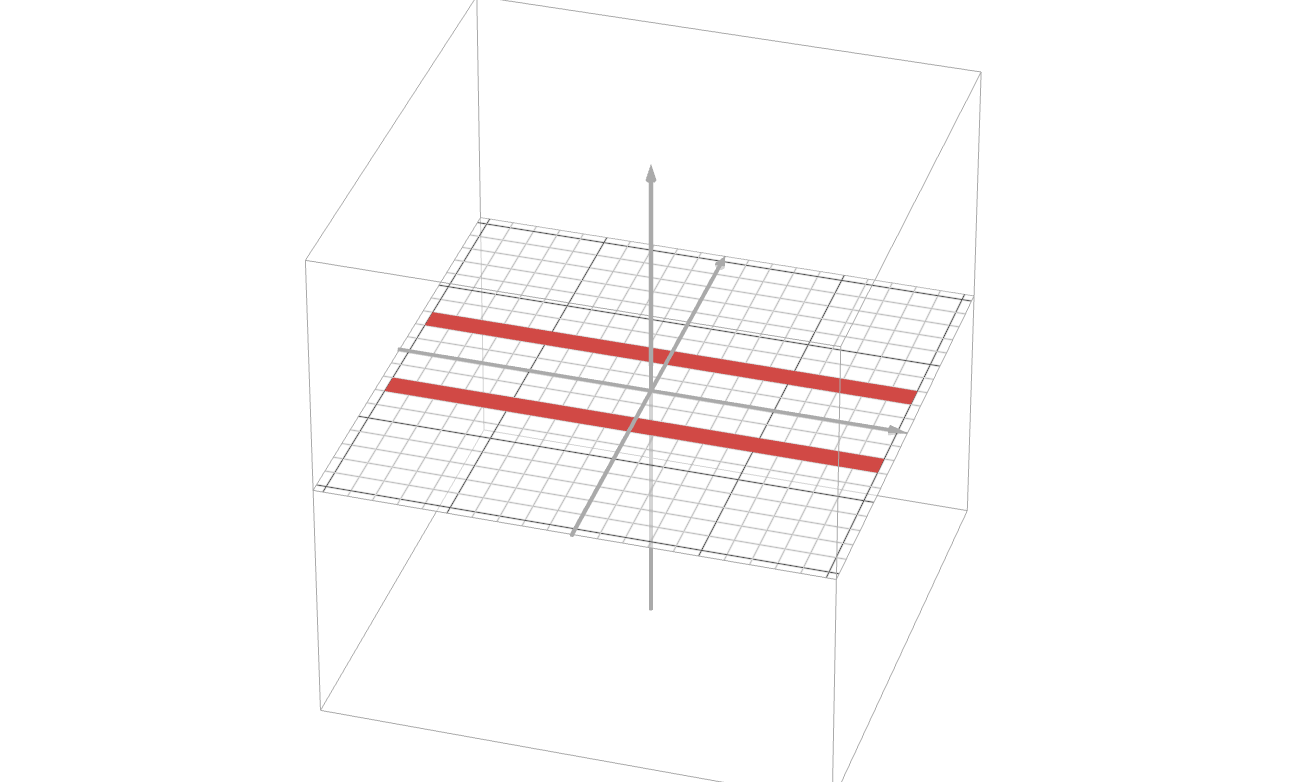}
    \label{fig:bifurcation_2d}
\end{subfigure}
\caption{\textbf{Topological Stratification.} Parallel, non-intersecting ribbon strata represent a macroscopic superposition of distinct sectors. Each stratum can independently carry distinct geometric twists depending on the local phase corrections of its respective sector.}
\label{fig:bifurcation}
\end{figure}

As illustrated in Fig.~\ref{fig:bifurcation}, the geometry stratifies the state into distinct parallel paths. This stratified layout allows the framed ribbon model to faithfully represent the generation of macroscopic superposition states as orthogonal tracks, providing a geometric language for superpositions of distinct topologies without violating their mutually exclusive nature.

\section{Cluster State in the Ribbon Frame}
\subsection{The Initial State in the Ribbon Frame}
\begin{figure}[t]
\centering
\begin{subfigure}{0.3\textwidth}
    \centering
    \includegraphics[width=\linewidth]{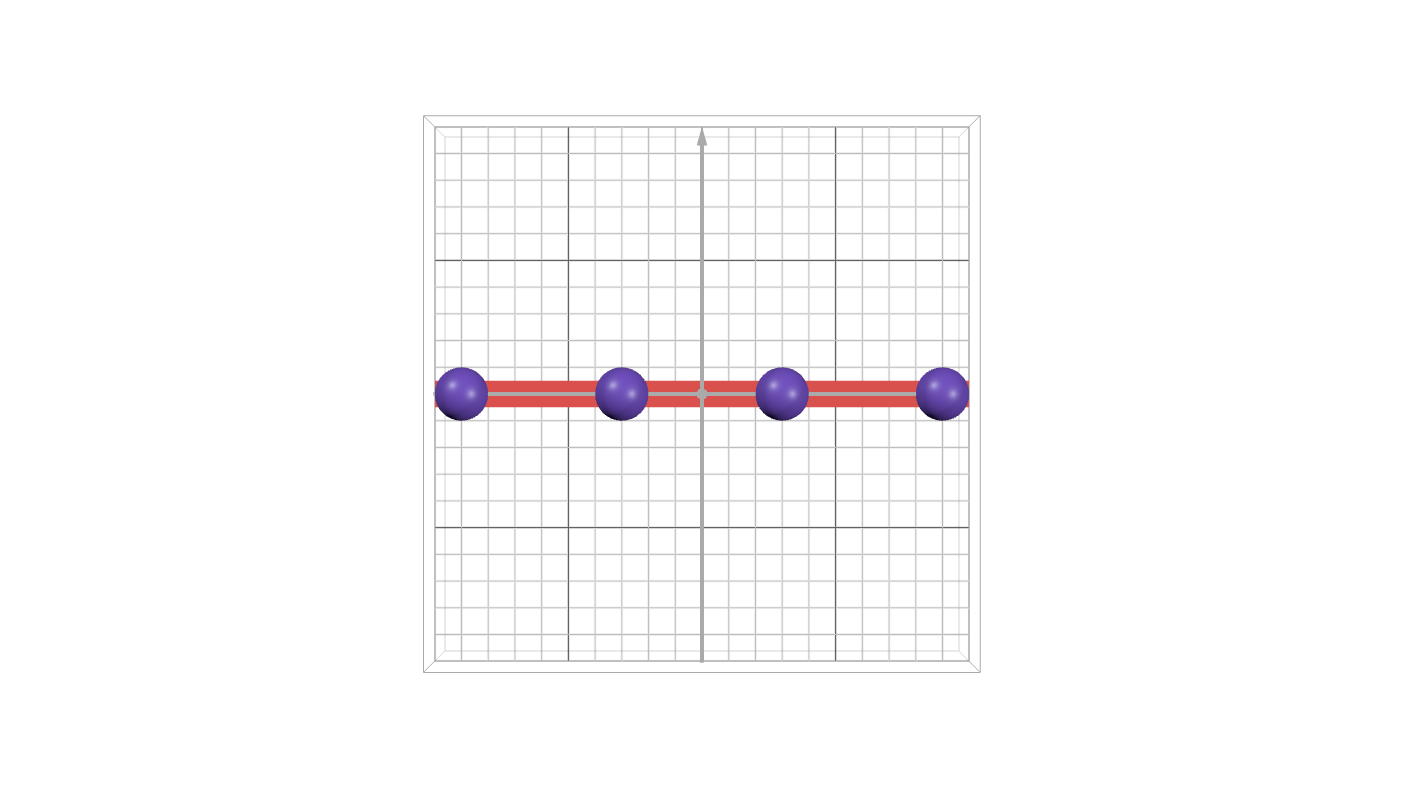}
    \label{fig:initial_ribbon_1}
\end{subfigure}
\hfill
\begin{subfigure}{0.3\textwidth}
    \centering
    \includegraphics[width=\linewidth]{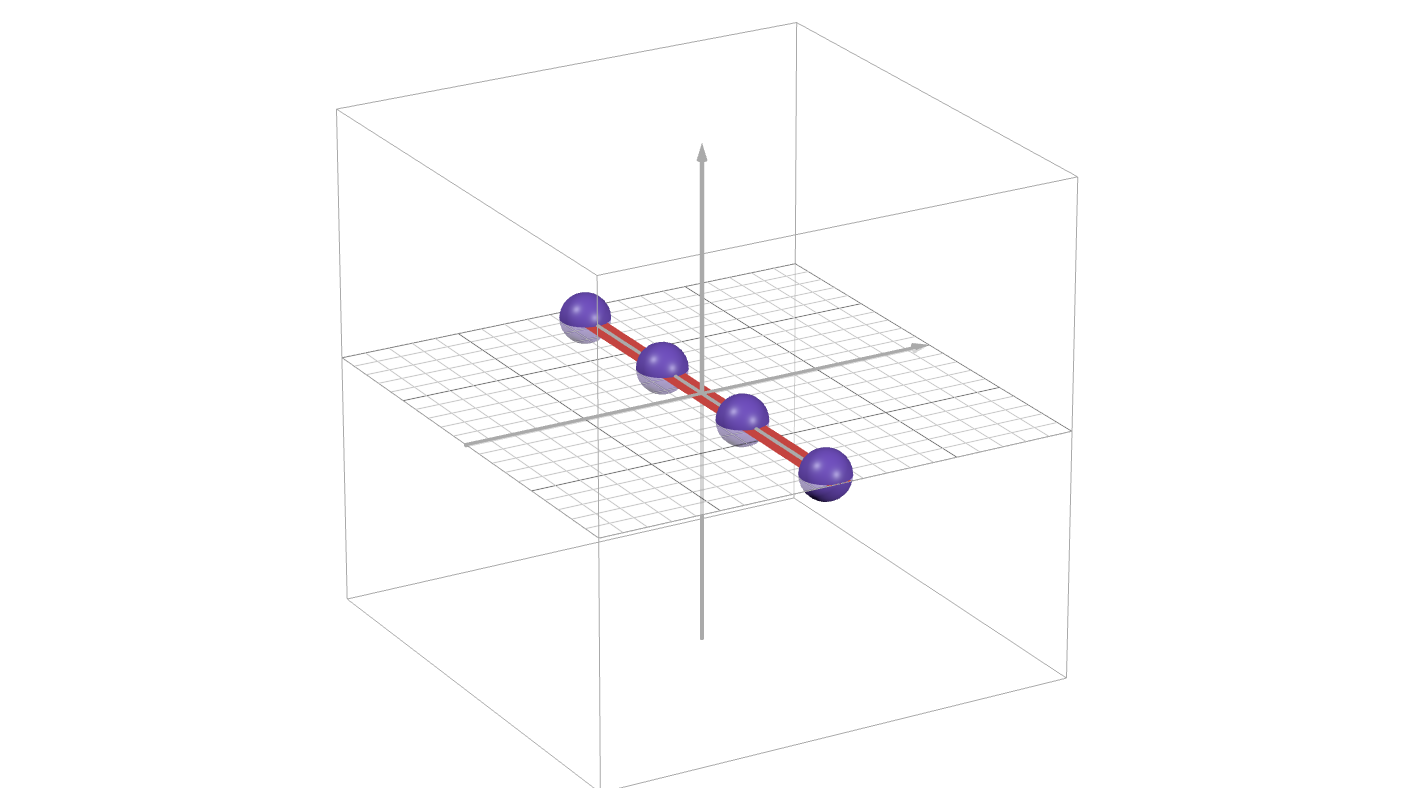}
    \label{fig:initial_ribbon_2}
\end{subfigure}
\caption{\textbf{The Flat Ribbon Chain.} The initial state $|C_N\rangle$ is represented as a linear array of spheres connected by flat, untwisted ribbons ($\theta=0^\circ$), reflecting the real-valued nature of the initial $CZ$ entanglement.}
\label{fig:initial_ribbon_chain}
\end{figure}

\noindent Before analyzing measurement dynamics, we must define the initial resource state $|C_N\rangle$ of Eq.(\ref{cluster}) within this topological framework. Recall that the standard LCS is generated by applying $CZ$ gates to a register of $|+\rangle$ states. The $CZ$ interaction corresponds to the Ising phase factor $(-1)^{x_i x_{i+1}}$. Crucially, this factor is purely real. In our Twist Dictionary, a real correlation corresponds to an untwisted connection. Therefore, we map the initial LCS $|C_N\rangle$ to a \textit{Flat Ribbon Chain}, illustrated in Fig. \ref{fig:initial_ribbon_chain}:
\begin{itemize}
    \item \textit{Spheres ($S$):} The $N$ qubits are represented as $N$ aligned spheres.
    \item \textit{Connecting Ribbons ($\mathcal{R}$):} Each adjacent pair $(i, i+1)$ is connected by a ribbon $\mathcal{R}_{i, i+1}$ initialized with a twist angle $\theta = 0^\circ$.
\end{itemize}
This \textit{zero-writhe} initialization is the topological dual to the standard graph state definition, where edges represent simple binary connectivity without complex phase. \textit{Zero-writhe} refers to the geometric condition that the ribbons representing entangling links in the initial cluster state carry no intrinsic twist or coiling i.e. the ribbon framing is flat everywhere before any measurement-induced operation. All subsequent measurement-induced twists (chiral $\pm 90^\circ$ or flipped $180^\circ$) are defined as deviations from this flat inertial frame.
% ==========================================
% FIGURE: INITIAL FLAT CHAIN - SIDE BY SIDE
% ==========================================

\subsection{Case I: Z-Basis (Topological Severance)}
\noindent As we know the $Z$-basis measurement is destructive, the ribbon topology is defined by its discontinuity.
\subsubsection{Measurement of end-qubit ($Q_1$)}
\begin{itemize}
    \item \textit{Outcome $|0\rangle_1$}
    Topology (\textsc{Pruning}:) The first ribbon segment is removed. The chain shortens ($N \to N-1$), as shown in Fig. \ref{fig:z_basis_pruning}. No twist is induced on the remaining bulk.
    % ==========================================
    % FIGURE: Z-BASIS END OUTCOME 0 - SIDE BY SIDE
    % ==========================================
    \begin{figure}[t]
    \centering
    \begin{subfigure}{0.3\textwidth}
        \centering
        \includegraphics[width=\linewidth]{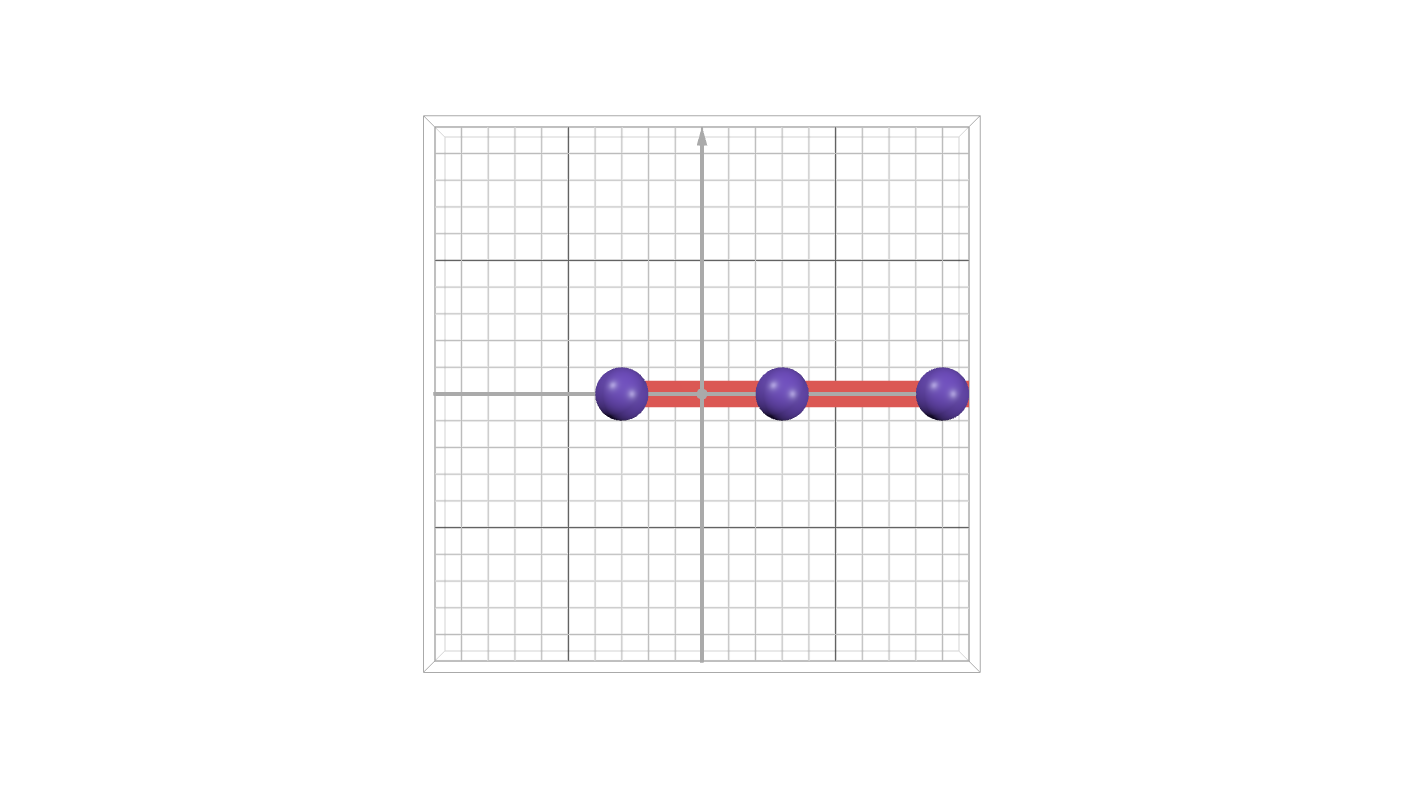}
        \label{fig:z_end_0_1}
    \end{subfigure}
    \hfill
    \begin{subfigure}{0.3\textwidth}
        \centering
        \includegraphics[width=\linewidth]{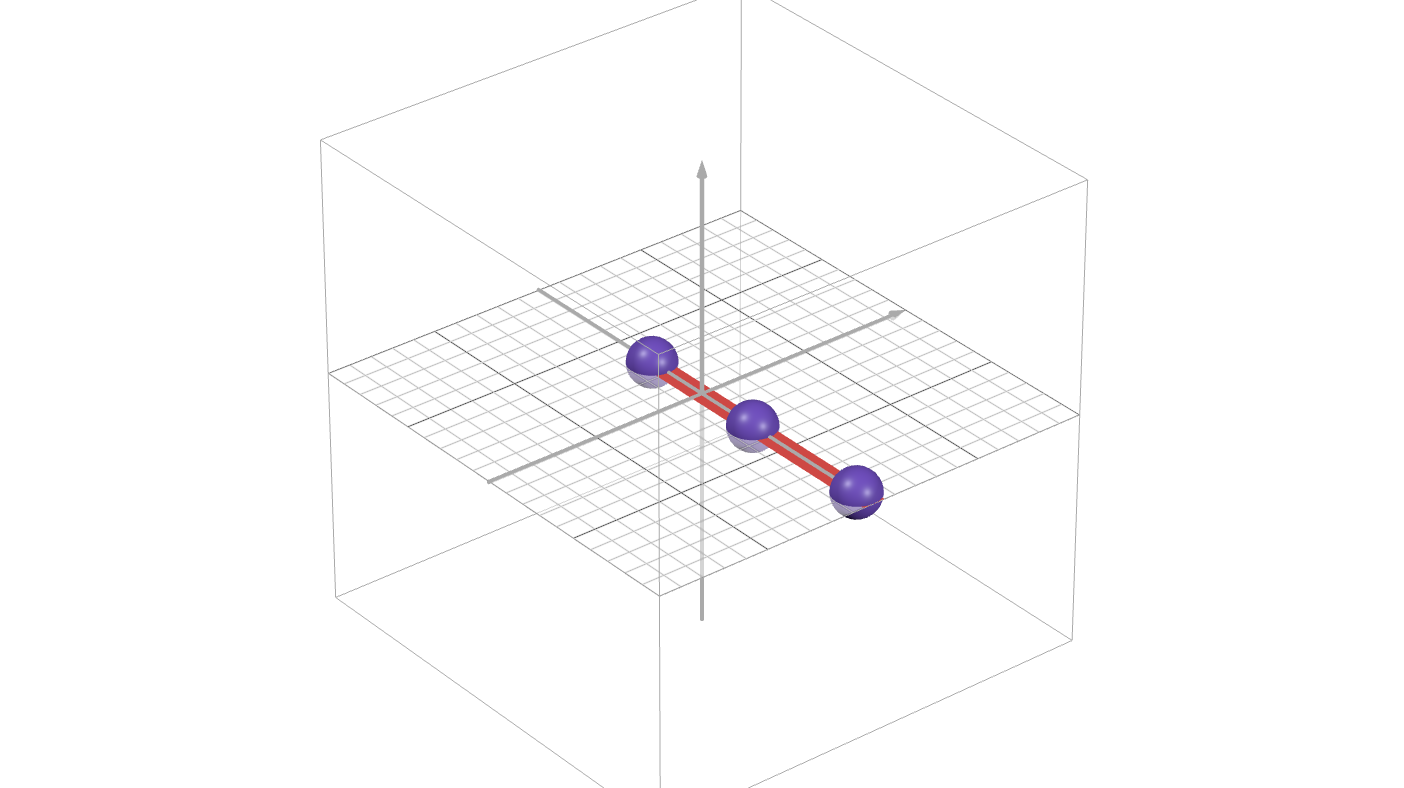}
        \label{fig:z_end_0_2}
    \end{subfigure}
    \caption{\textbf{$Z$-Basis Pruning (Outcome $|0\rangle_1$).} The measurement removes the first sphere and the ribbon $\mathcal{R}_{1,2}$. The remaining chain starts at $Q_2$ with no induced twist.}
    \label{fig:z_basis_pruning}
    \end{figure}    
    \item \textit{Outcome $|1\rangle_1$
    Topology} (\textsc{Pruning with Flip}:) The first ribbon is removed. The outcome induces a Pauli $Z$ on the qubit $Q_2$. In the ribbon frame, this corresponds to a $180^\circ$ flip of the next ribbon in the chain (between qubits $Q_2$ and $Q_3$), visualized in Fig. \ref{fig:z_basis_pruning_flip}.
    % ==========================================
    % FIGURE: Z-BASIS END OUTCOME 1 - SIDE BY SIDE
    % ==========================================
    \begin{figure}[t]
    \centering
    \begin{subfigure}{0.3\textwidth}
        \centering
        \includegraphics[width=\linewidth]{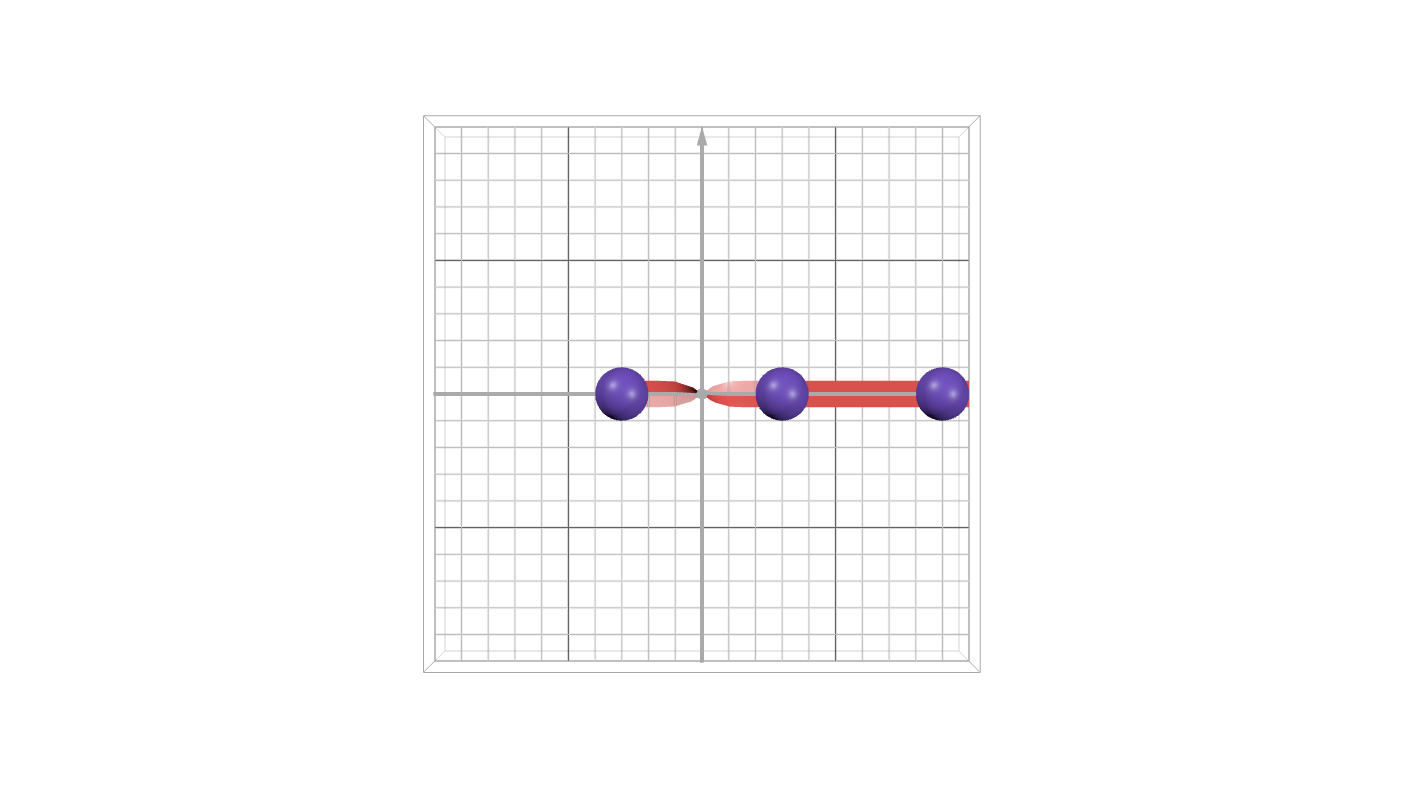}
        \label{fig:z_end_1_1}
    \end{subfigure}
    \hfill
    \begin{subfigure}{0.3\textwidth}
        \centering
        \includegraphics[width=\linewidth]{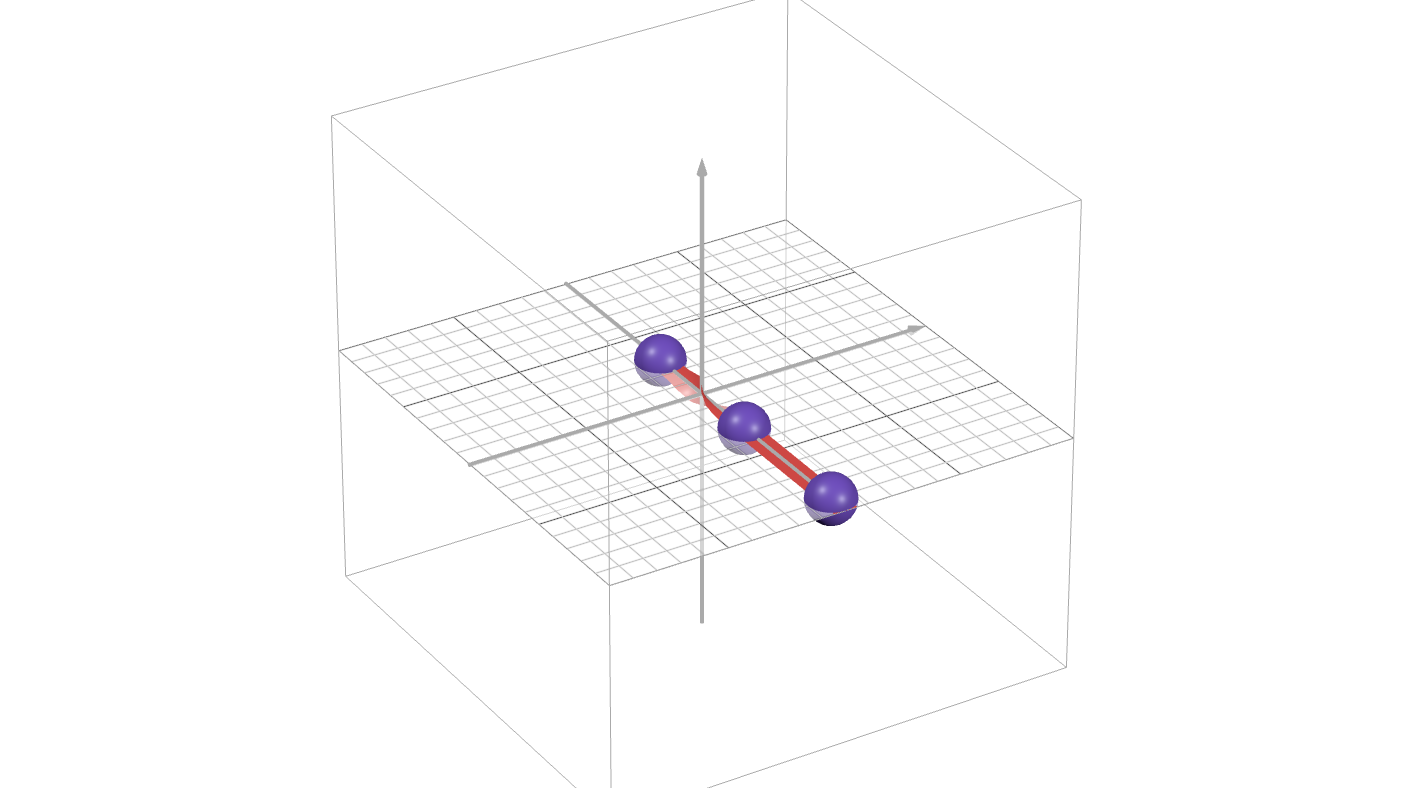}
        \label{fig:z_end_1_2}
    \end{subfigure}
    \caption{\textbf{Z-Basis Pruning with Flip (Outcome $|1\rangle_1$).} The measurement removes the qubit $Q_1$. The resulting Pauli $Z$ correction on the qubit $Q_2$ manifests as a $180^\circ$ flip on the \textit{subsequent} ribbon $\mathcal{R}_{2,3}$.}
    \label{fig:z_basis_pruning_flip}
    \end{figure}    
\end{itemize}

\subsubsection{Measurement of bulk-qubit ($Q_k$)}
\begin{itemize}
    \item \textit{Outcome $|0\rangle_k$
    Topology} (\textsc{Clean Cut}:) The ribbon at site $k$ is severed. The left and right segments are topologically disconnected ($R=1$), as depicted in Fig. \ref{fig:z_basis_severance}.
    % ==========================================
    % FIGURE: Z-BASIS BULK OUTCOME 0 - SIDE BY SIDE
    % ==========================================
    \begin{figure}[t]
    \centering
    \begin{subfigure}{0.3\textwidth}
        \centering
        \includegraphics[width=\linewidth]{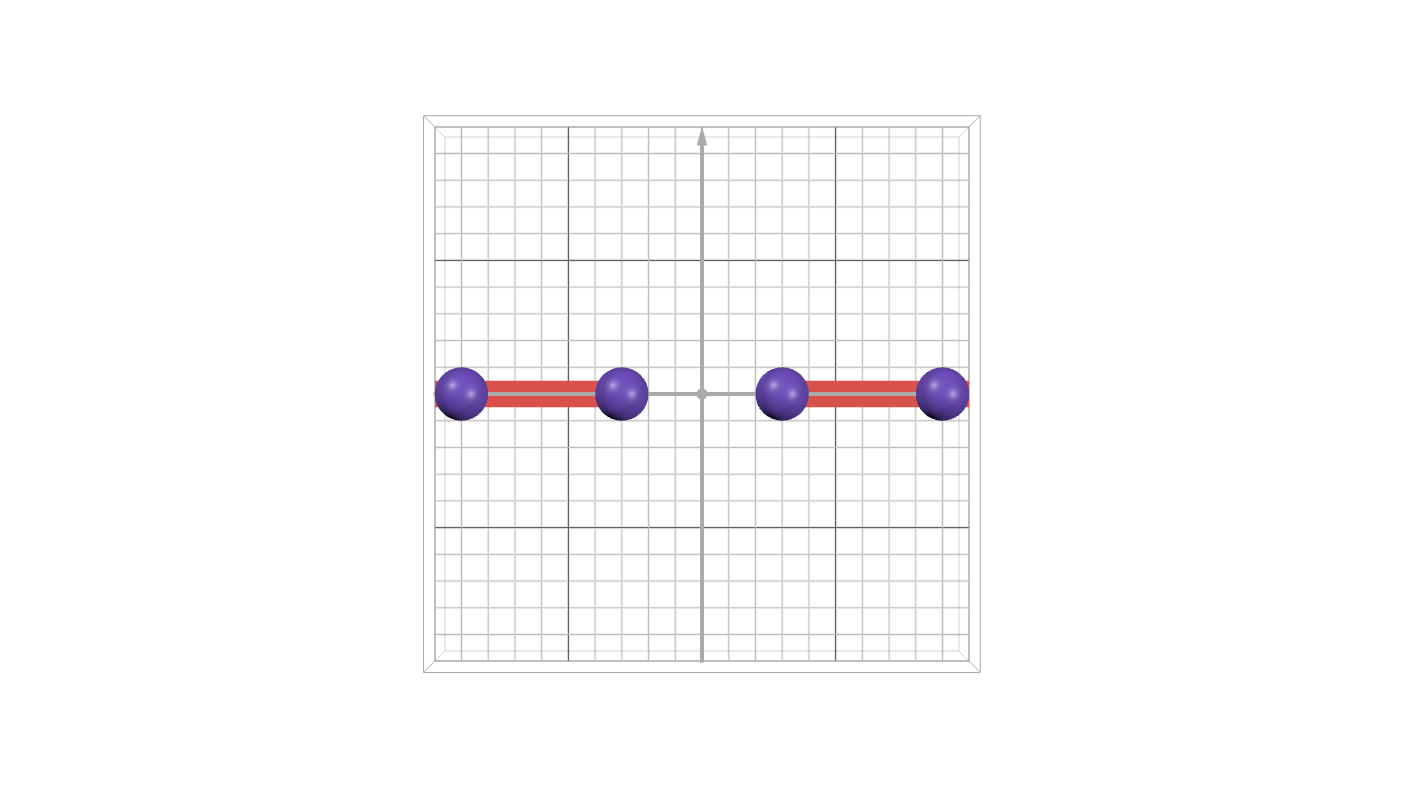}
        \label{fig:z_bulk_0_1}
    \end{subfigure}
    \hfill
    \begin{subfigure}{0.3\textwidth}
        \centering
        \includegraphics[width=\linewidth]{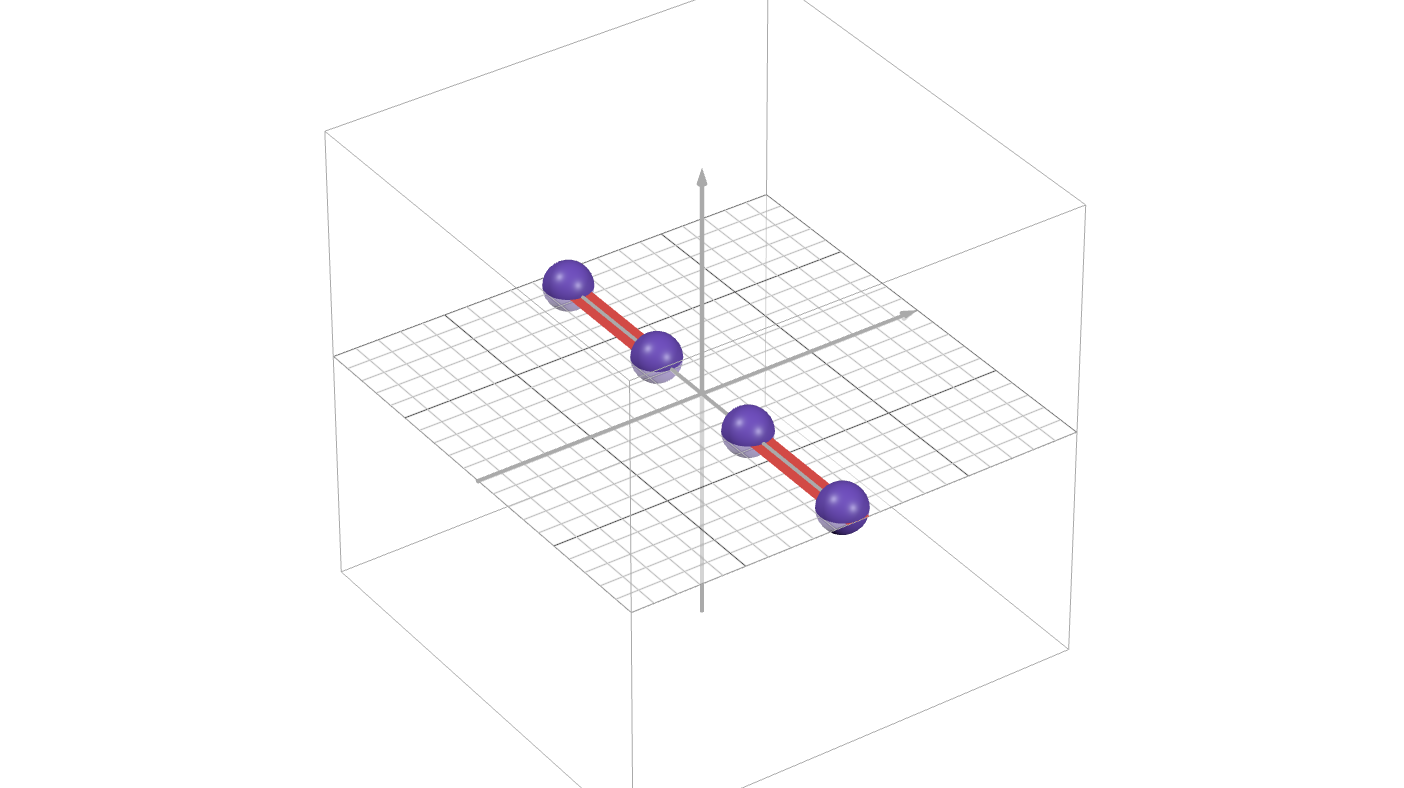}
        \label{fig:z_bulk_0_2}
    \end{subfigure}
    \caption{\textbf{Z-Basis Severance (Outcome $|0\rangle_k$).} The central sphere $Q_k$ and its attached ribbons are removed. The chain is physically cut into two disconnected, untwisted segments.}
    \label{fig:z_basis_severance}
    \end{figure}        
    \item \textit{Outcome $|1\rangle_k$
    Topology} (\textsc{Cut with Boundary Flip:}) The ribbon is severed. The outcome applies $Z$ operators to neighbours $Q_{k-1}$ and $Q_{k+1}$. This manifests as a $180^\circ$ flip on the adjacent ribbons connecting them to the rest of the chain, shown in Fig. \ref{fig:z_basis_boundary_flips}.
    % ==========================================
    % FIGURE: Z-BASIS BULK OUTCOME 1 - SIDE BY SIDE
    % ==========================================
    \begin{figure}[t]
    \centering
    \begin{subfigure}{0.3\textwidth}
        \centering
        \includegraphics[width=\linewidth]{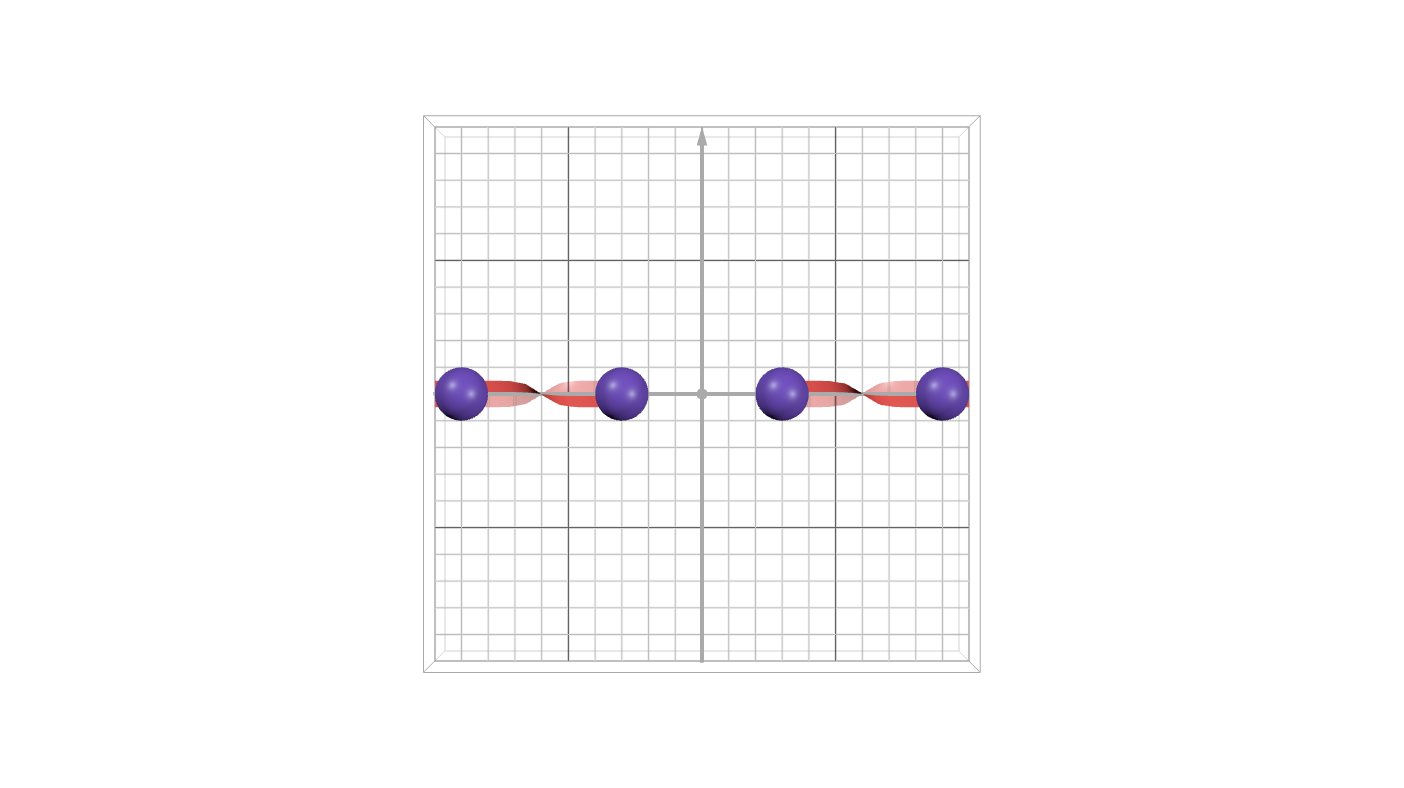}
        \label{fig:z_bulk_1_1}
    \end{subfigure}
    \hfill
    \begin{subfigure}{0.3\textwidth}
        \centering
        \includegraphics[width=\linewidth]{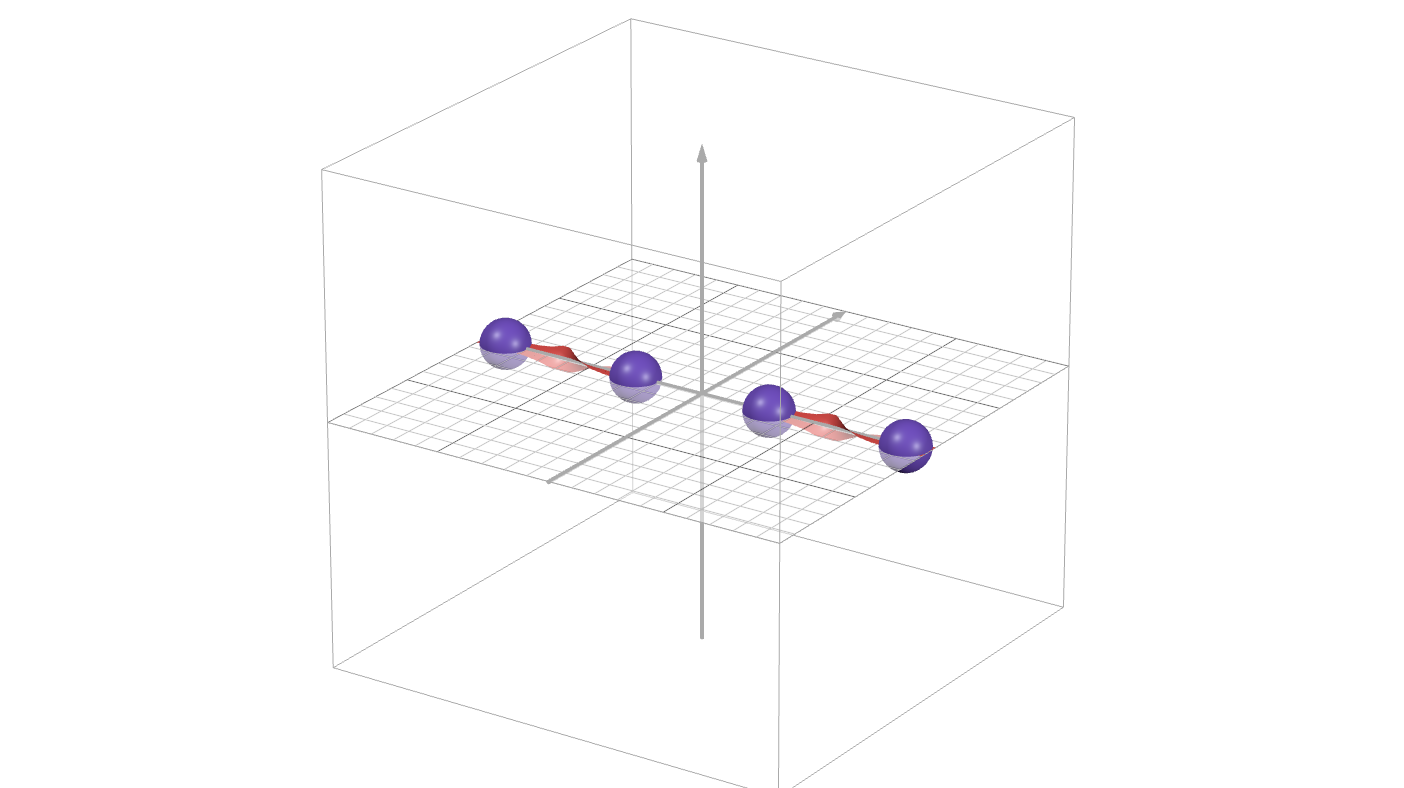}
        \label{fig:z_bulk_1_2}
    \end{subfigure}
    \caption{\textbf{Severance with Boundary Flips (Outcome $|1\rangle_k$).} The chain is cut at $Q_k$. The byproduct operators $Z_{k-1}Z_{k+1}$ induce $180^\circ$ flips on the ribbons adjacent to the cut ($\mathcal{R}_{k-2, k-1}$ and $\mathcal{R}_{k+1, k+2}$).}
    \label{fig:z_basis_boundary_flips}
    \end{figure}     
\end{itemize}

\subsection{Case II: $X$-Basis and Topological Stratification}
\noindent The $X$-basis measurement imposes strict parity constraints on its neighbours. At the boundaries, this results in sequence shortening. In the bulk, measuring a qubit in the transverse basis acts as a quantum switch that drives the post-measurement state into a macroscopic superposition of distinct topological sectors. 

\noindent \textit{Qualitative Meaning of Topological Stratification and its Geometric Nature.} To represent this superposition without violating the mutually exclusive, orthogonal realities of quantum branches, we introduce the concept of \textit{Topological Stratification}. Qualitatively, stratification means that instead of a single continuous 1D trajectory or a physically intersecting junction, the quantum state is layered into parallel, non-intersecting tracks. Each individual track represents a distinct, internally consistent branch of the macroscopic superposition. 

\noindent Is this stratification geometric? Yes, fundamentally so. The stratification is constructed geometrically in $\mathbb{R}^3$ by placing the mutually exclusive sectors onto parallel, separate ribbon strata that span across the measurement void (where $Q_k$ was deleted). Because these tracks never physically touch or cross, the geometry visually enforces the quantum mechanical orthogonality of the superposition branches, while allowing each separate stratum to host its own distinct geometric labels—such as $180^\circ$ half-twists ($\theta = 180^\circ$) representing outcome-dependent Pauli $Z$ corrections.

\subsubsection{Measurement of end-qubit ($Q_1$)}
\begin{itemize}
    \item \textit{Outcome $|+\rangle_1$
    Topology} (\textsc{Shortening}:) The measurement disentangles $Q_2$. The active boundary effectively skips to $Q_3$, illustrated in Fig. \ref{fig:x_basis_flat_shortening}. The logic flows forward without twist.
    % ==========================================
    % FIGURE: X-BASIS END OUTCOME +
    % ==========================================
    \begin{figure}[t]
    \centering
    \begin{subfigure}{0.3\textwidth}
        \centering
        \includegraphics[width=\linewidth]{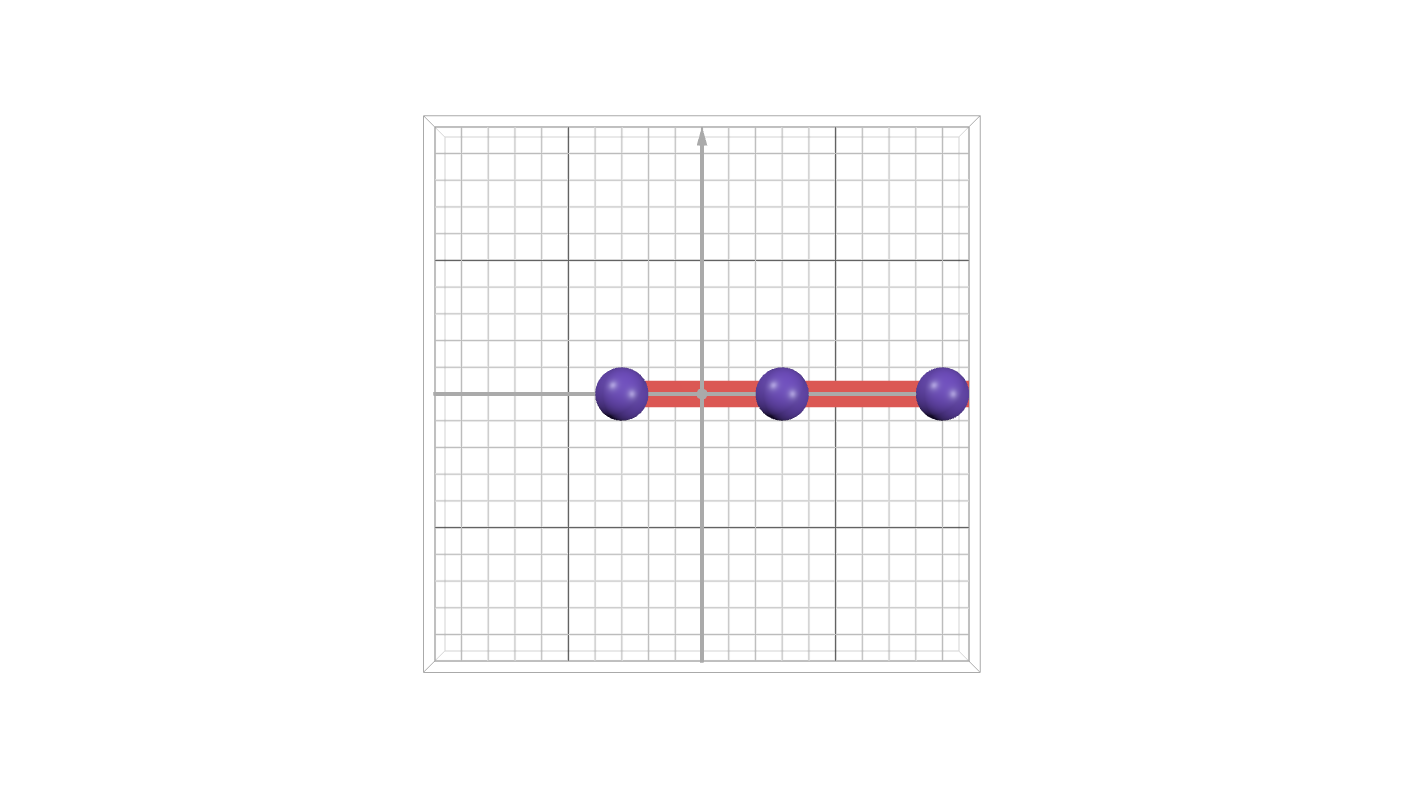}
        \label{fig:x_end_plus_1}
    \end{subfigure}
    \hfill
    \begin{subfigure}{0.3\textwidth}
        \centering
        \includegraphics[width=\linewidth]{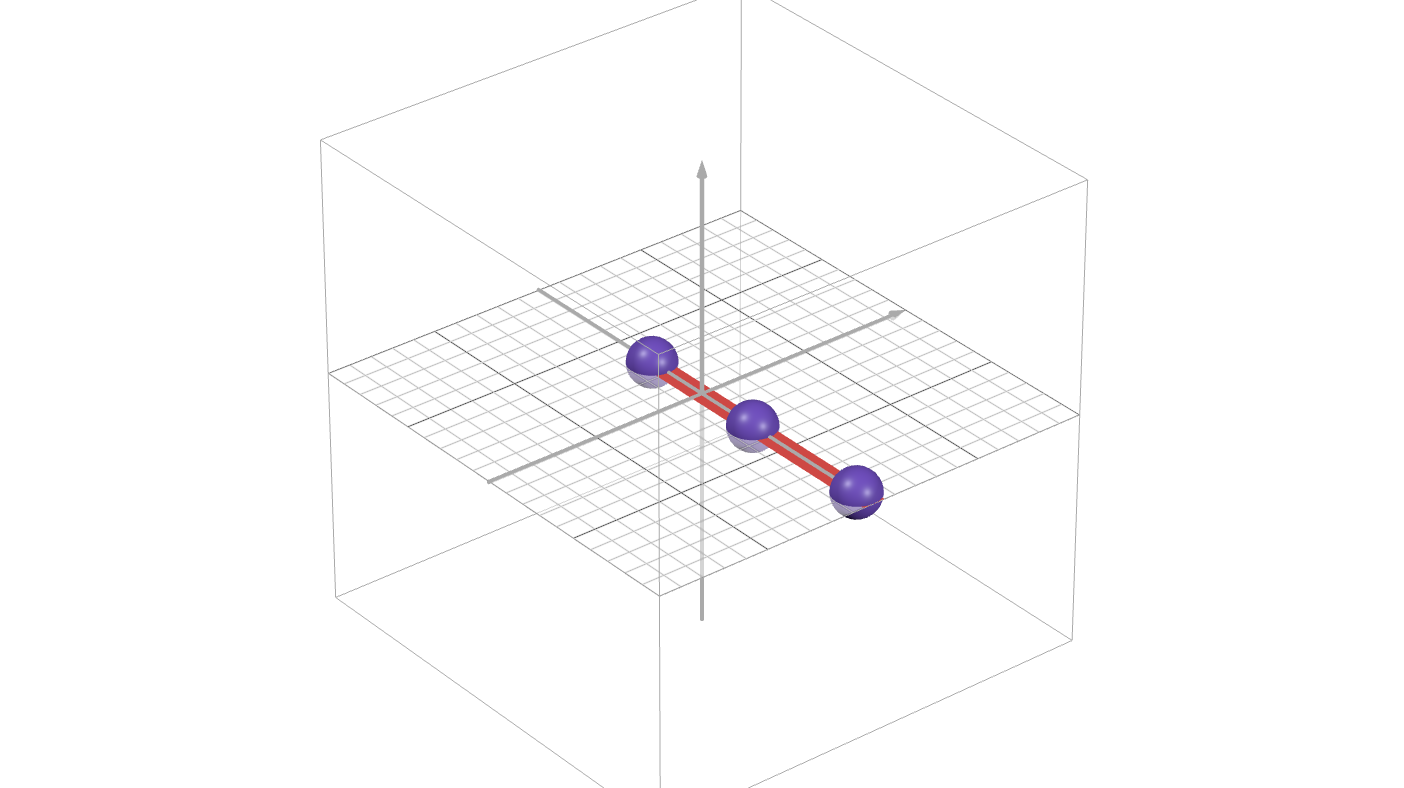}
        \label{fig:x_end_plus_2}
    \end{subfigure}
    \caption{\textbf{$X$-Basis Flat Shortening (Outcome $|+\rangle_1$).} $Q_1$ is removed and $Q_2$ is disentangled. The active boundary skips to $Q_3$, with the new boundary ribbon $\mathcal{R}_{3,4}$ remaining flat ($\theta=0^\circ$).}
    \label{fig:x_basis_flat_shortening}
    \end{figure}    
    
    \item \textit{Outcome $|-\rangle_1$
    Topology} (\textsc{Shortening with Flip:}) The boundary skips to $Q_3$, but the induced $Z$ operator creates a $180^\circ$ flip on the new boundary ribbon (Fig. \ref{fig:x_basis_flipped_shortening}).
    % ==========================================
    % FIGURE: X-BASIS END OUTCOME -
    % ==========================================
    \begin{figure}[t]
    \centering
    \begin{subfigure}{0.3\textwidth}
        \centering
        \includegraphics[width=\linewidth]{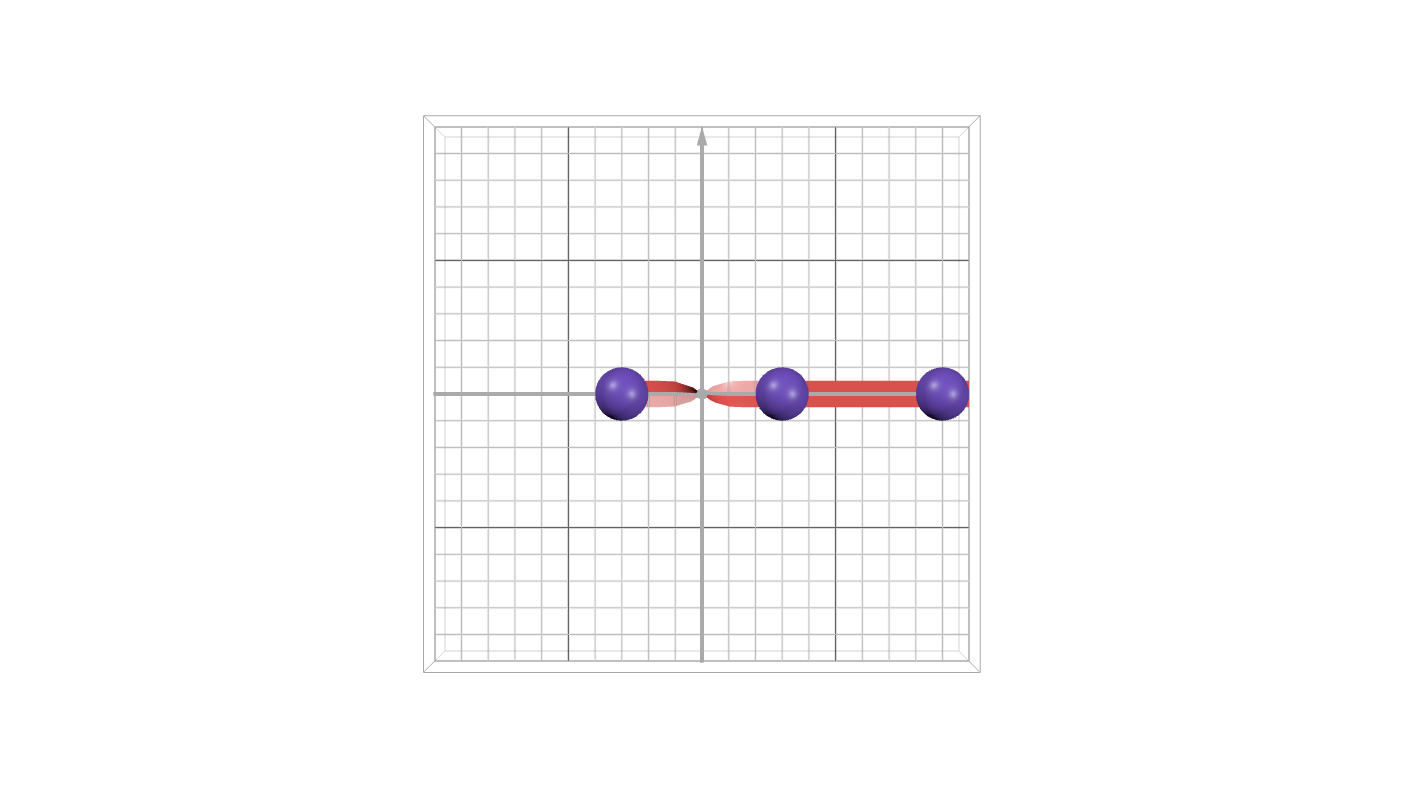}
        \label{fig:x_end_minus_1}
    \end{subfigure}
    \hfill
    \begin{subfigure}{0.3\textwidth}
        \centering
        \includegraphics[width=\linewidth]{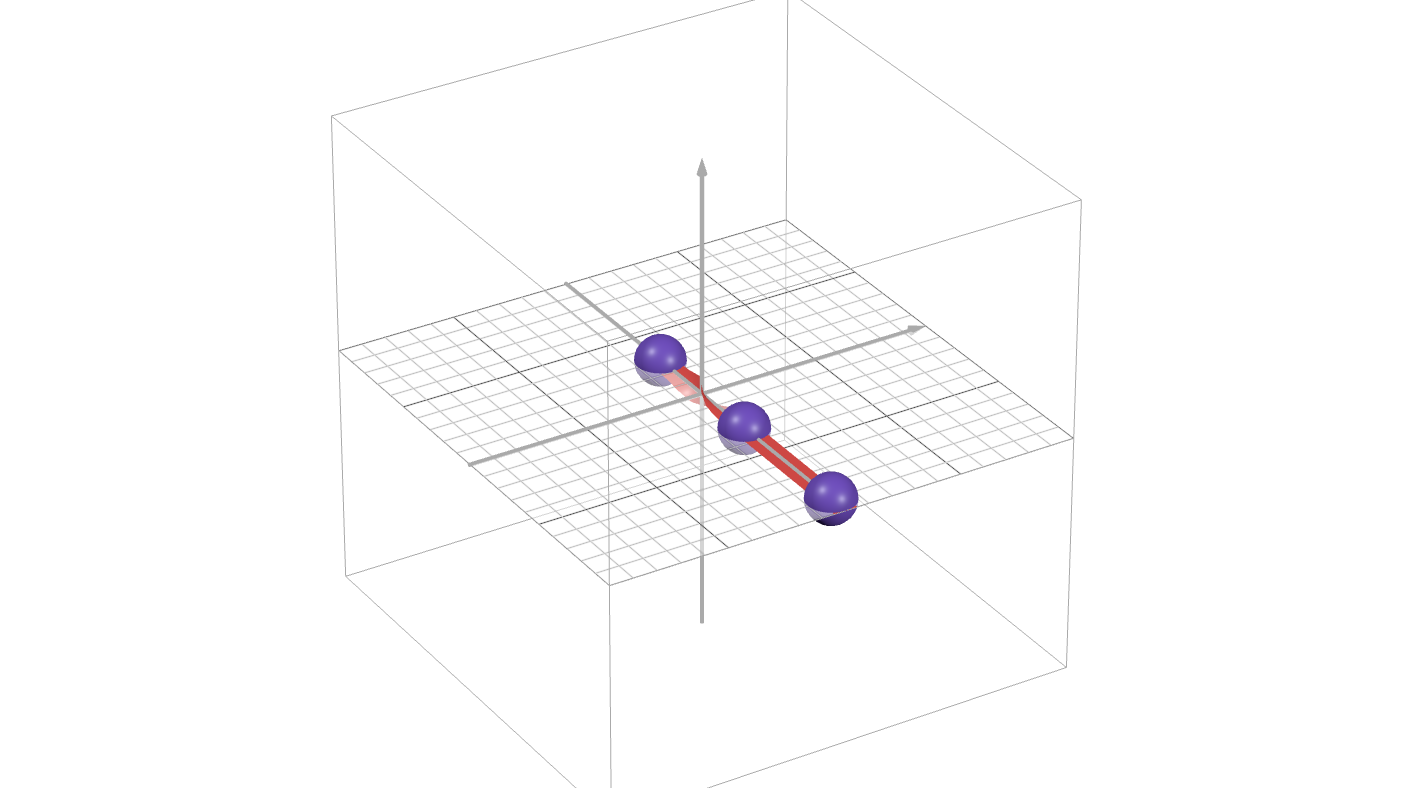}
        \label{fig:x_end_minus_2}
    \end{subfigure}
    \caption{\textbf{$X$-Basis Flipped Shortening (Outcome $|-\rangle_1$).} The boundary skips to $Q_3$. The negative parity outcome induces a $Z_3$ correction, causing the new boundary ribbon $\mathcal{R}_{3,4}$ to flip $180^\circ$.}
    \label{fig:x_basis_flipped_shortening}
    \end{figure}    
\end{itemize}

\subsubsection{Measurement of Bulk Qubit ($Q_k$)}
\begin{itemize}
    \item \textit{Outcome $|+\rangle_k$
    Topology} (\textsc{Correlated Stratification}:) The measurement stratifies the topology into two parallel, non-intersecting tracks representing the correlated parity sectors ($x_{k-1} = x_{k+1}$). Spanning across the measurement void, the upper track (representing the unperturbed sector $|\chi_0\rangle$) remains geometrically flat ($\theta = 0^\circ$) on both its left and right segments. Conversely, the lower track (representing the modified sector $|\chi_1\rangle = Z_{k-2}Z_{k+2}|\chi_0\rangle$) carries a full $180^\circ$ flip ($\theta = 180^\circ$) on both its outer left and outer right segments to encode the $Z_{k-2}$ and $Z_{k+2}$ Pauli corrections. See Fig.~\ref{fig:x_basis_flat_splice}.
    
    % ==========================================
    % FIGURE: X-BASIS BULK OUTCOME + 
    % ==========================================
    \begin{figure}[t]
    \centering
    \begin{subfigure}{0.3\textwidth}
        \centering
        \includegraphics[width=\linewidth]{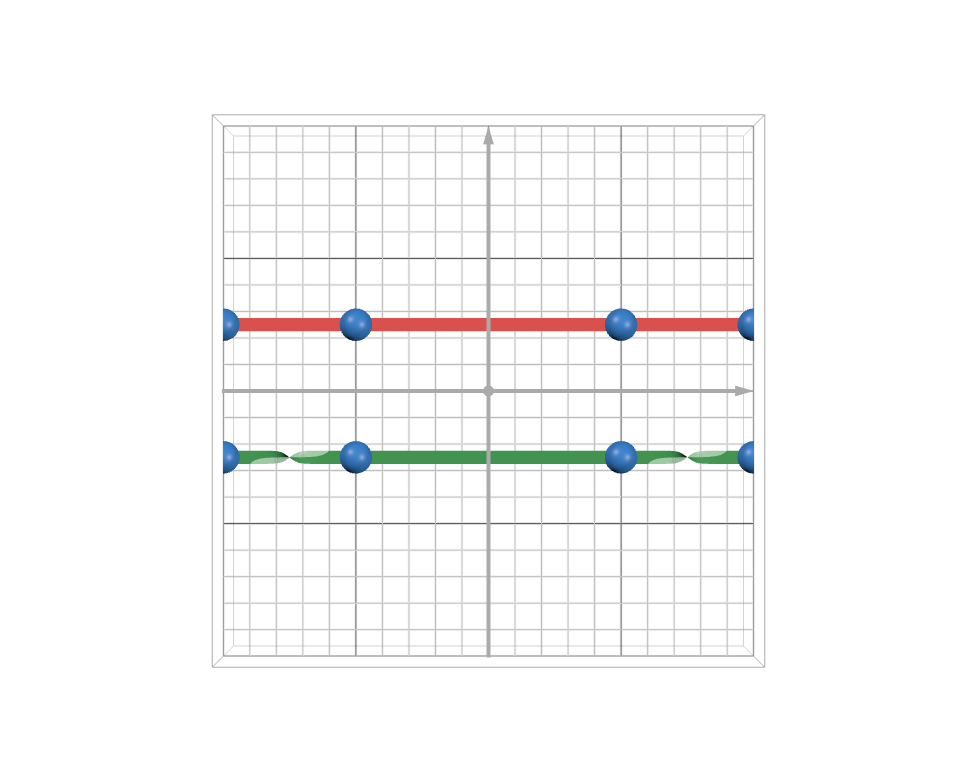}
        \label{fig:x_bulk_plus_1}
    \end{subfigure}
    \hfill
    \begin{subfigure}{0.3\textwidth}
        \centering
        \includegraphics[width=\linewidth]{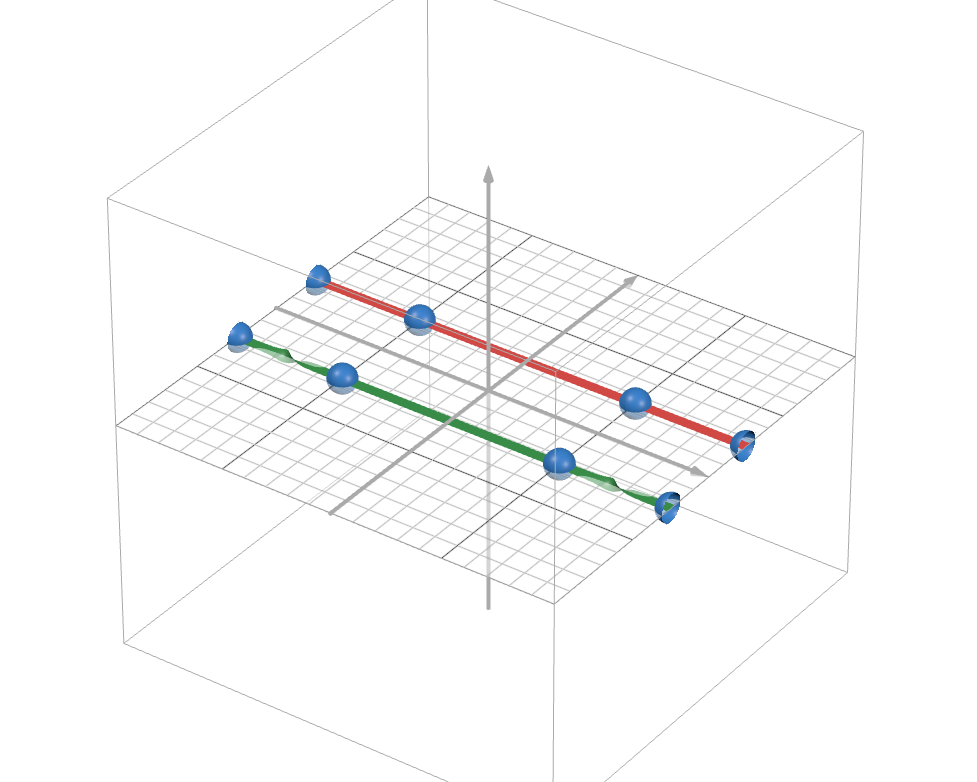}
        \label{fig:x_bulk_plus_2}
    \end{subfigure}
        \caption{\textbf{Correlated Stratification (Outcome $|+\rangle_k$).} The state is geometrically stratified into parallel tracks. The upper track remains flat, while the lower track carries $180^\circ$ topological flips on its outer segments to encode the local $Z_{k-2}$ and $Z_{k+2}$ corrections.}
        \label{fig:x_basis_flat_splice}
    \end{figure}

    \item \textit{Outcome $|-\rangle_k$
    Topology} (\textsc{Anti-Correlated Stratification}:) The topology stratifies into two parallel tracks representing the anti-correlated parity sectors ($x_{k-1} \neq x_{k+1}$), distributing the phase corrections asymmetrically. The upper track remains flat on the left but carries a $180^\circ$ flip on the right (encoding $Z_{k+2}$), while the lower track carries a $180^\circ$ flip on the left (encoding $Z_{k-2}$) and remains flat on the right. See Fig.~\ref{fig:x_basis_flipped_splice}.

    % ==========================================
    % FIGURE: X-BASIS BULK OUTCOME - 
    % ==========================================
    \begin{figure}[t]
    \centering
    \begin{subfigure}{0.3\textwidth}
        \centering
        \includegraphics[width=\linewidth]{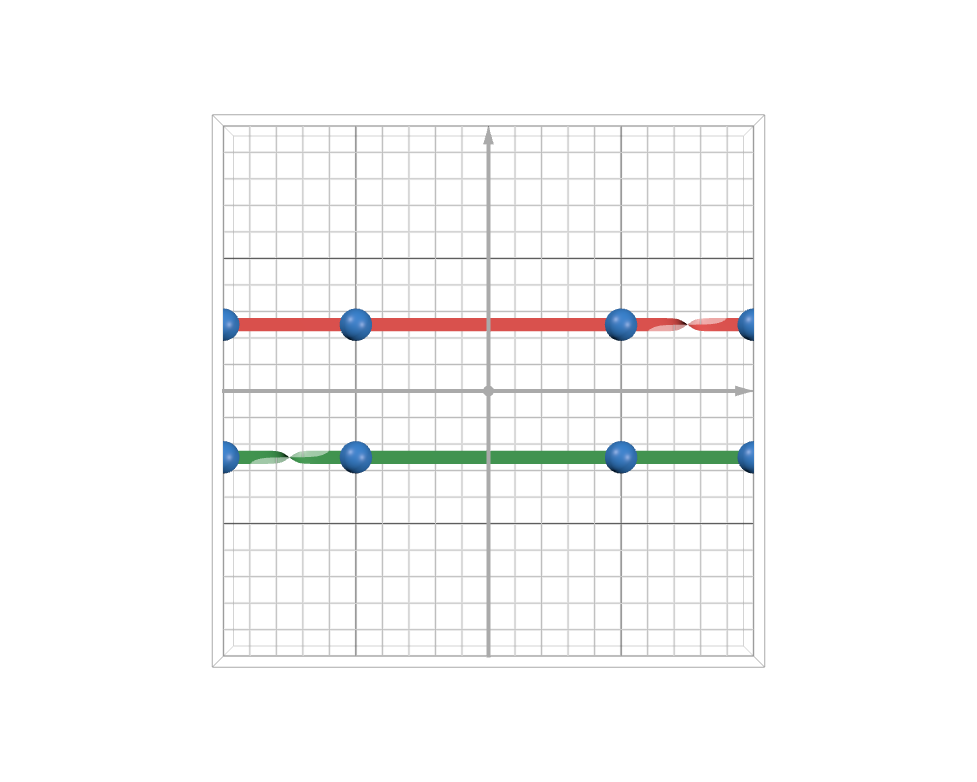}
        \label{fig:x_bulk_minus_1}
    \end{subfigure}
    \hfill
    \begin{subfigure}{0.3\textwidth}
        \centering
        \includegraphics[width=\linewidth]{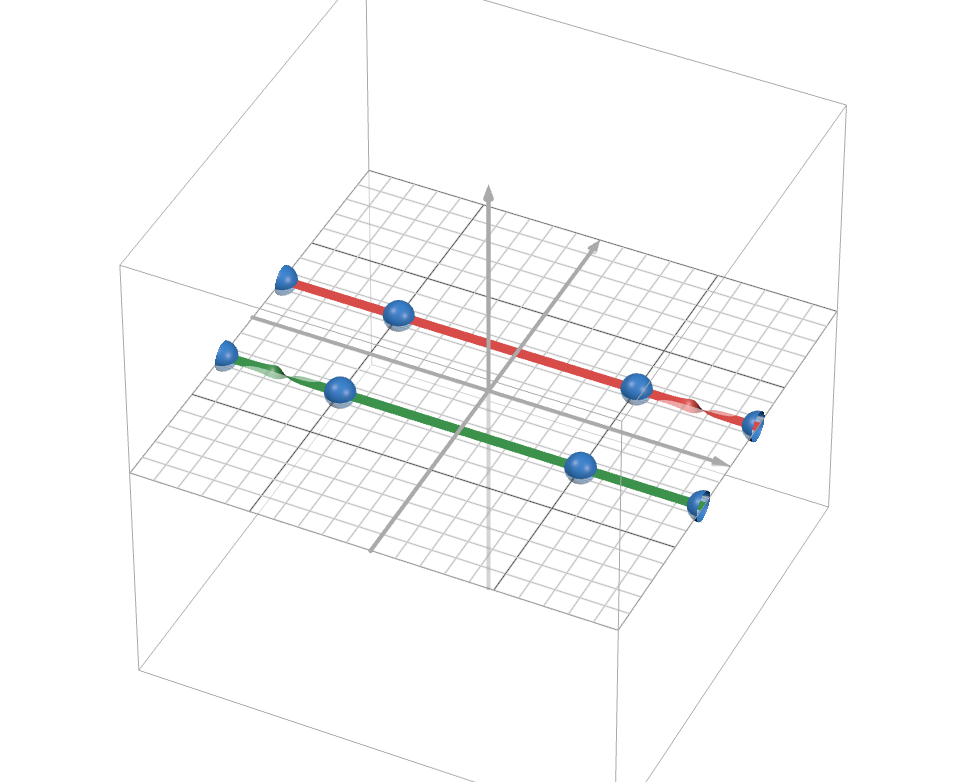}
        \label{fig:x_bulk_minus_2}
    \end{subfigure}
        \caption{\textbf{Anti-Correlated Stratification (Outcome $|-\rangle_k$).} The parallel tracks encode the anti-correlated sectors with asymmetric twist placement. The upper track twists only on the right, and the lower track twists only on the left.}
        \label{fig:x_basis_flipped_splice}
    \end{figure}
\end{itemize}

\subsection{Case III: $Y$-Basis (Chiral Splicing)}
\noindent The $Y$-basis measurement embeds complex phase via \textit{Twisted Splicing}. Unlike the stratified superposition generated by the $X$-basis, the $Y$-basis maintains a single, distinct 1D topological continuity.
\subsubsection{Measurement of end-qubit ($Q_1$)}
\begin{itemize}
    \item \textit{Outcome $|+i\rangle_1$
 Topology} \textsc{Right-Handed Boundary Twist:} The chain shortens, but the open end of the ribbon at $Q_2$ is twisted $+90^\circ$ (Clockwise), as visualized in Fig. \ref{fig:y_basis_rh_boundary_twist}. This represents the application of the $S$ gate.
    % ==========================================
    % FIGURE: Y-BASIS END OUTCOME +i
    % ==========================================
    \begin{figure}[t]
    \centering
    \begin{subfigure}{0.3\textwidth}
        \centering
        \includegraphics[width=\linewidth]{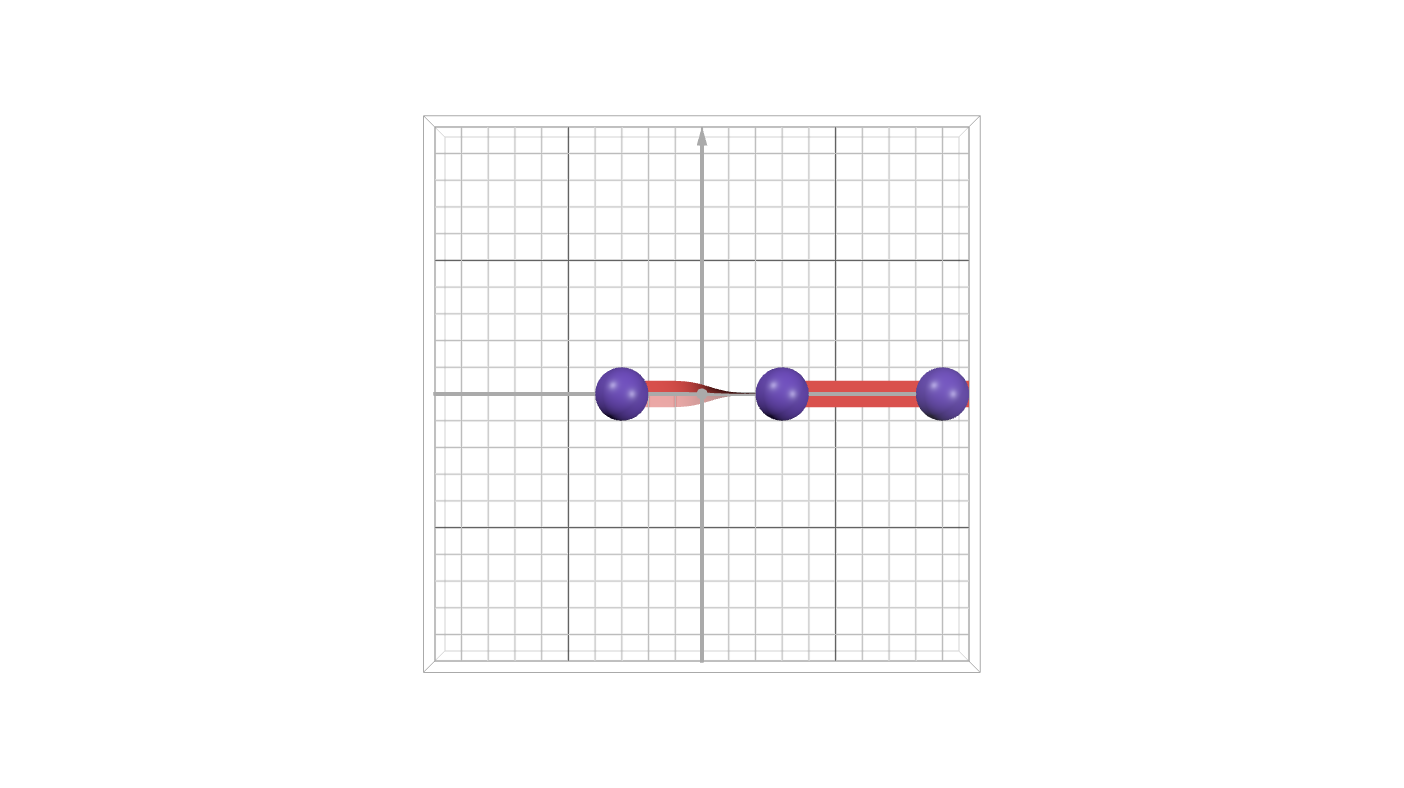}
        \label{fig:y_end_plusi_1}
    \end{subfigure}
    \hfill
    \begin{subfigure}{0.3\textwidth}
        \centering
        \includegraphics[width=\linewidth]{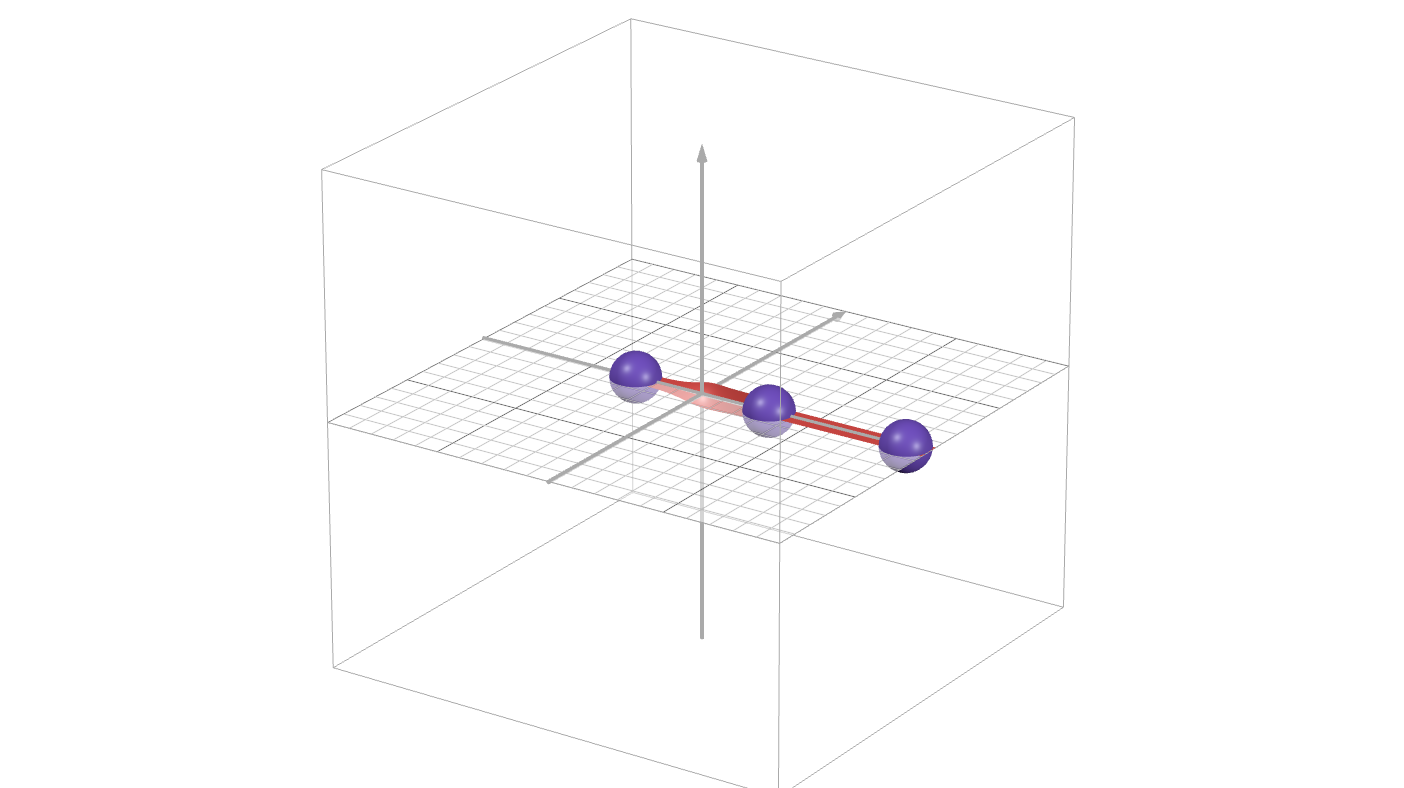}
        \label{fig:y_end_plusi_2}
    \end{subfigure}
    \caption{\textbf{Right-Handed Boundary Twist (Outcome $|+i\rangle_1$)} The boundary moves to $Q_2$. The open end of the ribbon $\mathcal{R}_{2,3}$ acquires a $+90^\circ$ clockwise twist, visualized as the right edge crossing \textit{OVER} the left.}
    \label{fig:y_basis_rh_boundary_twist}
    \end{figure}    
    \item \textit{Outcome $|-i\rangle_1$
    Topology} \textsc{Left-Handed Boundary Twist:} The chain shortens, and the ribbon at $Q_2$ is twisted $-90^\circ$ (Counter-Clockwise), shown in Fig. \ref{fig:y_basis_lh_boundary_twist}. This represents the $S^\dagger$ gate.
    % ==========================================
    % FIGURE: Y-BASIS END OUTCOME -i
    % ==========================================
    \begin{figure}[t]
    \centering
    \begin{subfigure}{0.3\textwidth}
        \centering
        \includegraphics[width=\linewidth]{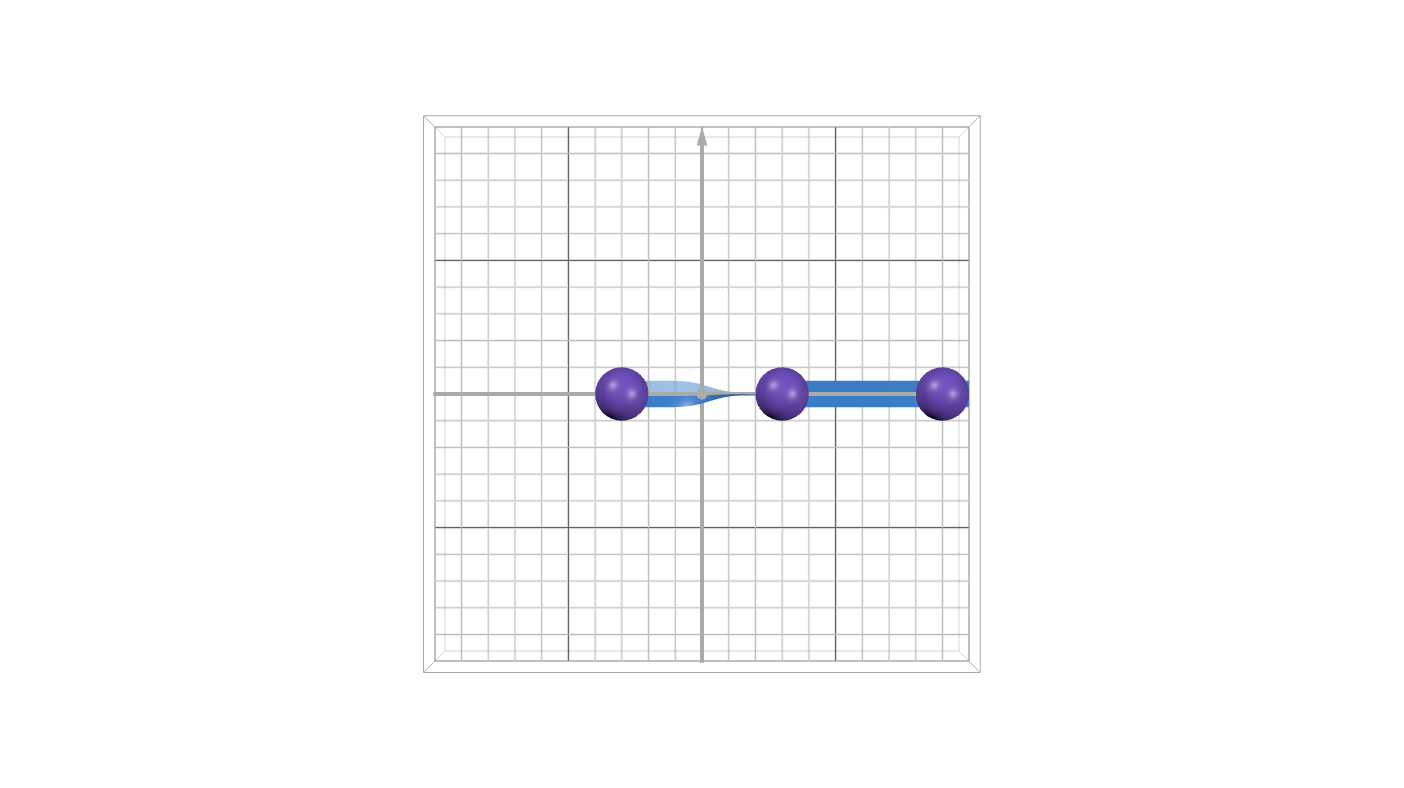}
        \label{fig:y_end_minusi_1}
    \end{subfigure}
    \hfill
    \begin{subfigure}{0.3\textwidth}
        \centering
        \includegraphics[width=\linewidth]{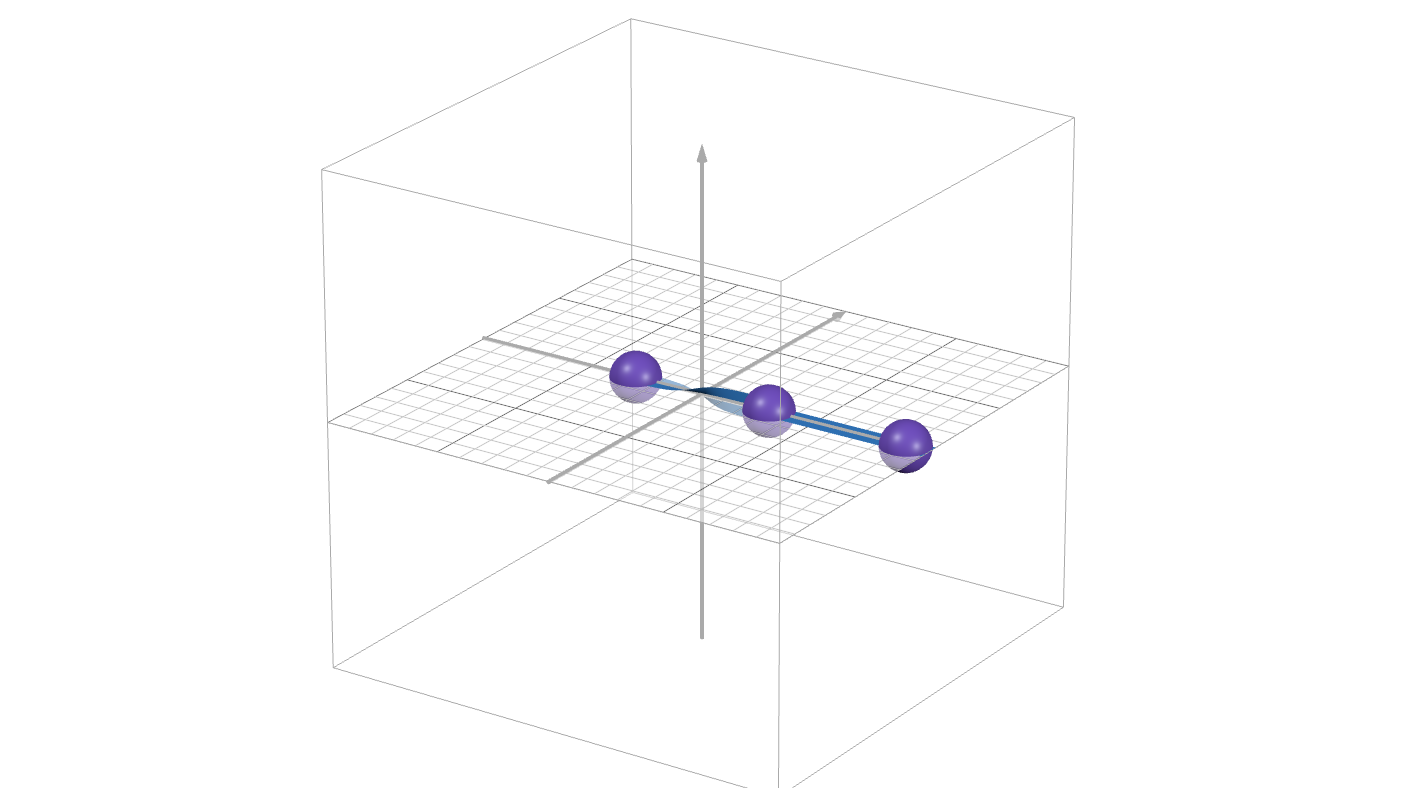}
        \label{fig:y_end_minusi_2}
    \end{subfigure}
    \caption{\textbf{Left-Handed Boundary Twist (Outcome $|-i\rangle_1$).} The boundary moves to $Q_2$. The ribbon $\mathcal{R}_{2,3}$ acquires a $-90^\circ$ counter-clockwise twist, visualized as the right edge crossing \textbf{UNDER} the left.}
    \label{fig:y_basis_lh_boundary_twist}
    \end{figure}    
\end{itemize}

\subsubsection{Measurement of bulk-qubit ($Q_k$)}
\begin{itemize}
    \item \textit{Outcome $|+i\rangle_k$
   Topology} \textsc{Right-Handed Splice ($\theta = +90^\circ$):} Neighbours $Q_{k-1}$ and $Q_{k+1}$ are fused. The connecting ribbon has a single continuous chiral $+90^\circ$ twist (Right-Over-Left), encoding the phase $+i$ (Fig. \ref{fig:y_basis_rh_twisted_splice}).
    % ==========================================
    % FIGURE: Y-BASIS BULK OUTCOME +i
    % ==========================================
    \begin{figure}[t]
    \centering
    \begin{subfigure}{0.3\textwidth}
        \centering
        \includegraphics[width=\linewidth]{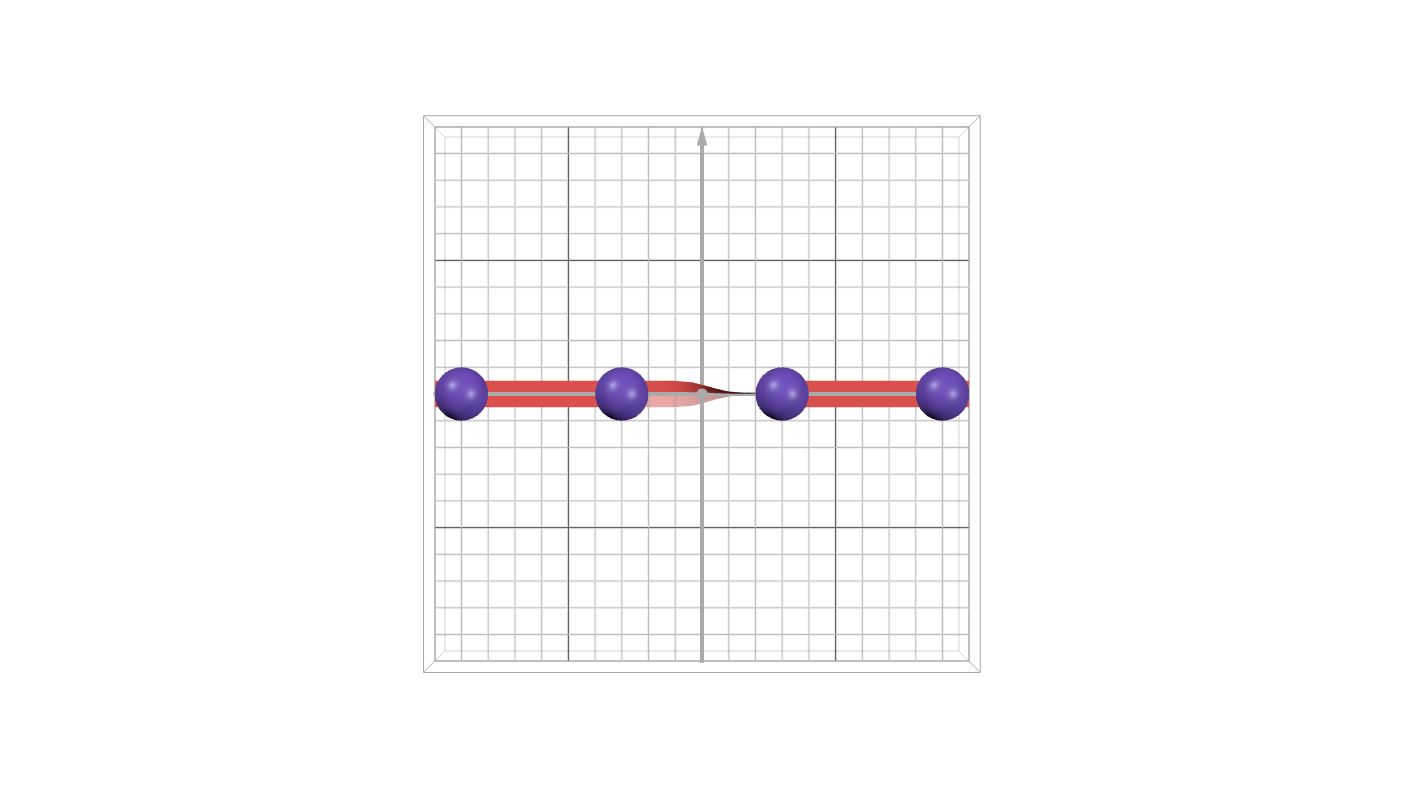}
        \label{fig:y_bulk_plusi_1}
    \end{subfigure}
    \hfill
    \begin{subfigure}{0.3\textwidth}
        \centering
        \includegraphics[width=\linewidth]{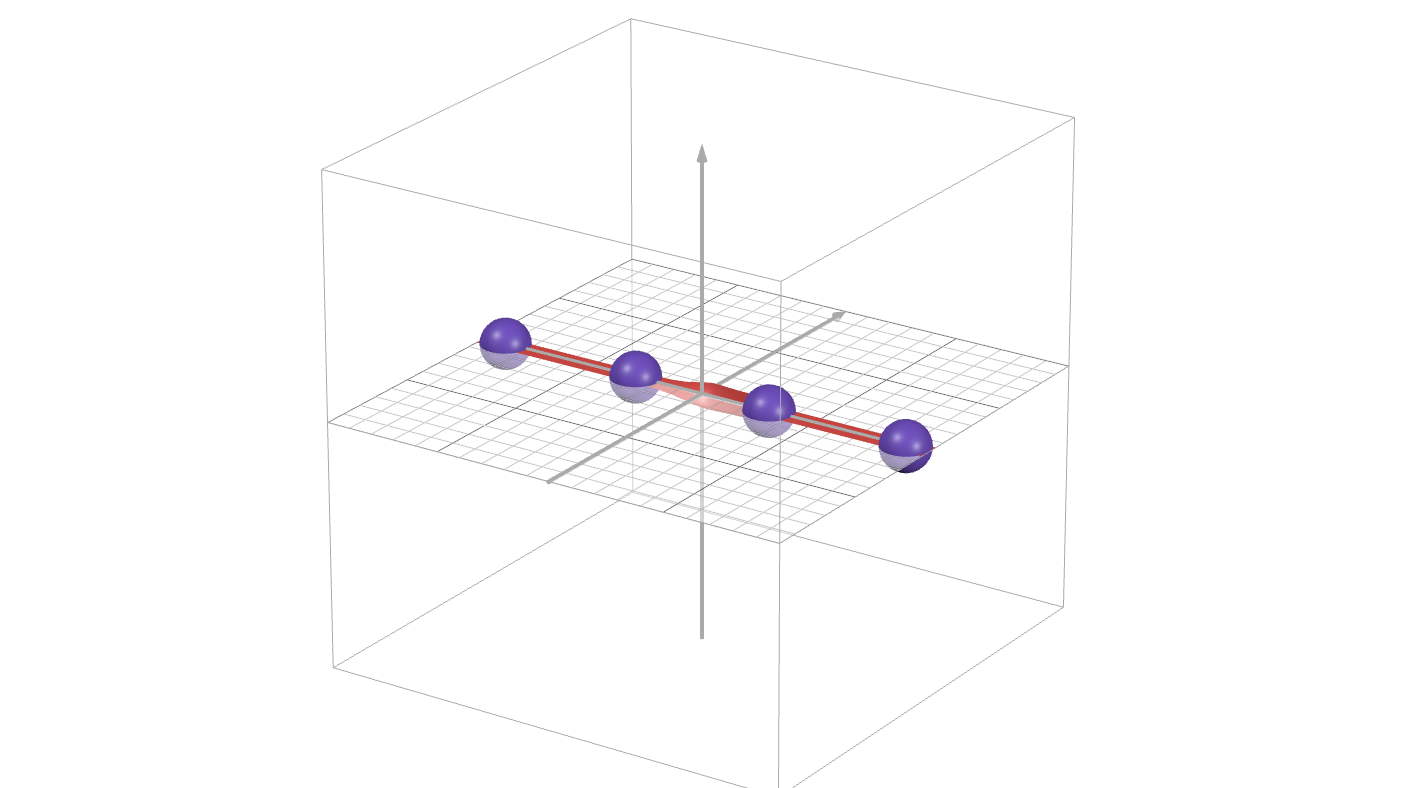}
        \label{fig:y_bulk_plusi_2}
    \end{subfigure}
    \caption{\textbf{Right-Handed Twisted Splice (Outcome $|+i\rangle_k$).} Neighbors $Q_{k-1}$ and $Q_{k+1}$ are fused into a single continuous chain. The connecting ribbon exhibits a $+90^\circ$ chiral twist, encoding the complex phase $+i$.}
    \label{fig:y_basis_rh_twisted_splice}
    \end{figure}       
    \item \textit{Outcome $|-i\rangle_k$
  Topology} \textsc{Left-Handed Splice ($\theta = -90^\circ$):} Neighbors are fused. The connecting ribbon has a single continuous chiral $-90^\circ$ twist (Right-Under-Left), encoding the phase $-i$, as depicted in Fig. \ref{fig:y_basis_lh_twisted_splice}.
    % ==========================================
    % FIGURE: Y-BASIS BULK OUTCOME -i
    % ==========================================
    \begin{figure}[t]
    \centering
    \begin{subfigure}{0.3\textwidth}
        \centering
        \includegraphics[width=\linewidth]{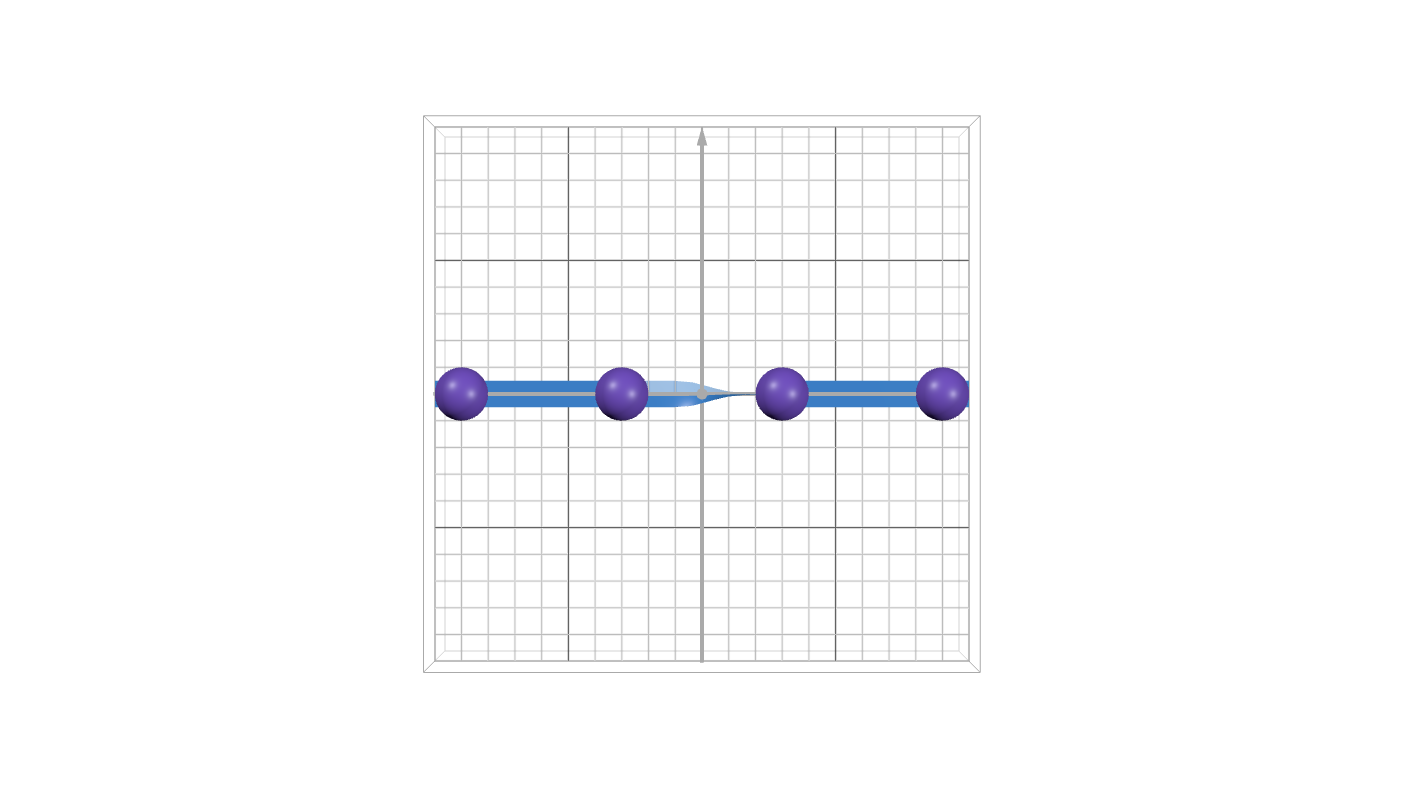}
        \label{fig:y_bulk_minusi_1}
    \end{subfigure}
    \hfill
    \begin{subfigure}{0.3\textwidth}
        \centering
        \includegraphics[width=\linewidth]{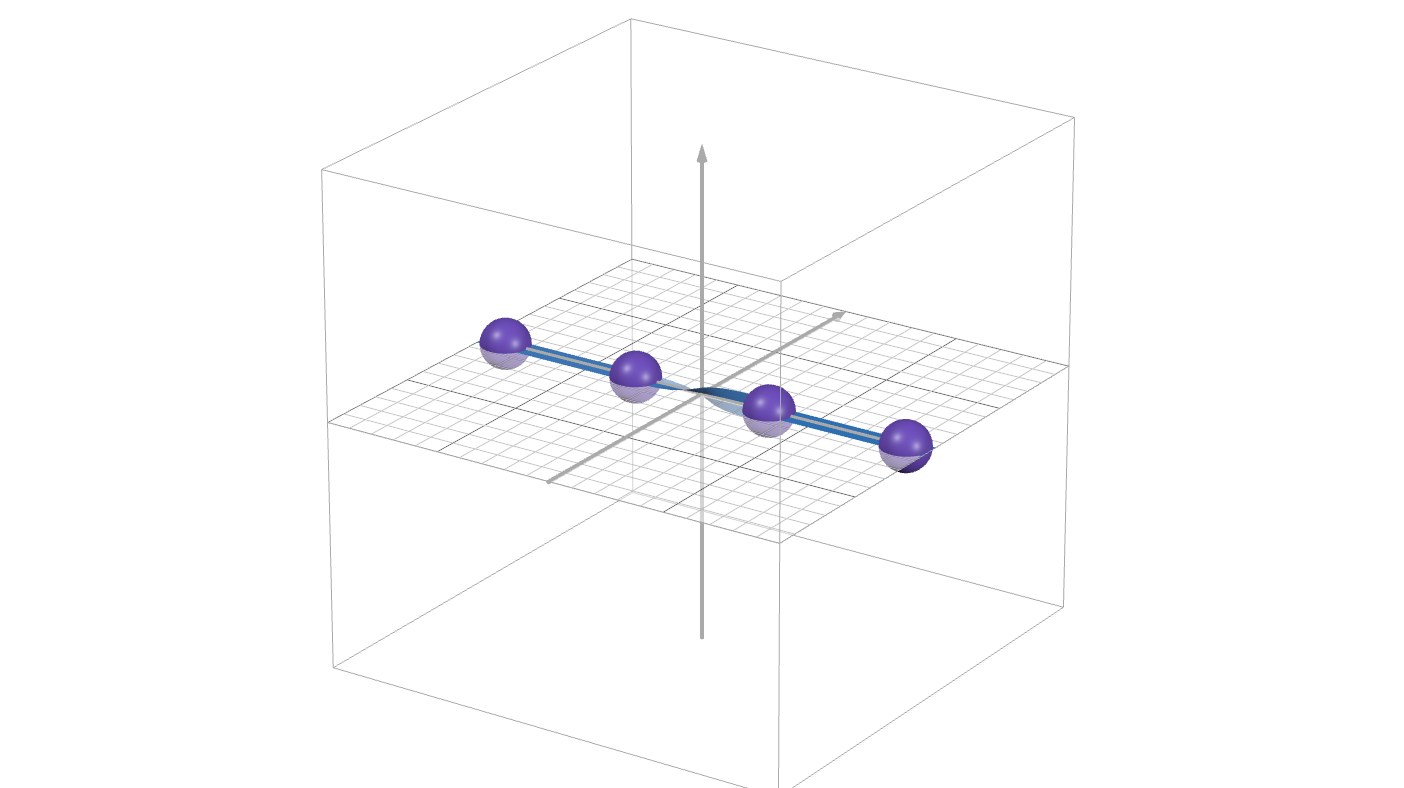}
        \label{fig:y_bulk_minusi_2}
    \end{subfigure}
    \caption{\textbf{Left-Handed Twisted Splice (Outcome $|-i\rangle_k$).} Neighbors are fused with a $-90^\circ$ chiral twist. The change in crossing (Under vs. Over) visually distinguishes this from the $|+i\rangle$ outcome.}
    \label{fig:y_basis_lh_twisted_splice}
    \end{figure}       
\end{itemize}
\section{Conclusion}
\noindent We have established a complete geometric classification of single-shot single-qubit Pauli measurements on one-dimensional linear cluster states. Our analysis reveals three fundamentally distinct measurement-induced transformations, as summarized in Table~\ref{tab:summary_classification}.

\begin{table*}[t]
\centering
\small
\renewcommand{\arraystretch}{1.4}
\begin{tabular}{@{}lcc@{}}
\toprule
\textbf{Basis} & \textbf{Topological Operation} & \textbf{Schmidt Rank (Bulk)} \\
\midrule
$Z$ & Severance & R=1 (disconnected) \\
$X$ & Parallel Stratification & R=2 (superposition of parallel tracks) \\
$Y$ & Twisted Splicing & R=2 (single continuous chain) \\
\bottomrule
\end{tabular}
\caption{\textbf{Synthesis of Measurement Classification.} The triality of measurement operations emerges from our topological analysis. Structural geometry clearly distinguishes all three bases: $Z$ severs the chain, $Y$ preserves 1D connectivity with a single chiral twist, and $X$ stratifies the state into a macroscopic superposition of parallel, non-intersecting tracks.}
\label{tab:summary_classification}
\end{table*}

\noindent This classification provides a geometric interpretation of measurement-based quantum computation (MBQC) through the \textit{measurement-cutting correspondence}, which maps local projective measurements to topological surgery operations on a Hopf chain representation of the cluster state.\\\\ 
\noindent We have shown three things: (i) $Z$-basis measurements act as \textit{topological severance}, physically disconnecting the chain (Schmidt rank $R=1$) and shortening boundaries without introducing complex phases. (ii) $X$-basis measurements induce \textit{topological stratification}. Rather than preserving a single continuous 1D path or forming an intersecting junction, the measurement acts as a quantum switch, layering the linear chain into parallel, non-intersecting tracks that represent a macroscopic superposition of highly correlated topological sectors. (iii) $Y$-basis measurements perform \textit{twisted splicing}. Uniquely, they maintain a single continuous 1D connectivity ($R=2$) while imprinting chiral complex phases ($\pm i$) that modulate the flow of quantum information. 

The central limitation of conventional topological models is their \textit{phase and superposition blindness}: while they faithfully capture general connectivity changes -- and, in particular, already distinguish the $Z$-, $X$-, and $Y$-basis mechanisms from one another by unframed shape (Sec.~IX) -- they cannot distinguish, \emph{within} a fixed basis, which of the two possible outcomes was actually recorded: e.g.\ the handedness of the induced twist for the two $Y$-basis outcomes, or the assignment of the Pauli-$Z$ correction between the two strata for the two $X$-basis outcomes. We resolved this ambiguity by introducing a \textit{framed ribbon model} that utilizes non-intersecting parallel geometries for superpositions, alongside geometric \textit{twists} to encode quantum phases: flat ribbons ($0^\circ$) for real positive correlations, \textit{flipped ribbons} ($180^\circ$) for real negative Pauli-$Z$ corrections, and \textit{chiral twists} ($\pm90^\circ$) for complex phases $\pm i$. This framing provides unambiguous geometric labels for all single-shot measurement outcomes and establishes a geometric language for phase-sensitive quantum correlations, subject to the caveat of Sec.~IX that these labels are operationally fixed rather than isotopy invariant.\\\\
\noindent \textit{Limitations.} Three restrictions should be stated plainly. First, the analysis covers single-shot measurements only. No composition rule for sequential measurements is established here, and no geometric analogue of classical feedforward is proposed; a naive additivity of twists under successive $Y$-basis measurements does not hold, because the second measurement acts on a state that already carries $S^{\pm1}$ dressings and the effective measurement basis is thereby rotated. We therefore make no claim to a unified interpretation of MBQC, and restrict the contribution to the classification of single-shot Pauli measurements on $1$D linear cluster states. Second, the twist angles are geometric labels, not topological invariants (Sec.~IX and Appendix~D). Third, framing encodes only the diagonal part of a by-product operator; a by-product with a non-diagonal Pauli component—as arises for the $X$-basis bulk outcome $|-\rangle_k$ when expressed relative to $|\Psi_+\rangle$ rather than in the graph frame $|G'\rangle$—is not representable by framing alone.\\\\
\noindent Several promising research directions emerge from this framework, including: (a) generalizing the framed ribbon model to 2D and 3D cluster states \cite{r2003}, which may reveal richer topological structures—potentially involving braided ribbons, multi-layered networks, or more complex framed surfaces—associated with multi-qubit measurements and universal MBQC patterns (\textit{Extension to Higher Dimensions}); (b) incorporating time ordering, classical feedforward, and adaptive measurement strategies into the topological picture, which could yield fully geometric descriptions of complete quantum circuits and potentially reveal new invariants related to computational depth or complexity (\textit{Dynamic and Adaptive Circuits}); (c) the relationship between ribbon framing, parallel tracking, local Clifford operations, and fault-tolerant logical gates, which suggests potential links with topological error correction \cite{kitaev2003,nayak2008} -- whether any genuine framed-link invariant, as opposed to the operational labels used here, can be attached to such schemes is an open question (\textit{Connections to Topological Quantum Computation}); (d) applying this framework to non-stabilizer resources (magic states, hypergraph states), which may provide geometric insights into contextuality, quantum advantage, and the role of complex phases in quantum computational supremacy (\textit{Beyond Stabilizer States}); and (e) developing software tools to visualize measurement sequences and stratified superposition tracks using our topological framework, which could enhance understanding of MBQC dynamics and provide intuitive educational resources (\textit{Experimental Visualization}).\\\\
\noindent In summary, our work proposes framed parallel ribbons as a physically transparent \emph{representation} of phase-sensitive quantum correlations generated by single-shot Pauli measurements on one-dimensional cluster states. By bridging quantum information theory with geometric topology, this approach offers new perspectives on measurement-based quantum computation and suggests that geometric methods may play an increasingly central role in understanding and designing quantum information protocols.
\vskip 0.5cm
{\bf Declaration of competing interest} The authors declare that they have no known competing financial interests or personal relationships that could have appeared to influence the work reported in this paper.
\vskip 0.5cm
{\bf Data availability statement} No external data is required for this research work.
%%%%%%%%%%%%%%%%%%%%%%%%%%%%%%%%%%%%%%APPENDIX%%%%%%%%%%%%%%%%%%%%%%%%%%%%%%%%%%%%%%%%%%%%%%%%%%%%%%%%%%%%%%%%%%%%%%%%%%%%%%%%
\newpage
\appendix
\phantomsection
\addcontentsline{toc}{section}{Appendices}

\section{$Z$-Basis Measurement on a 1D Cluster State}
\noindent We present a detailed derivation of projective measurements in the $Z$ basis for a 1D cluster state. End-qubit and bulk-qubit measurements are treated separately. All calculations are carried out using the computational-basis definition of the cluster state with explicit summation over binary variables.

\subsection{Definition of the Cluster State}
The $N$-qubit 1D cluster state is
\begin{equation}
\label{eq:cluster_def_Z}
|C_N\rangle = \frac{1}{2^{N/2}} \sum_{x_1,\dots,x_N \in \{0,1\}} (-1)^{\sum_{i=1}^{N-1} x_i x_{i+1}} |x_1 x_2 \cdots x_N\rangle,
\end{equation}
where all additions in the exponent are understood modulo $2$.

The $Z$-basis eigenstates are
\begin{equation}
|0\rangle, \qquad |1\rangle,
\end{equation}
with projectors
\begin{equation}
P_0 = |0\rangle\langle 0|, \qquad P_1 = |1\rangle\langle 1|.
\end{equation}

\subsection{Measurement on an end-qubit}
We first consider measurement of the end qubit $1$ in the $Z$ basis.

\subsubsection{Initial State and Decomposition}
Separate the $x_1$-dependent term in the cluster phase:
\begin{equation}
\sum_{i=1}^{N-1} x_i x_{i+1} = x_1 x_2 + \sum_{i=2}^{N-1} x_i x_{i+1}.
\end{equation}
Substituting into Eq.~\eqref{eq:cluster_def_Z},
\begin{align}
|C_N\rangle &= \frac{1}{2^{N/2}} \sum_{x_2,\dots,x_N} (-1)^{\sum_{i=2}^{N-1} x_i x_{i+1}} \nonumber\\
&\quad \times \sum_{x_1=0}^1 (-1)^{x_1 x_2} |x_1\rangle_1 |x_2\cdots x_N\rangle.
\end{align}

\subsubsection{Outcome $|0\rangle_1$}
Applying the projector $\langle 0 |_1$ fixes $x_1=0$:
\begin{align}
|\Psi_0\rangle &= \langle 0 |_1 |C_N\rangle \nonumber\\
&= \frac{1}{2^{N/2}} \sum_{x_2,\dots,x_N} (-1)^{\sum_{i=2}^{N-1} x_i x_{i+1}} |x_2\cdots x_N\rangle.
\end{align}
This is exactly the $(N-1)$-qubit cluster state:
\begin{equation}
|\Psi_0\rangle = |C_{N-1}\rangle.
\end{equation}

\subsubsection{Outcome $|1\rangle_1$}
Applying the projector $\langle 1 |_1$ fixes $x_1=1$:
\begin{align}
|\Psi_1\rangle &= \langle 1 |_1 |C_N\rangle \nonumber\\
&= \frac{1}{2^{N/2}} \sum_{x_2,\dots,x_N} (-1)^{x_2} (-1)^{\sum_{i=2}^{N-1} x_i x_{i+1}} |x_2\cdots x_N\rangle.
\end{align}
The factor $(-1)^{x_2}$ is exactly the action of a local $Z$ operator on qubit $2$. Thus,
\begin{equation}
|\Psi_1\rangle = Z_2 \, |C_{N-1}\rangle.
\end{equation}

\subsection{Measurement on a bulk-qubit}
We now consider measurement of a bulk qubit $k$ with neighbours $k-1$ and $k+1$.

\subsubsection{Initial State and Decomposition}
Split the quadratic phase as
\begin{equation}
\sum_{i=1}^{N-1} x_i x_{i+1} = \Phi_{\mathrm{rest}} + x_{k-1}x_k + x_k x_{k+1},
\end{equation}
where $\Phi_{\mathrm{rest}}$ contains all terms not involving $x_k$. Then,
\begin{align}
|C_N\rangle &= \frac{1}{2^{N/2}} \sum_{\mathbf{x}_{\neq k}} (-1)^{\Phi_{\mathrm{rest}}} \nonumber\\
&\quad \times \sum_{x_k=0}^1 (-1)^{x_k(x_{k-1}+x_{k+1})} |x_k\rangle_k |\mathbf{x}_{\neq k}\rangle.
\end{align}

\subsubsection{Outcome $|0\rangle_k$}
Projecting onto $|0\rangle_k$ sets $x_k=0$:
\begin{align}
|\Psi_0\rangle &= \langle 0 |_k |C_N\rangle \nonumber\\
&= \frac{1}{2^{N/2}} \sum_{\mathbf{x}_{\neq k}} (-1)^{\Phi_{\mathrm{rest}}} |\mathbf{x}_{\neq k}\rangle.
\end{align}
This corresponds to two disconnected cluster states on the left and right of site $k$:
\begin{equation}
|\Psi_0\rangle = |C_{k-1}\rangle_L \otimes |C_{N-k}\rangle_R.
\end{equation}

\subsubsection{Outcome $|1\rangle_k$}
Projecting onto $|1\rangle_k$ sets $x_k=1$:
\begin{align}
|\Psi_1\rangle &= \langle 1 |_k |C_N\rangle \nonumber\\
&= \frac{1}{2^{N/2}} \sum_{\mathbf{x}_{\neq k}} (-1)^{x_{k-1}+x_{k+1}} (-1)^{\Phi_{\mathrm{rest}}} |\mathbf{x}_{\neq k}\rangle.
\end{align}
Using the relation
\begin{equation}
(-1)^{x_{k-1}+x_{k+1}} = (-1)^{x_{k-1}}(-1)^{x_{k+1}},
\end{equation}
this factor corresponds to local $Z$ operators acting on both neighbours. Hence,
\begin{equation}
|\Psi_1\rangle = (Z_{k-1} \otimes Z_{k+1}) \, \big( |C_{k-1}\rangle_L \otimes |C_{N-k}\rangle_R \big).
\end{equation}

\newpage
\section{$X$-Basis Measurement on a 1D Cluster State}
\noindent We present a detailed derivation of projective measurement in the $X$ basis for a 1D cluster state, treating end-qubit and bulk-qubit measurements separately. All calculations are performed using the computational-basis definition of the cluster state and explicit summation over binary variables.

\subsection{Definition of the Cluster State}
The $N$-qubit 1D cluster state is defined as
\begin{equation}
\label{eq:cluster_def}
|C_N\rangle = \frac{1}{2^{N/2}} \sum_{x_1,\dots,x_N \in \{0,1\}} (-1)^{\sum_{i=1}^{N-1} x_i x_{i+1}} |x_1 x_2 \cdots x_N\rangle,
\end{equation}
where all additions in the exponent are understood modulo $2$.

The $X$-basis eigenstates are
\begin{equation}
|+\rangle = \frac{|0\rangle + |1\rangle}{\sqrt{2}}, \qquad |-\rangle = \frac{|0\rangle - |1\rangle}{\sqrt{2}},
\end{equation}
with overlaps
\begin{equation}
\langle \pm | x \rangle = \frac{1}{\sqrt{2}}(\pm 1)^x, \qquad x\in\{0,1\}.
\end{equation}

\subsection{Measurement on an end-qubit}
We first consider measurement of the end qubit $1$ in the $X$ basis.

\subsubsection{Initial State and Decomposition}
Separating the $x_1$-dependent term in the cluster phase,
\begin{equation}
\sum_{i=1}^{N-1} x_i x_{i+1} = x_1 x_2 + \sum_{i=2}^{N-1} x_i x_{i+1}.
\end{equation}
Substituting into Eq.~\eqref{eq:cluster_def},
\begin{align}
|C_N\rangle &= \frac{1}{2^{N/2}} \sum_{x_2,\dots,x_N} (-1)^{\sum_{i=2}^{N-1} x_i x_{i+1}} \nonumber\\
&\quad \times \sum_{x_1=0}^1 (-1)^{x_1 x_2} |x_1\rangle_1 |x_2\cdots x_N\rangle.
\end{align}

\subsubsection{Outcome $|+\rangle_1$}
Applying the projector $\langle + |_1$,
\begin{align}
|\Psi_+\rangle &= \langle + |_1 |C_N\rangle \nonumber\\
&= \frac{1}{2^{(N+1)/2}} \sum_{x_2,\dots,x_N} (-1)^{\sum_{i=2}^{N-1} x_i x_{i+1}} \nonumber\\
&\quad \times \sum_{x_1=0}^1 (+1)^{x_1} (-1)^{x_1 x_2} |x_2\cdots x_N\rangle.
\end{align}
Evaluate the inner sum explicitly:
\begin{equation}
\sum_{x_1=0}^1 (+1)^{x_1}(-1)^{x_1 x_2} = \begin{cases} 2, & x_2=0 \\ 0, & x_2=1 \end{cases}
\end{equation}
Thus, the outcome $|+\rangle_1$ pins $x_2=0$, and the surviving phase no longer contains the bond $x_2x_3$:
\begin{equation}
\label{eq:B8}
|\Psi_+\rangle \propto \sum_{x_3,\dots,x_N} (-1)^{\sum_{i=3}^{N-1} x_i x_{i+1}} |0\,x_3\cdots x_N\rangle.
\end{equation}
Equation~\eqref{eq:B8} fixes qubit $2$ in the product state $|0\rangle$, so the residual state is \emph{not} an $(N-1)$-qubit cluster state on qubits $2,\dots,N$: qubit $2$ separates from the entangled subsystem of qubits $3,\dots,N$,
\begin{equation}
\label{eq:B9new}
|\Psi_+\rangle = |0\rangle_2 \otimes |C_{N-2}\rangle_{3\dots N}.
\end{equation}

\subsubsection{Outcome $|-\rangle_1$}
Applying the projector $\langle - |_1$,
\begin{align}
|\Psi_-\rangle &= \langle - |_1 |C_N\rangle \nonumber\\
&= \frac{1}{2^{(N+1)/2}} \sum_{x_2,\dots,x_N} (-1)^{\sum_{i=2}^{N-1} x_i x_{i+1}} \nonumber\\
&\quad \times \sum_{x_1=0}^1 (-1)^{x_1} (-1)^{x_1 x_2} |x_2\cdots x_N\rangle.
\end{align}
The inner sum gives
\begin{equation}
\sum_{x_1=0}^1 (-1)^{x_1}(-1)^{x_1 x_2} = \begin{cases} 0, & x_2=0 \\ 2, & x_2=1 \end{cases}
\end{equation}
Hence $x_2=1$ is pinned, and the bond $x_2x_3$ contributes a residual factor $(-1)^{x_3}$:
\begin{align}
\label{eq:B12}
|\Psi_-\rangle &\propto \sum_{x_3,\dots,x_N} (-1)^{x_3}(-1)^{\sum_{i=3}^{N-1} x_i x_{i+1}} |1\,x_3\cdots x_N\rangle.
\end{align}
By the same argument as above, qubit $2$ is fixed in $|1\rangle$ and the factor $(-1)^{x_3}$ is a local $Z$ on qubit $3$:
\begin{equation}
\label{eq:B13new}
|\Psi_-\rangle = |1\rangle_2 \otimes Z_3\,|C_{N-2}\rangle_{3\dots N}.
\end{equation}

\subsection{Measurement on a bulk-qubit}
We now measure a bulk qubit $k$ with neighbours $k-1$ and $k+1$.

\subsubsection{Initial State and Decomposition}
To expose the boundary dependencies explicitly, we split the quadratic phase into the segments not touching $k-1, k, k+1$, and the cross-terms that connect them to the measurement site:
\begin{align}
\sum_{i=1}^{N-1} x_i x_{i+1} &= \Phi_L + \Phi_R + x_{k-2}x_{k-1} \nonumber\\
&\quad + x_{k-1}x_k + x_k x_{k+1} + x_{k+1}x_{k+2},
\end{align}
where $\Phi_L = \sum_{i=1}^{k-3} x_i x_{i+1}$ is the phase of the left segment (qubits $1\dots k-2$) and $\Phi_R = \sum_{i=k+2}^{N-1} x_i x_{i+1}$ is the phase of the right segment (qubits $k+2\dots N$).

The cluster state decomposes as:
\begin{align}
|C_N\rangle &= \frac{1}{2^{N/2}} \sum_{\mathbf{x}_{\neq k}} (-1)^{\Phi_L + \Phi_R + x_{k-2}x_{k-1} + x_{k+1}x_{k+2}} \nonumber\\
&\quad \times \sum_{x_k=0}^1 (-1)^{x_k(x_{k-1}\oplus x_{k+1})} |x_k\rangle_k |\mathbf{x}_{\neq k}\rangle.
\end{align}

\subsubsection{Outcome $|+\rangle_k$}
Applying $\langle + |_k$, the inner sum over $x_k$ evaluates to:
\begin{equation}
\sum_{x_k=0}^1 (+1)^{x_k} (-1)^{x_k(x_{k-1}\oplus x_{k+1})} = 2\,\delta_{x_{k-1}, x_{k+1}}.
\end{equation}
This is a stringent selection rule, not a phase restoration. It outright eliminates configurations where $x_{k-1} \neq x_{k+1}$. Setting $x_{k-1} = x_{k+1} \equiv b$, the remaining cross-terms $x_{k-2}x_{k-1} = x_{k-2}b$ and $x_{k+1}x_{k+2} = b x_{k+2}$ survive.

The residual state naturally branches based on the shared binary variable $b$:
\begin{align}
|\Psi_+\rangle &\propto \big(|C_{k-2}\rangle_{1..k-2} \otimes |0\rangle_{k-1}\big) \nonumber\\
&\qquad \otimes \big(|0\rangle_{k+1} \otimes |C_{N-k-1}\rangle_{k+2..N}\big) \nonumber\\
&\quad + \big(Z_{k-2}|C_{k-2}\rangle_{1..k-2} \otimes |1\rangle_{k-1}\big) \nonumber\\
&\qquad \otimes \big(|1\rangle_{k+1} \otimes Z_{k+2}|C_{N-k-1}\rangle_{k+2..N}\big).
\end{align}

Normalizing the state yields a macroscopic branching structure controlled by the completely correlated neighbours:
\begin{equation}
|\Psi_+\rangle = \frac{1}{\sqrt{2}} \Big[ |0\rangle_{k-1}|0\rangle_{k+1}|\chi_0\rangle + |1\rangle_{k-1}|1\rangle_{k+1}|\chi_1\rangle \Big],
\end{equation}
where the conditional topological sectors are given by:
\begin{align}
|\chi_0\rangle &= |C_{k-2}\rangle_{1..k-2} \otimes |C_{N-k-1}\rangle_{k+2..N}, \\
|\chi_1\rangle &= Z_{k-2}|C_{k-2}\rangle_{1..k-2} \otimes Z_{k+2}|C_{N-k-1}\rangle_{k+2..N}.
\end{align}
This correlation between the boundaries ($b=0$ and $b=1$) provides the direct algebraic basis for representing the $X$-measurement as a geometric stratification into two parallel branches.

\subsubsection{Outcome $|-\rangle_k$}
Applying $\langle - |_k$, the inner sum evaluates to:
\begin{equation}
\sum_{x_k=0}^1 (-1)^{x_k} (-1)^{x_k(x_{k-1}\oplus x_{k+1})} = 2\,\delta_{x_{k-1} \neq x_{k+1}}.
\end{equation}
The outcome $|-\rangle_k$ selects the anti-correlated parity sector. Setting $x_{k-1} \equiv b$ forces $x_{k+1} = 1-b \equiv \bar{b}$. The surviving cross-terms become $x_{k-2}b$ and $x_{k+2}\bar{b}$.

Following the same decomposition, the state branches into the opposite parity configurations:
\begin{align}
|\Psi_-\rangle &= \frac{1}{\sqrt{2}} \Big[ |0\rangle_{k-1}|1\rangle_{k+1} \big(|C_{k-2}\rangle \otimes Z_{k+2}|C_{N-k-1}\rangle \big) \nonumber\\
&\qquad + |1\rangle_{k-1}|0\rangle_{k+1} \big(Z_{k-2}|C_{k-2}\rangle \otimes |C_{N-k-1}\rangle\big) \Big],
\end{align}
where the subsystem labels have been omitted for brevity. As with the positive parity outcome, the state is a highly correlated superposition of two macroscopically distinct topological sectors, confirming that the bulk $X$-measurement induces a geometric stratification into two parallel, non-intersecting branches rather than a simple 1D splice.

\newpage
\section{$Y$-Basis Measurement on a 1D Cluster State}
\noindent We present a detailed derivation of projective measurement in the $Y$ basis for a 1D cluster state, treating end-qubit and bulk-qubit measurements separately. All calculations use the computational-basis definition of the cluster state and explicit summation over binary variables.

\subsection{Definition of the Cluster State}
The $N$-qubit 1D cluster state is
\begin{equation}
\label{eq:cluster_def_Y}
|C_N\rangle = \frac{1}{2^{N/2}} \sum_{x_1,\dots,x_N\in\{0,1\}} (-1)^{\sum_{i=1}^{N-1} x_i x_{i+1}} |x_1 x_2 \cdots x_N\rangle,
\end{equation}
where all sums in the exponent are taken modulo $2$.

The $Y$-basis eigenstates are
\begin{equation}
|y_\pm\rangle = \frac{|0\rangle \pm i|1\rangle}{\sqrt{2}}.
\end{equation}
Note that the bra is the conjugate of the ket, $\langle y_\pm| = (\langle 0| \mp i\langle 1|)/\sqrt{2}$, so the overlap carries the \emph{opposite} sign:
\begin{equation}
\label{eq:C3new}
\langle y_\pm | x \rangle = \frac{1}{\sqrt{2}} (\mp i)^x, \qquad x\in\{0,1\}.
\end{equation}
This sign convention is used consistently in all the steps below.

\subsection{Measurement on an end-qubit}
We first consider measurement of the end qubit $1$ in the $Y$ basis.

\subsubsection{Initial State and Decomposition}
Separating the $x_1$-dependent term,
\begin{equation}
\sum_{i=1}^{N-1} x_i x_{i+1} = x_1 x_2 + \sum_{i=2}^{N-1} x_i x_{i+1}.
\end{equation}
Thus,
\begin{align}
|C_N\rangle &= \frac{1}{2^{N/2}} \sum_{x_2,\dots,x_N} (-1)^{\sum_{i=2}^{N-1} x_i x_{i+1}} \nonumber\\
&\quad \times \sum_{x_1=0}^1 (-1)^{x_1 x_2} |x_1\rangle_1 |x_2\cdots x_N\rangle.
\end{align}

\subsubsection{Outcome $|y_+\rangle_1$}
Applying the projector $\langle y_+ |_1$,
\begin{align}
|\Psi_+\rangle &= \langle y_+ |_1 |C_N\rangle \nonumber\\
&= \frac{1}{2^{(N+1)/2}} \sum_{x_2,\dots,x_N} (-1)^{\sum_{i=2}^{N-1} x_i x_{i+1}} \nonumber\\
&\quad \times \sum_{x_1=0}^1 (-i)^{x_1} (-1)^{x_1 x_2} |x_2\cdots x_N\rangle.
\end{align}
The inner sum evaluates to
\begin{equation}
\sum_{x_1=0}^1 (-i)^{x_1} (-1)^{x_1 x_2} = \begin{cases} 1 - i, & x_2 = 0 \\ 1 + i, & x_2 = 1 \end{cases}
\end{equation}
Using the identity $1-i(-1)^{a} = \sqrt{2}\,e^{-i\pi/4}\, i^{\,a}$, which holds for both $a=0$ and $a=1$, the inner sum equals $\sqrt{2}\,e^{-i\pi/4}\, i^{\,x_2}$. Factoring out the global phase $\sqrt{2}e^{-i\pi/4}$ and renormalising, the residual factor $i^{\,x_2}$ is a local $S$ gate on qubit $2$. Hence,
\begin{equation}
|\Psi_+\rangle = S_2 \, |C_{N-1}\rangle.
\end{equation}

\subsubsection{Outcome $|y_-\rangle_1$}
Applying the projector $\langle y_- |_1$,
\begin{align}
|\Psi_-\rangle &= \langle y_- |_1 |C_N\rangle \nonumber\\
&= \frac{1}{2^{(N+1)/2}} \sum_{x_2,\dots,x_N} (-1)^{\sum_{i=2}^{N-1} x_i x_{i+1}} \nonumber\\
&\quad \times \sum_{x_1=0}^1 (i)^{x_1} (-1)^{x_1 x_2} |x_2\cdots x_N\rangle.
\end{align}
The inner sum gives
\begin{equation}
\sum_{x_1=0}^1 (i)^{x_1} (-1)^{x_1 x_2} = \begin{cases} 1 + i, & x_2 = 0 \\ 1 - i, & x_2 = 1 \end{cases}
\end{equation}
Using $1+i(-1)^{a} = \sqrt{2}\,e^{i\pi/4}\,(-i)^{\,a}$, valid for $a=0,1$, the inner sum equals $\sqrt{2}\,e^{i\pi/4}(-i)^{\,x_2}$. After normalization, this corresponds to an $S^\dagger$ gate on qubit $2$:
\begin{equation}
|\Psi_-\rangle = S_2^\dagger \, |C_{N-1}\rangle.
\end{equation}

\subsection{Measurement on a bulk-qubit}
We now consider measurement of a bulk qubit $k$ with neighbours $k-1$ and $k+1$.

\subsubsection{Initial State and Decomposition}
Split the quadratic phase:
\begin{equation}
\sum_{i=1}^{N-1} x_i x_{i+1} = \Phi_{\mathrm{rest}} + x_{k-1}x_k + x_k x_{k+1},
\end{equation}
where $\Phi_{\mathrm{rest}}$ excludes all terms involving $x_k$. Then,
\begin{align}
|C_N\rangle &= \frac{1}{2^{N/2}} \sum_{\mathbf{x}_{\neq k}} (-1)^{\Phi_{\mathrm{rest}}} \nonumber\\
&\quad \times \sum_{x_k=0}^1 (-1)^{x_k(x_{k-1}\oplus x_{k+1})} |x_k\rangle_k |\mathbf{x}_{\neq k}\rangle.
\end{align}

\subsubsection{Outcome $|y_+\rangle_k$}
Applying $\langle y_+ |_k$,
\begin{align}
|\Psi_+\rangle &= \langle y_+ |_k |C_N\rangle \nonumber\\
&= \frac{1}{2^{(N+1)/2}} \sum_{\mathbf{x}_{\neq k}} (-1)^{\Phi_{\mathrm{rest}}} \nonumber\\
&\quad \times \sum_{x_k=0}^1 (-i)^{x_k} (-1)^{x_k(x_{k-1}\oplus x_{k+1})} |\mathbf{x}_{\neq k}\rangle.
\end{align}
The inner sum evaluates to
\begin{equation}
\label{eq:C15}
 1 - i(-1)^{x_{k-1}\oplus x_{k+1}}.
\end{equation}
The identity
\begin{equation}
1 - i(-1)^a = \sqrt{2}\, e^{-i\pi/4}\, i^{\,a},\qquad a\in\{0,1\},
\end{equation}
which is readily checked at both values of $a$. Writing $a = x_{k-1}\oplus x_{k+1}$ and using the exclusive-or splitting identity
\begin{equation}
i^{\,x\oplus y} = i^{\,x}\, i^{\,y}\, (-1)^{xy},\qquad x,y\in\{0,1\},
\end{equation}
the factor $i^{\,x_{k-1}\oplus x_{k+1}}$ separates into local phases $i^{\,x_{k-1}}$ and $i^{\,x_{k+1}}$, i.e., local $S$ gates on both neighbours, together with the genuine quadratic term $(-1)^{x_{k-1}x_{k+1}}$, which is the new $CZ$ bond $k-1\leftrightarrow k+1$. Unlike the $X$-basis case of Appendix~B.3, no parity constraint is imposed here, so the splice is exact.

After normalization,
\begin{equation}
|\Psi_+\rangle = (S_{k-1} \otimes S_{k+1}) \,|C_{N-1}^{(k-1\leftrightarrow k+1)}\rangle.
\end{equation}

\subsubsection{Outcome $|y_-\rangle_k$}
Applying $\langle y_- |_k$,
\begin{align}
|\Psi_-\rangle &= \langle y_- |_k |C_N\rangle \nonumber\\
&= \frac{1}{2^{(N+1)/2}} \sum_{\mathbf{x}_{\neq k}} (-1)^{\Phi_{\mathrm{rest}}} \nonumber\\
&\quad \times \sum_{x_k=0}^1 (i)^{x_k} (-1)^{x_k(x_{k-1}\oplus x_{k+1})} |\mathbf{x}_{\neq k}\rangle.
\end{align}
The inner sum gives
\begin{equation}
\label{eq:C18}
1 + i(-1)^{x_{k-1}\oplus x_{k+1}} = \sqrt{2}\,e^{i\pi/4}\,(-i)^{\,x_{k-1}\oplus x_{k+1}}.
\end{equation}
With $(-i)^{\,x\oplus y} = (-i)^{x}(-i)^{y}(-1)^{xy}$, this produces the inverse local phases, corresponding to $S^\dagger$ gates, together with the same $CZ$ bond $k-1\leftrightarrow k+1$:
\begin{equation}
|\Psi_-\rangle = (S_{k-1}^\dagger \otimes S_{k+1}^\dagger) \,|C_{N-1}^{(k-1\leftrightarrow k+1)}\rangle.
\end{equation}

\section{Framed ribbon and chirality}
\noindent A framed ribbon $\mathcal{R}_{ij}$ connecting qubits $Q_i$ and $Q_j$ is an oriented, embedded strip $I\times [-\epsilon,\epsilon] $ in $\mathbb{R}^3$ (Euclidean 3D space). Here $I$ parametrizes the core curve from $Q_i$ to $Q_j$, and the $[-\epsilon,\epsilon]$ fibers define the framing (a continuous choice of normal direction along the core). The framing is equivalently specified by a unit normal vector field $n(s)$ along the core curve $\gamma(s)$, $s\in [0,1]$. The ribbon's two edges are $\gamma(s) \pm \epsilon\: n(s)$. The twist angle $\theta \in \mathbb{R}$ is defined as the total accumulated rotation of the framing vector $n(s)$ about the tangent direction $t(s)$ as the ribbon is traversed from $Q_i$ to $Q_j$. For a ribbon lying in a plane without intrinsic torsion, $\theta$ is simply the net rotation angle of the ribbon's cross-section. In this work we have considered only discrete, geometrically stabilized framing corresponding to measurement outcomes $\theta \in \lbrace 0^\circ, \pm 90^\circ, 180^\circ\rbrace \ (\mathrm{mod}\ 360^\circ)$. These are explicitly not topological invariants (as they are not preserved under all continuous deformations, and quarter twists are not integer framings), but rather geometric labels that are operationally fixed by the quantum measurement protocol and by the by-product operator it induces. This point is stated in the abstract, in Sec.~IX and in the conclusion. This quantization of twist reflects the discrete nature of measurement outcomes in quantum mechanics. The chirality of a $90^\circ$ twist is defined with respect to a fixed orientation of the cluster chain (from left to right). A right-handed $+90^\circ$ twist corresponds to the right edge of the ribbon crossing over the left edge when viewed from above; a left-handed $-90^\circ$ twist corresponds to the right edge crossing under the left edge.
%\subsection{Summary}
%
%\begin{itemize}
%\item Y-basis measurement deletes the measured qubit.
%\item End-qubit measurement shortens the chain and induces a local $S$ or $S^\dagger$.
%\item Bulk-qubit measurement splices the chain and induces $S^{\pm1}$ gates on neighbours.
%\item The two outcomes differ by complex conjugation.
%\item Y-measurement introduces local chirality absent in X-basis measurement.
%\end{itemize}

\end{document}